\documentclass[a4paper,fleqn,usenatbib]{mnras}
\usepackage{newtxtext,newtxmath}

\usepackage[T1]{fontenc}
\usepackage{ae,aecompl}
\usepackage{xcolor}
\usepackage[normalem]{ulem}		

\usepackage{graphicx}	 
\usepackage{amsmath} 

\usepackage{amssymb} 
\usepackage{soul} 
\usepackage[bottom]{footmisc}

\usepackage{hyperref}

\title[PIC jet simulations]{3D PIC Simulations for Relativistic Jets with a Toroidal Magnetic Field}

\author[Meli, A. et al.]{Athina Meli,$^{1,2}$\thanks{E-mail: ameli@ncat.edu}
Kenichi Nishikawa,$^{3}$
Christoph K\"ohn,$^{4}$ 
Ioana Du\c{t}an,$^{5}$
Yosuke Mizuno,$^{6,7}$
\newauthor Oleh Kobzar,$^{8}$
Nicholas MacDonald,$^{9}$
Jos\'e L. G\'omez$^{10}$  and Kouichi Hirotani$^{11}$
\\
$^{1}$College of Science and Technology, North Carolina A\&T State University, North Carolina, NC 27411, USA\\
$^{2}$Space Sciences \& Technologies for Astrophysics Research (STAR) Institute Universite 
de Liege, Sart Tilman, 4000 Li{\'{e}}ge, Belgium\\
$^{3}$Department of Physics, Chemistry and Mathematics, Alabama A\&M University, Normal, 
AL 35762, USA\\
$^{4}$Technical University of Denmark, National Space Institute (DTU Space), Elektrovej 328, 
2800 Kgs Lyngby, Denmark\\
$^{5}$Institute of Space Science, Atomi\c{s}tilor 409, RO-077125 Bucharest-M\u{a}gurele, Romania \\
$^{6}$Institute for Theoretical Physics, Goethe University, D-60438 Frankfurt am Main, Germany\\
$^{7}$Tsung-Dao Lee Institute, Shanghai Jiao Tong University, Shanghai, 200240, China \\
$^{8}$Faculty of Materials Engineering and Physics, Cracow University of Technology, PL-30084 Krakow, Poland\\
$^{9}$Max-Planck-Institut f\"ur Radioastronomie, Auf dem H\"ugel 69, D-53121 Bonn, Germany\\
$^{10}$Instituto de Astrof\'isica de Andaluc\'ia, CSIC, Apartado 3004, 18080 Granada, Spain\\
$^{11}$Taiwan Institute of Astronomy and Astrophysics, Academia Sinica, Taipei 10617, Taiwan, Republic of China\\
}

\date{Accepted  Received ; in original form }

\pubyear{2021}

\usepackage{hyperref}

\begin{document}
\label{firstpage}
\pagerange{\pageref{firstpage}--\pageref{lastpage}}
\maketitle

\begin{abstract}

{\rm We have investigated how kinetic instabilities such as the Weibel instability (WI), the mushroom instability (MI), and the kinetic Kelvin-Helmholtz instability (kKHI) are excited in jets without and with a toroidal magnetic field, and how such instabilities contribute to particle acceleration. In this work we use a new jet injection scheme where an electric current is self-consistently generated at the jet orifice by the jet particles which produce the toroidal magnetic field. We perform five different simulations for a sufficiently long time to examine the non-linear effects of the jet evolution. We inject unmagnetized e$^{\pm}$ and e$^{-}$ - p$^{+}$ (mp/me = 1836), as well as magnetized e$^{\pm}$ and  
e$^{-}$ - i$^{+}$ (mi/me = 4) jets with a top-hat jet density profile into an unmagnetized ambient plasmas of the same species. We show that WI, MI, and kKHI excited at the linear stage, generate a non-oscillatory x-component of the electric field accelerating and decelerating electrons. We find that the two different jet compositions (e$^{\pm}$ and  e$^{-}$ - i$^{+}$)  display different instability modes respectively. Moreover, the magnetic field in the non-linear stage generated by different instabilities is dissipated and reorganized into new topologies. A 3D magnetic field topology depiction indicates possible reconnection sites in the non-linear stage where the particles are significantly accelerated by the dissipation of the magnetic field associated to a possible reconnection event.}

\end{abstract}

\begin{keywords}
acceleration of particles,  relativistic processes, instabilities, shock waves, galaxies: jets
\end{keywords}



\section{Introduction}

Relativistic astrophysical jets are ubiquitous in astrophysical systems. Most collimated relativistic jets 
extend between several thousands up to millions of parsecs \cite[e.g.,][]{blandford2019} and have been 
observationally associated with the activity of central black holes in Active Galactic Nuclei 
\cite[AGN, e.g.][] {EHT19,EHT-3C279}. It is also theorized that relativistic jets may also occur in Gamma-ray Bursts (GRBs) \cite[e.g,][]{ruiz2018}. 
The formation and powering of these astrophysical jets are highly complex phenomena involving 
relativistic plasmas and twisted magnetic fields which are organized in such a manner as to ultimately launch an outflow from a central
compact source.

For almost two decades, Particle-in-Cell (PIC) models have been used to study unmagnetized and magnetized relativistic jets that 
interact with the interstellar medium. These investigations have offered tremendous insight into the instabilities, turbulence, 
and shocks that can develop in the out-flowing plasma, leading to particle acceleration and the production of nonthermal radiation
\cite[e.g.,][]{silva2003,nishikawa2003,jaroschek2005,spitkovsky2008a,spitkovsky2008b, 
dieckmann2008,nishikawa2009,giannios2009,Pino10,uzdensky2011,granot2012,mcKinney2012,sironi2013,
sironi2015,Ardaneh16,Kadowaki18,Kadowaki19,Christie19,Fowler19}.

Relativistic jets interact with the plasma environment of an astrophysical source and subsequently 
instabilities occur which are responsible for the acceleration of particles \cite[e.g.,][]{nishikawa2021}.
Beam-plasma instabilities are a key physical process in
many astrophysical phenomena, therefore extensive investigations
of various instabilities have been performed (e.g., Bret 2009;
Bret et al. 2010). In the context of this work we discuss possible
instabilities in the simulation system used and we note that the
filamentation instability and the two-stream instability are not
the same instability. The non-resonant filamentation instability
is mostly magnetic and drives waves with wavevectors that
are almost perpendicular to the plasma flow direction. The current
filaments are seen in the early linear stage as shown in the
movies provided, and in particular in the unmagnetized cases.
The two-stream instability or Buneman instability is a resonant
electrostatic instability, which triggers the growth of wavevectors
that are almost parallel to the plasma flow direction. Both
instabilities have growth rates that scale quite differently with
the Lorentz factor of the flow velocity (e.g., Bret et al. 2010). For
the magnetized cases, different modes such as oblique and Bell’s
ones can be excited (e.g., Bret 2009; Bret et al. 2010). On that
matter further studies are necessary to recognize these excited
modes in the linear stage. In the present work we use the term
"Weibel instability" (WI) for possible filamentation instability and
other modes.
The WI results in particle acceleration
\cite[e.g.,][]{silva2003,nishikawa2003,jaroschek2005,spitkovsky2008a,spitkovsky2008b, 
dieckmann2008,nishikawa2009}. 

{\rm Other instabilities such as the kinetic Kelvin-Helmhotz (kKHI) and the mushroom instability (MI) are driven by the velocity- shear at the boundary between the jet and the ambient medium in 2D and 3D systems without external magnetic fields  
\cite[e.g.,][]{alves2012,nishikawa2013,liang2013a,liang2013b,grismayer2013,nishikawa2014,alves2015}. 
It should be noted here that the kKHI is generated along the jet direction, in contrast to the MI, which is excited in the direction perpendicular to the jet and their growth rates depend on the relative velocity of the particles \cite[e.g.,][]{alves2012,alves2015}.  For an $e^{\pm}$ jet
\cite[e.g.,][]{liang2013a,liang2013b,nishikawa2014,alves2015}, both kKHI and MI generate an AC magnetic field whilst an electron-ion jet \cite[e.g.,][]{alves2012,nishikawa2014} generates a DC magnetic field. For the case of $m_{\rm i}/m_{\rm e} = 4$ the growth rate of the MI is reduced comparing to the case of $m_{\rm p}/m_{\rm e} = 1836$, and a quasi-AC electric field is generated with slightly accelerated jet ions.}

{\rm
PIC simulation studies of the evolution of cylindrical jets without and with a helical magnetic-field topology have been performed in the 
past \cite[e.g.,][]{nishikawa2016a,nishikawa2016b,nishikawa2019,nishikawa2021}. The present investigation focuses on the nature of particle acceleration in these relativistic plasma flows with a toroidal magnetic field.
}

An additional possible mechanism of particle acceleration in jets is magnetic reconnection. In this process the 
magnetic topology is rearranged and the magnetic energy is converted into thermal and kinetic particle energy.
Magnetic reconnection is observed in solar and planetary magnetospheric plasmas. It is also 
often assumed to be an important mechanism of particle acceleration in extragalactic environments such as 
AGN (active galactic nuclei) and GRB (gamma-ray bursts) jets \cite[e.g,][]{Drenkhahn2002, Pino05, uzdensky2011, zhang2011, granot2012, mcKinney2012,giannios2010, 
komissarov2012, giannios2013,sironi2015, Pino18, Kadowaki18, Kadowaki19,
Christie19,Fowler19,Zhang2018b}.

Magnetic reconnection has commonly been studied with PIC simulations using the so-called Harris model 
in a slab geometry. It was observed to produce a significant particle acceleration
\cite[e.g.,][]{zenitani2005,oka2008, daughton2011, kagan2013, wendel2013, karimabadi14, sironi2014, guo2015, guo2016a, guo2016b}. These studies however cannot be applied directly to astrophysical relativistic jets, 
since observations \citep{hawley2015,gabuzda2019}, as well as MHD modelling \citep{tchekhovskoy2015}, suggest that the magnetic-field
topology is pre-dominantly helical; i.e., it
consists of toroidal and poloidal magnetic field components.
 
Global 3D PIC modeling of relativistic jets allows for a self-consistent investigation of the complex kinetic processes
occurring in the jet and the surrounding medium. These processes can reveal electron-scale short-wavelength
instabilities, their saturation and associated phenomena. Such studies have 
been first performed for unmagnetized jets
\citep{nishikawa2016a}. PIC simulations of relativistic jets containing helical magnetic fields were, for 
the first time, presented by \citet{nishikawa2016b}. These initial studies addressed the early, linear growth of kinetic instabilities in the electron-ion and electron-positron jets. However, such simulations were limited by the size of the computational box. 

The present study involves a much larger jet radius and longer simulation 
times than previous works \cite[e.g.,][]{nishikawa2016a, nishikawa2016b, nishikawa2017}, allowing for a non-linear evolution of 
the jets with a toroidal magnetic field. It is designed to address the following key questions: 
\begin{enumerate}
\item How does a toroidal magnetic field affect the growth of kKHI, MI, and WI within the jet and in the jet-ambient plasma boundary? 

\item How do jets composed of electrons and positrons 

and jets composed of electrons an ions evolve in the presence of a large-scale toroidal magnetic field?

\item How and where are particles accelerated in jets with different plasma compositions?
\end{enumerate}

Since the magnetic field structure and particle composition of relativistic jets is still not well understood, this systematic 
study of e$^{-}$ - i$^{+}$ (e$^{-}$ - p$^{+}$) and e$^{\pm}$ jets containing a toroidal field helps to provide an advanced 
and detailed understanding of the magnetic field evolution, the generation of instabilities, possible reconnection events, and 
the particle acceleration applicable in the environments of AGN and GRB jets. It is important to note that the differences 
in the magnetic field morphologies between jets composed of e$^{-}$ - i$^{+}$ (e$^{-}$ - p$^{+}$) and e$^{\pm}$ could leave significant imprints on
the polarized emission from AGN jets and GRBs. Particularly, circular polarization 
(measured as the Stokes parameter $V$) in the continuum radio emission from AGN jets provides  a powerful diagnostic tool of magnetic structures 
and particle composition because, unlike linear polarization, circular polarization is expected to remain almost completely unmodified by external 
screens \cite[e.g.,][]{osu2013,MacDonald2021}. 

It is important to note, as discussed in \cite{nishikawa2020} for e$^{-}$ -\, p$^{+}$ 
jets, that our simulations do not address the large-scale plasma flows of macroscopic 
parsec-scale jets studied by relativistic magnetohydrodynamic (RMHD) simulations. Instead they explore the relevant kinetic-scale physics within relativistic 
jet plasmas, which cannot be studied with RMHD simulations. Although in PIC simulations we need to resolve the electron Debye length,
the simulation sizes which are feasible at the present time are limited (for more details see \cite{MacDonald2021}).  Our study, therefore, 
is complementary to 
RMHD models and yields important insights into 
the kinetic processes at work in relativistic astrophysical jets \citep[see also][]{nishikawa2020,nishikawa2021,meli21}.

This paper is organized as follows: After we describe the simulation set-up in Section 2, the main differences between electron-positron 
and electron-ion (electron-proton) jets are shown at the linear stage and at the fully developed non-linear stage in Section 3. 
In particular, we discuss the  kinetic instabilities in the linear and non-linear stage in subsection 3.1. In subsection 3.2 we present results about acceleration and discuss the patterns of electron acceleration and deceleration in comparison with the structure of the electromagnetic field. We show the three-dimensional magnetic field evolution in subsection 3.4. Lastly, in subsection 3.4 we present results about the role of the non-linear stage, the instabilities growth and their role to consequent acceleration. In Section 4, we summarize and discuss our conclusions.

\begin{figure}
\hspace{0.8cm}
\includegraphics[scale=0.4,angle=0]{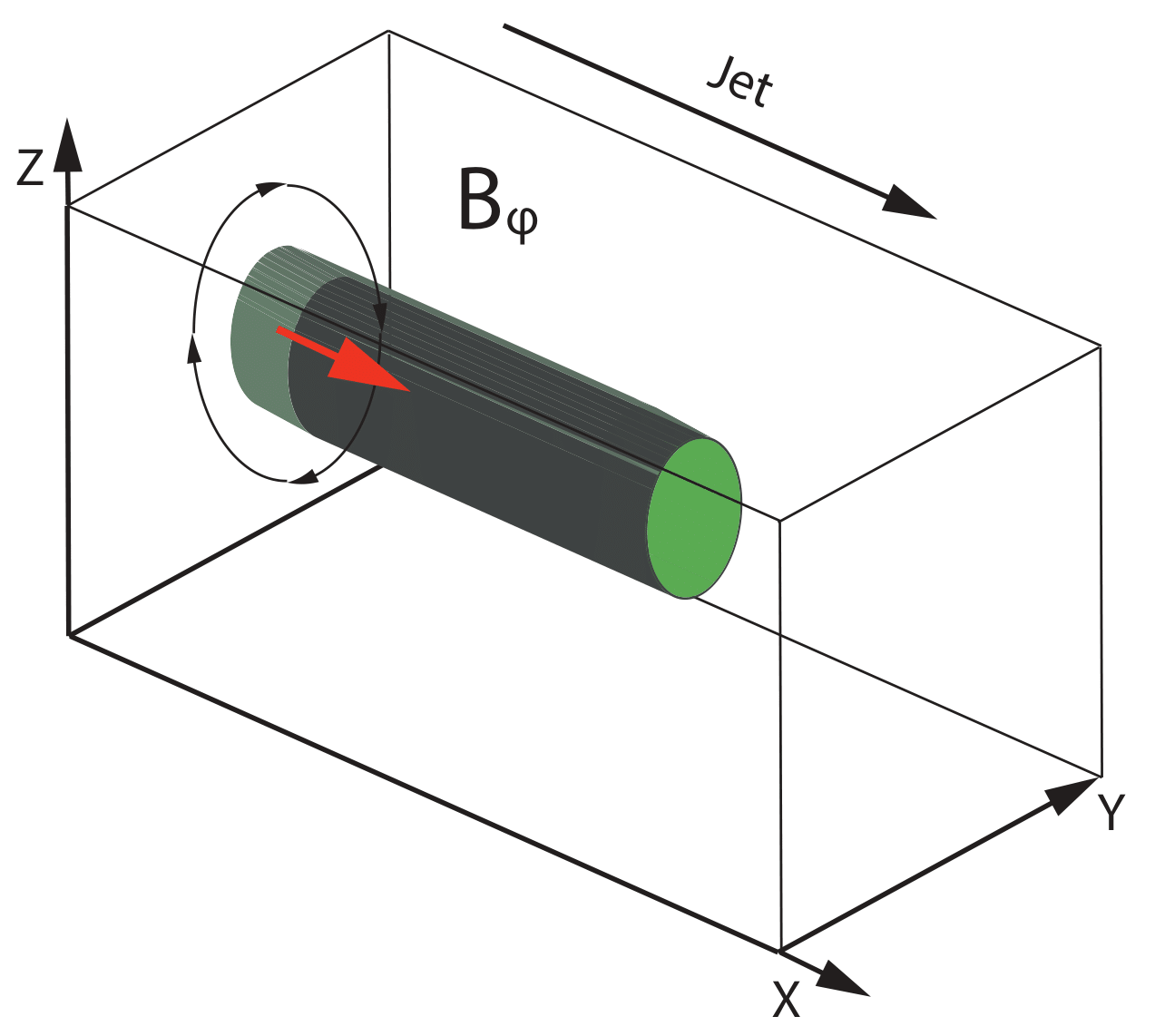}
\caption{A schematic of the jet injection scheme with a toroidal magnetic field ($B_{\phi}$). 
The jet electrons and positrons/ions/protons are injected such that a current (indicated by the red arrow) is generated to support 
the toroidal magnetic field.}
\label{jet_inj}
\end{figure}

\section{Simulation set-up}
\vspace*{-0.0cm}

Our 3D PIC code is a modified version of the relativistic electromagnetic PIC code
TRISTAN \citep{tristan} with MPI-based parallelization \citep{niemiec_2008,nishikawa2009}. 
The numerical grid is set to $(L_{x}, L_{y}, L_{z}) = (1285\Delta, 789\Delta, 789\Delta)$ and is twice as long as that in our 
previous simulation studies \citep{nishikawa2016b,nishikawa2017,nishikawa2019}. Here $\Delta =1$ is 
the size of an individual grid cell. Open boundaries are used on the surfaces at $x/\Delta=0$ and $x/\Delta=1285$, whilst periodic 
boundary conditions are implemented along the transverse directions $y$ and $z$.  
Since the jet is located in the center of the simulation box far from the boundaries, the effect of periodic boundaries is negligible.


\subsection{New jet injection scheme and simulation parameters}

In this work, we use a new jet injection scheme.  
The system needs to be neutral and in equilibrium, therefore we fill the ambient plasma with electrons and positrons (ions $m_{\rm i}/m_{\rm e} = 4$) 
at the same positions randomly, and then we inject jet particles (electrons, positrons and ions). All particles are self-consistently pushed in the system.
We inject a cylindrical jet into the ambient plasma which is at rest.  The jet then
 propagates 
in the $x$-direction, with a toroidal magnetic field (Eq. \ref{hmfcar}),  as schematically shown in Fig.~\ref{jet_inj}. 
The jet is injected at $x = 100\Delta$ in the center of the $y$ - $z$ plane at $(y_{\rm jc}=381\Delta,\,z_{\rm jc}=381\Delta$, and propagates in the
$x$-direction. Its radial width in cylindrical coordinates 
is $r_{\rm jet} = 100\Delta$.  In this study, we apply only a toroidal magnetic field 

\begin{eqnarray}
\vec {B}_{\phi}(r)=\frac{B_{0}(r/a)}{[1 + (r/a)^2]}.
\end{eqnarray}
The poloidal field component, $B_{\rm x}$ \citep[e.g.,][]{nishikawa2020}, is not included. 
$B_{\phi}$ has a peak amplitude at $r=a$ where $a$ is the characteristic radius. 
In Cartesian coordinates, the  corresponding field components, $B_{\rm y}$ and $B_{\rm z}$, are calculated as
\begin{eqnarray}
B_{y}(y, z) =  \frac{(z-z_{\rm jc})B_{0}}{a[1 + (r/a)^2]}, \, \, \ \     
B_{z}(y, z) =  -\frac{(y-y_{\rm jc})B_{0}}{a[1 + (r/a)^2]}.
\label{hmfcar}
\end{eqnarray}
Equation~(\ref{hmfcar}) describes the magnetic field of left-handed polarity for positive $B_{0}$. In this study we assume $a=50\Delta$. 

With respect to the electric field and its associated current, initially there is no current, and so  $\partial E / \partial t = $curl$\ {\bf B} = {\bf J}$.  
Since we need to generate a current based on the applied toroidal magnetic field $B_{\phi}$ 
as defined above, we apply a current

\begin{equation}
J_x =\frac{1}{r}  \frac{\partial(r/a) B_0 /[1 + (r/a)^2]}{\partial r} = \frac{2B_0} {a[1 + (r/a)^2]^2}.
\label{Jx}
\end{equation}

For the toroidal magnetic field outside of the jet we multiply  Equation~(\ref{hmfcar}) with a damping function:
\begin{equation}
\Theta(r- r_{\rm jet})\,= \frac{r_{\rm jet}}{r}, \,\,\, \, \, {\rm where}\,\,\, r > r_{\rm jet}. 
\end{equation}
Since the current is situated only inside the jet, the toroidal magnetic field outside the jet decays sharply, which help us utilise a smaller 
simulation box.

We assume a top-hat density profile for both jets. Although the shape of a real jet is far more complex, 
the present results are the first step in a series of advanced numerical investigations including an implementation of 
a Gaussian (Lorentzian) (non top-hat) profile, which is closer to jet density profiles generated by General RMHD (GRPIC) 
simulations \citep[e.g.,][]{nishikawa2021}. 


First, we calculate the velocities of the jet particles based on $\mathbf{J}(r)= \nabla\, \times \mathbf{B}$ in the
jet frame using Eq. (\ref{Jx}). 
After applying a Lorentz transformation we get the velocity of the jet electrons ($v_{\rm e,x}'$) and ions ($v_{\rm i,x}'$) in 
the jet frame:
\begin{eqnarray}
v_{\rm e,x}'\!\!\!\!\!&=&\!\!\!\!\!\frac{1}{e\,n_{\rm e}\,}\, \frac{m_{\rm i}}{m_{\rm i}+m_{\rm e}}\,\, \frac{2B_0}{a[1+(r/a)^2]^{2}} \nonumber \\
v_{\rm i,x}'\!\!\!\!\!&=&\!\!\!\!\!-\frac{1}{e\,n_{\rm e}\,}\, \frac{m_{\rm e}}{m_{\rm i}+m_{\rm e}}\,\, \frac{2B_0}{a[1+(r/a)^2]^{2}}
\label{eq9}
\end{eqnarray}
where  $n_{\rm e}$: jet electron density, $e$: electron charge, $m_{\rm e}$: electron mass, and   $m_{\rm i}$: ion mass.

We now transform the jet-frame drift velocities back to the simulation frame. The relative velocity between the frames is 
$v_{\rm j}/c =\beta_0=\sqrt{1-\Gamma_0^{-2}}\simeq 0.9977753$ ($c =1$ ). 
Based on the Lorentz transformation $v_{\rm e,x}$ in the simulation frame:
\begin{eqnarray}
v_{\rm e,x}\!\!\!\!\!&=&\!\!\!\!\!\frac{v_{\rm e,x}'+\beta_0\,c}{1+\beta_0\frac{v_{\rm e,x}'}{c}}
\label{eq11}
\end{eqnarray}
Then the velocities of the jet particles are Lorentz-transformed to the simulation frame. In the simulation frame the jet electrons 
propagate faster than positrons (ions, protons), which generates a negative current in the jet 
(clockwise as viewed from the jet head). In order to sustain the current
in the jet, a toroidal magnetic field is gradually applied at the jet orifice using $B(t) =B_{0}*0.5*(\tanh(t_{\rm ramp}-3.0)+1.0)$ where    
$t_{\rm ramp}={\rm float}(nstep)*0.012$ ($nstep$: simulation step),
located at $x/\Delta=100-102$, 
and a motional electric field is established, $\mathbf{E}_{\rm mot} = - \mathbf{v}_{\rm j} 
\times \mathbf{B}$. Here, $\mathbf{v}_{\rm j}={v}_{\rm j,x}\mathbf{\hat{x}}$, where ${v}_{\rm j,x}$ is the $x$-component of the jet velocity.
In this way one avoids non-linear effects emerging from the constantly applied magnetic field in the simulation frame where unnatural 
banding and currents in the centre of the jet might occur \citep[e.g.,][]{nishikawa2020}. 

\begin{figure*}
\hspace*{2.6cm} {\bf unmagnetized e$^{\pm}$ jet} \hspace*{1.5cm} (a)  \hspace*{3.5cm} {\bf unmagnetized e$^{-}$ - p$^{+}$ jet}
\hspace*{1.5cm} (b) 

\includegraphics[scale=0.45,angle=0]{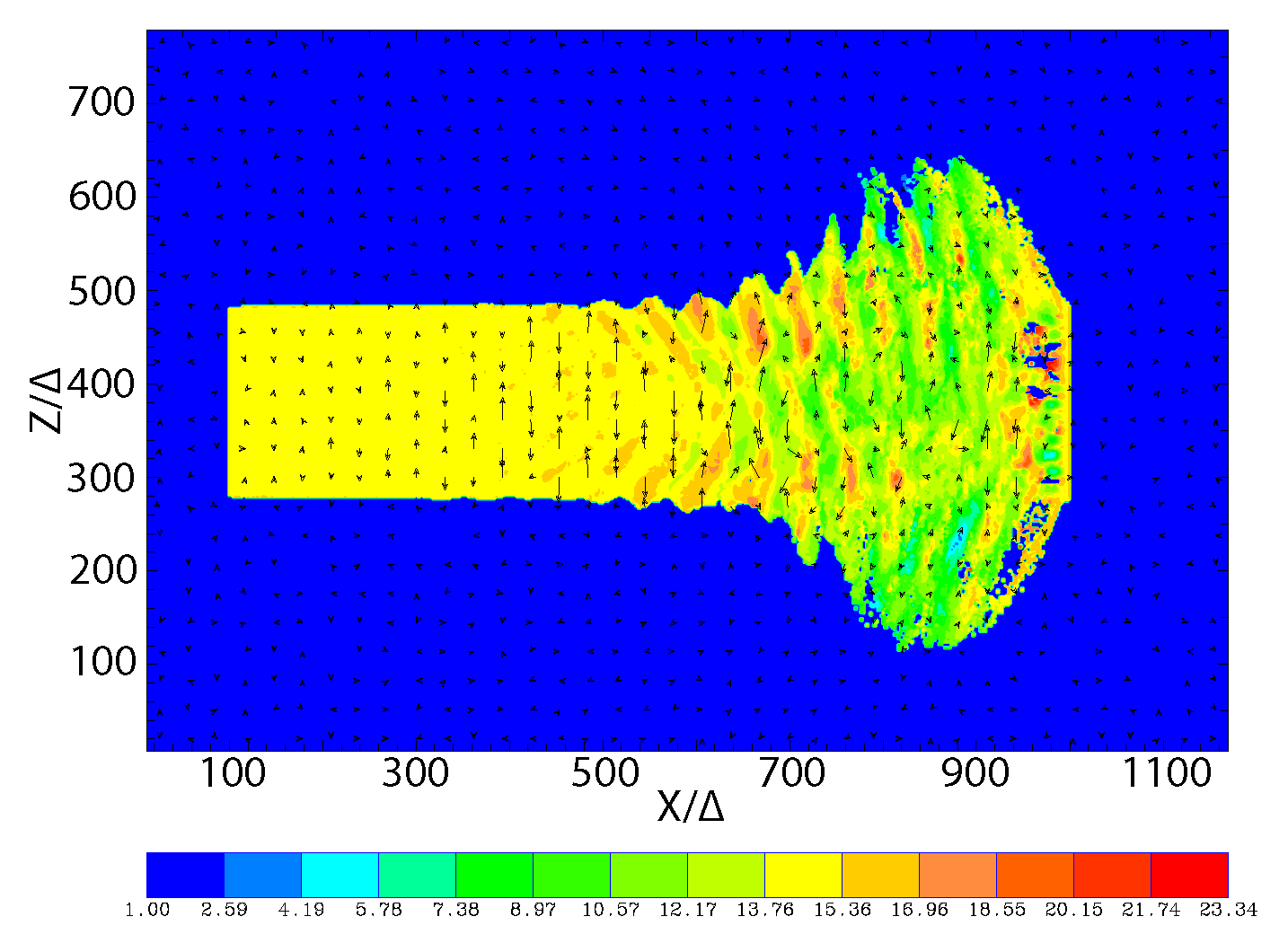}
\includegraphics[scale=0.45,angle=0]{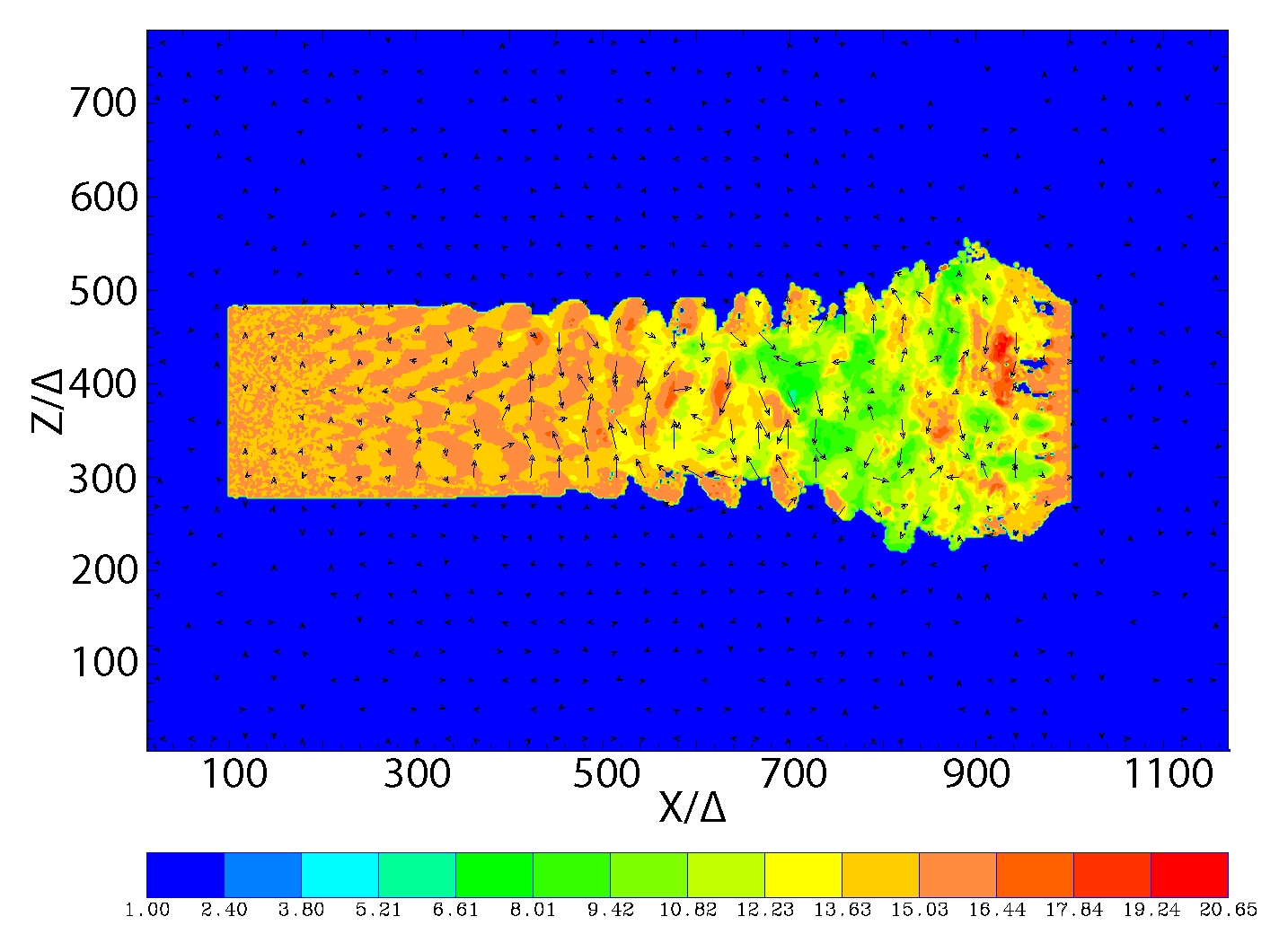}

\hspace*{2.6cm} {\bf magnetized e$^{\pm}$ jet} \hspace*{1.8cm} (c)  \hspace*{3.3cm} {\bf magnetized e$^{-}$ - i$^{+}$ jet}
\hspace*{1.7cm} (d) 

\includegraphics[scale=0.45,angle=0]{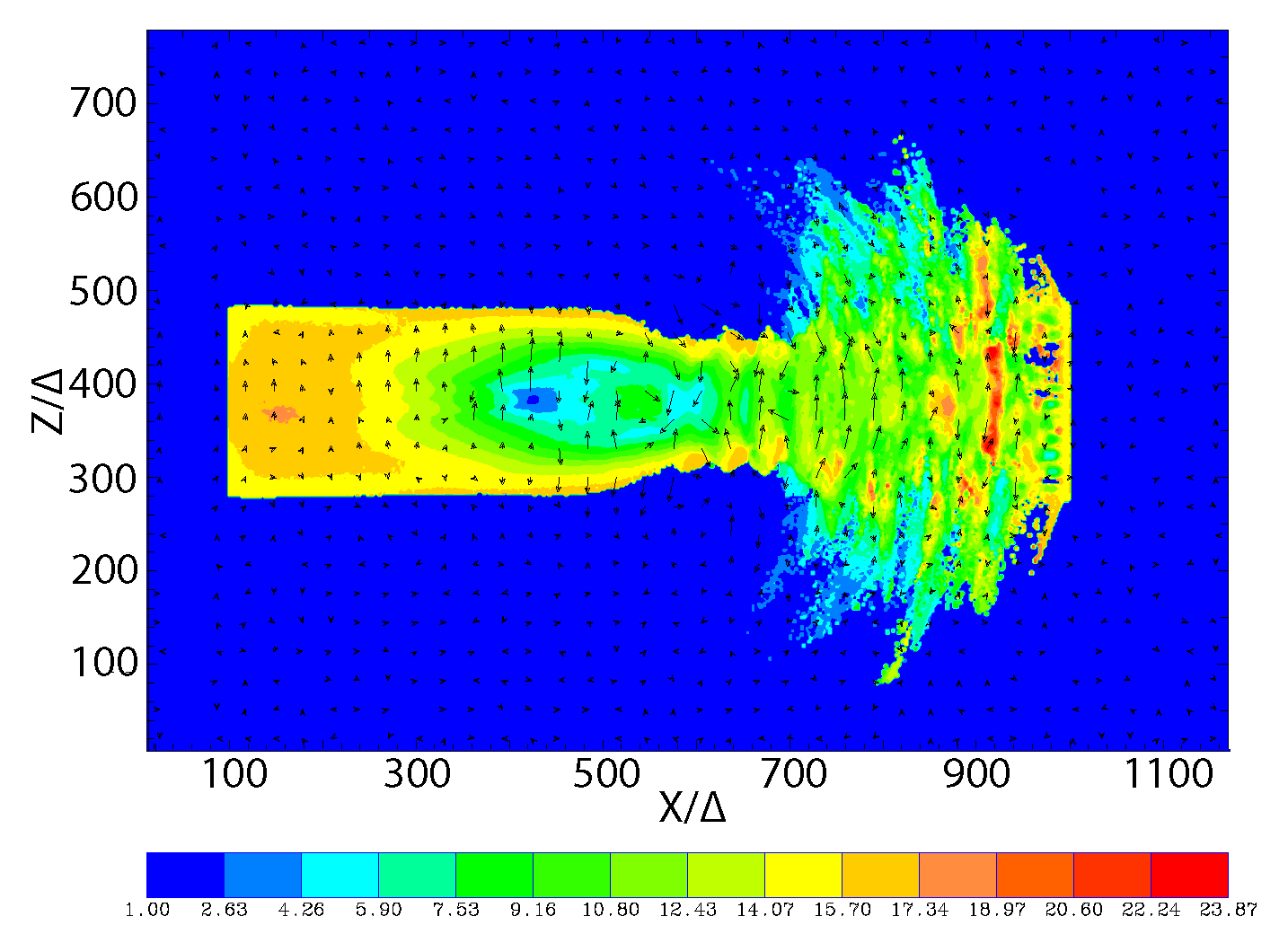}
\includegraphics[scale=0.45,angle=0]{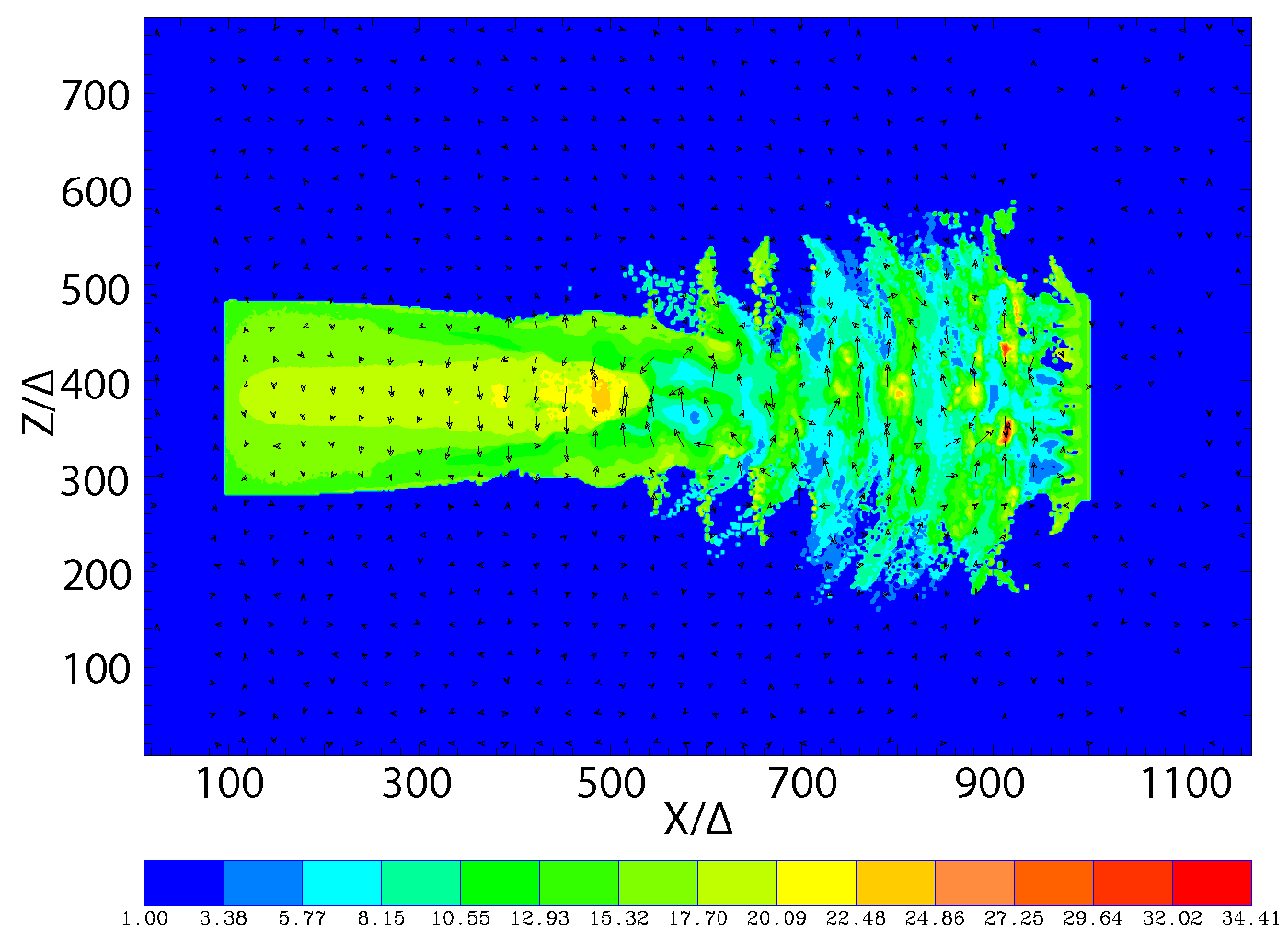}

\caption{Color maps of the Lorentz factor of the jet electrons at $y/\Delta =381$ (shown below the panels) for  
e$^{\pm}$ jets (left panels) and the e$^{-}$ - p$^{+}$  jet (b) the e$^{-}$ - i$^{+}$  jet (d) 
with $r_{\rm jet} = 100\Delta$ at time $t = 900\,\omega_{\rm pe}^{-1}$. 
Panels (a) and (b) show unmagnetized jets and panels (c) and (d) the jets with the toroidal magnetic field.
Black arrows show the in-plane magnetic field $(B_x,B_z)$. The maximum is (a): 23.34, (b): 20.65, (c): 23.87, and (d): 34.41. The minimum is 1.0 for all panels.} 
\label{Lorentz}
\end{figure*}

In these simulations, the jet Lorentz factor is set to $\gamma_{\rm jt}=15$. We perform simulations for both unmagnetized and
magnetized jets, while keeping the ambient medium unmagnetized. In the case of a magnetized jet, the jet is initially moderately 
magnetized, that is the jet's magnetic-field amplitude 
$B_{0}=0.5$, corresponds to a plasma magnetization \mbox{$\sigma = B_{\rm 0}^{2}/(n_{\rm e}m_{\rm
e}\gamma_{\rm jt}c^{2}) = 1.73\times 10^{-2}$}, where $c$ is the speed of light, $m_{\rm e}$ is the electron rest mass 
and $n_{\rm e}$ the electron density. In order to investigate 
the non-linear stage of the jet's evolution, 
we follow the jet for a sufficiently long time of $t_{\rm max}=900\omega_{\rm pe}^{-1}$, where $\omega_{\rm pe} = (e^{2}n_{\rm amb}/(\epsilon_0 m_{\rm e}))^{1/2}$ 
is the electron plasma frequency.

Both the jet and the ambient plasma are composed of electrons and ions (protons) or of electrons and positrons. 
The initial number densities measured in the simulation frame are $n_{\rm jt}= 8$ and 
$n_{\rm amb} = 12$  in the jet and in the ambient plasma, respectively. 
The Debye length for the ambient electrons is $\lambda_{\rm D}=0.5\Delta$ and the electron skin depth is $\lambda_{\rm se} = c/\omega_{\rm pe} = 10.0\Delta$.
The thermal speed of jet electrons is $v_{\rm jt,th,e} = 0.014c$ in the jet frame 
whilst in the ambient plasma it is $v_{\rm am,th,e} = 0.05c$. 
The thermal speed of ions is smaller by a factor of $(m_{\rm p}/m_{\rm e})^{1/2}\approx 42$ for unmagnetized (magnetized) 
e$^{-}$ - p$^{+}$ 
jet, and $(m_{\rm i}/m_{\rm e})^{1/2}\approx 2$ for magnetized e$^{-}$ - i$^{+}$ jet.  It should be noted that for the 
unmagnetized e$^{-}$ - p$^{+}$ jet the large mass ratio contributes to a stronger growth of the
MI. For that reason we reduce the mass ratio to control the growth of 
MI, that is from $m_{\rm p}/m_{\rm e} =1836$ to $m_{\rm i}/m_{\rm e} =4$.
Although in most PIC simulation studies mass ratios of 16, 25 or larger are used  \citep[e.g.,][]{bretdieckmann10}, in this study 
we chose to use a smaller mass 
ratio ($m_{\rm i}/m_{\rm e} = 4$). This is because we need to avoid anomalous effects (when using $m_{\rm p}/m_{\rm e} = 1836$) 
in the evolution of the relativistic  jet (which is not shown), given the small radius we apply in the present simulations. 
At the same time with this small mass ratio of ion-to-electron, we can compare the evolution differences
to a pair jet. 



\section{Results of simulations of a jet with a toroidal magnetic field}

\begin{figure*}

\hspace*{2.6cm} {\bf unmagnetized e$^{\pm}$ jet} \hspace*{1.6cm} (a)  \hspace*{3.7cm} {\bf unmagnetized e$^{-}$ - p$^{+}$ jet}
\hspace*{1.7cm} (b) 

\includegraphics[scale=0.48,angle=0]{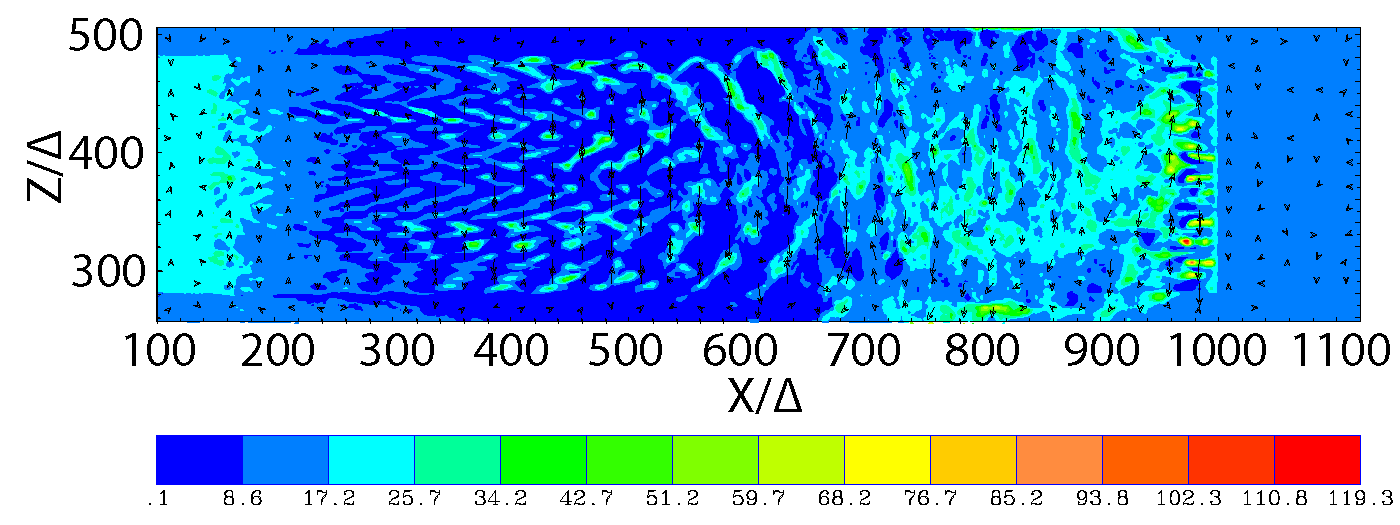}
\includegraphics[scale=0.48,angle=0]{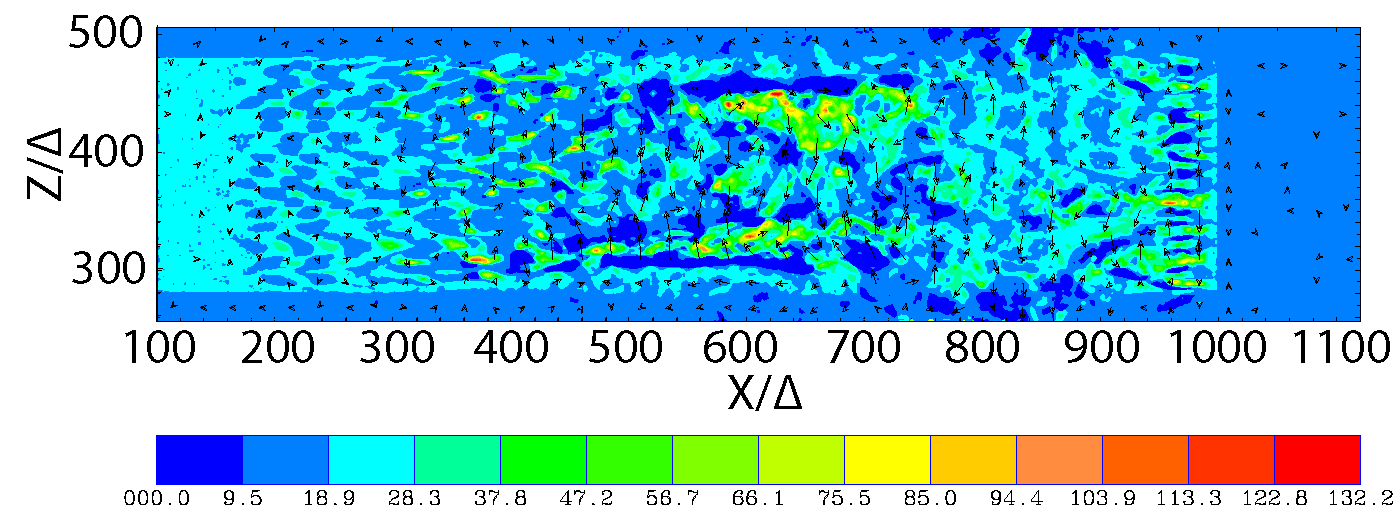}

\hspace*{2.8cm} {\bf magnetized e$^{\pm}$ jet} \hspace*{1.8cm} (c)  \hspace*{3.8cm} {\bf magnetized e$^{-}$ - i$^{+}$ jet}
\hspace*{1.7cm} (d) 
\includegraphics[scale=0.48,angle=0]{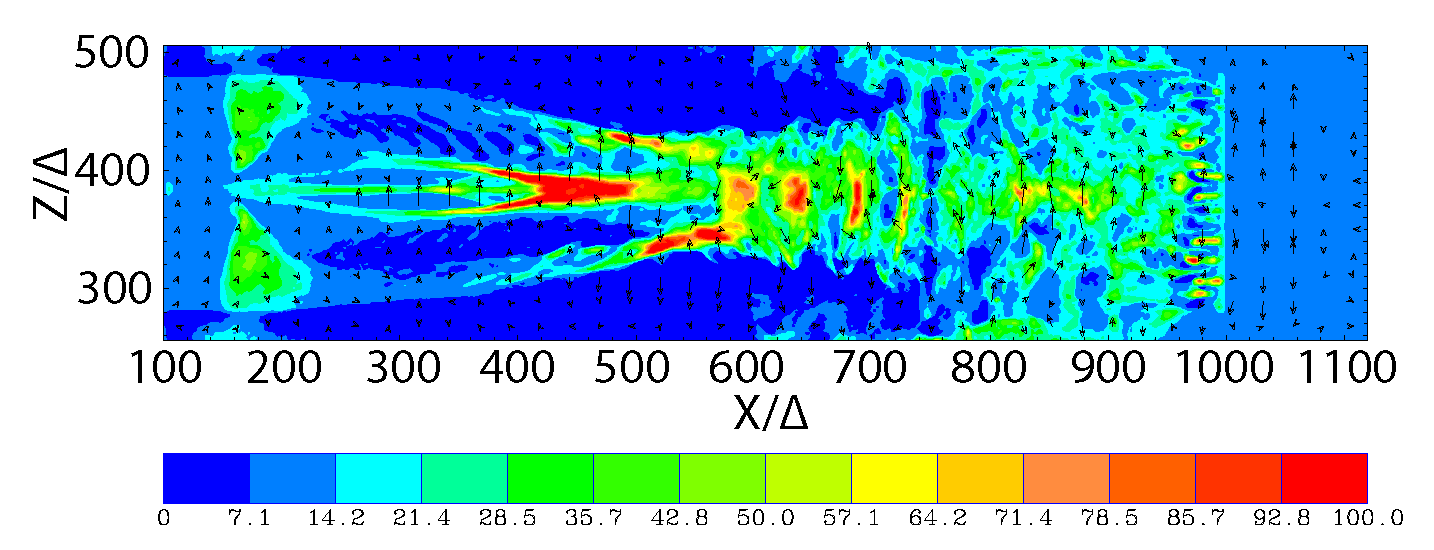}
\includegraphics[scale=0.48,angle=0]{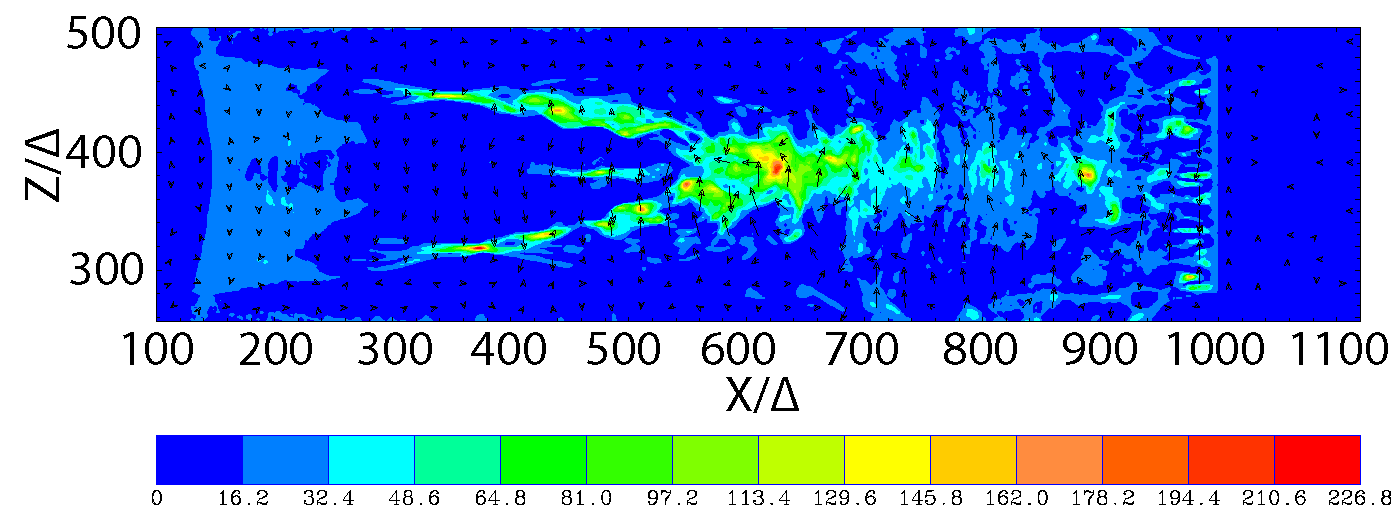}
\caption{Color maps of the jet electron density with black arrows depicting the magnetic field components in the 
$x$ - $z$ plane, at   $t=900\, \omega_{\rm pe}^{-1}$.
As in Fig.~\ref{Lorentz}, left panels show results for e$^{\pm}$ jets and the right panels for e$^{-}$ - i$^{+}$ jets. Upper panels (a) and (b) show unmagnetized jets and lower panels (c) 
and (d) jets with a toroidal magnetic field. Note that the maps display the jet structures in a region of the computational box, $250<x/\Delta<500$, that  
fully encompasses the injected jet with $r_{\rm jet} = 100\Delta$.  The jet electron density at injection is $n_{\rm jt}= 8$.
The maximum is (a): 119.3, (b): 132.2, (c): 100.0, and (d): 226.8. The minimum is essentially 0.0 for all panels.}
\label{ede4}
\end{figure*}

\begin{figure*}

\hspace*{2.6cm} {\bf unmagnetized e$^{\pm}$ jet} \hspace*{1.6cm} (a)  \hspace*{3.7cm} {\bf unmagnetized e$^{-}$ - p$^{+}$ jet}
\hspace*{1.7cm} (b) 
\includegraphics[scale=0.48,angle=0]{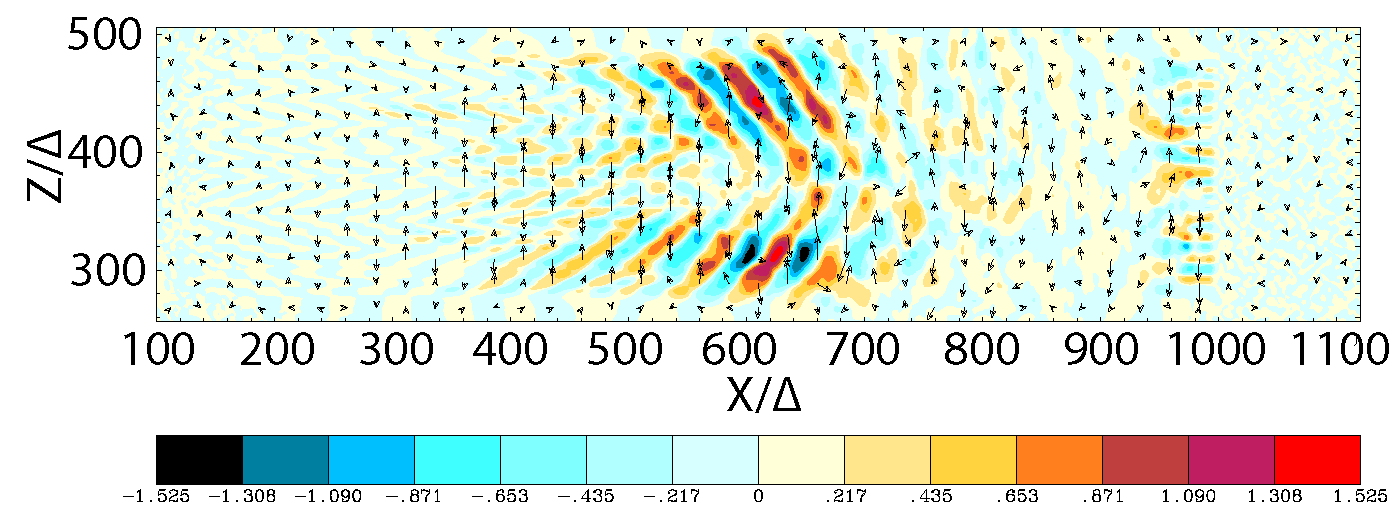}
\includegraphics[scale=0.48,angle=0]{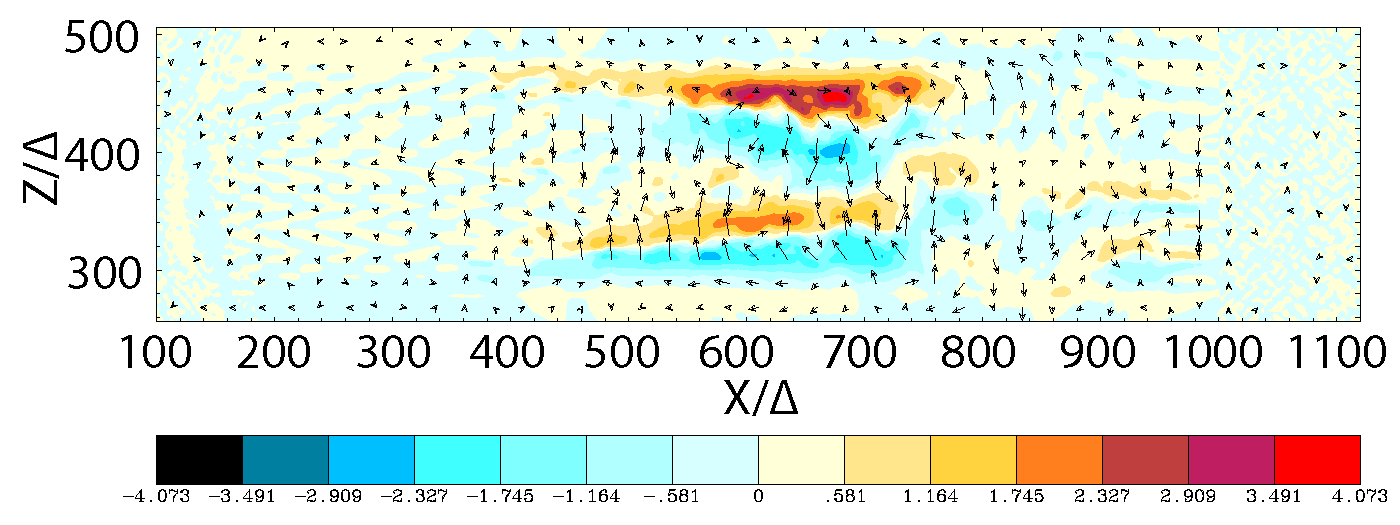}

\hspace*{2.8cm} {\bf magnetized e$^{\pm}$ jet} \hspace*{1.8cm} (c)  \hspace*{3.8cm} {\bf magnetized e$^{-}$ - i$^{+}$ jet}
\hspace*{1.7cm} (d) 
\includegraphics[scale=0.48,angle=0]{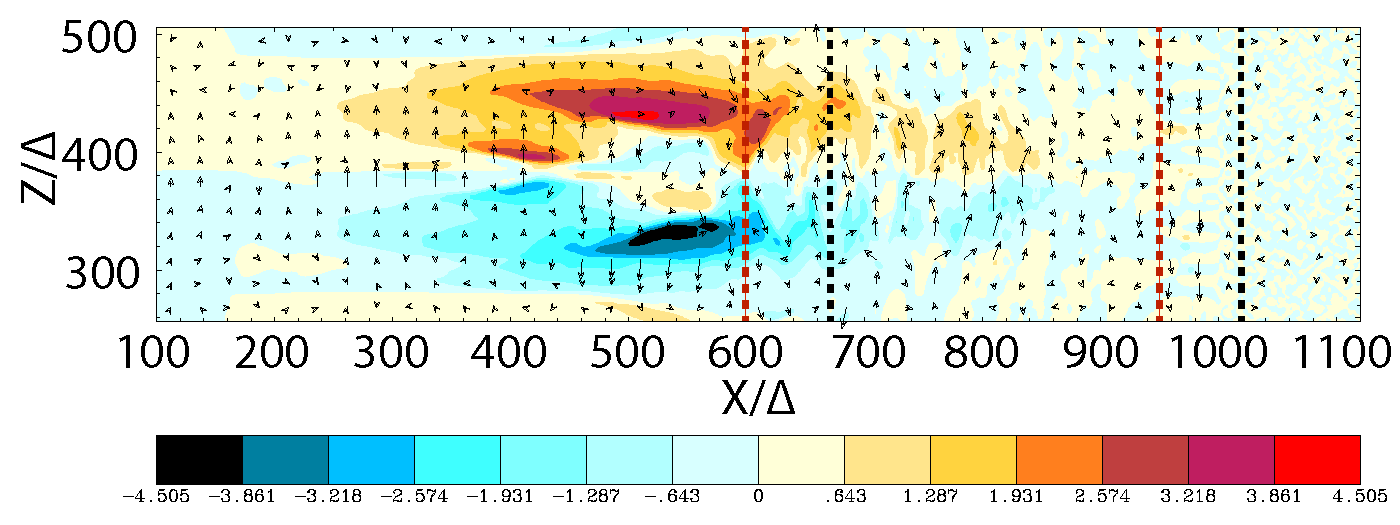}
\includegraphics[scale=0.48,angle=0]{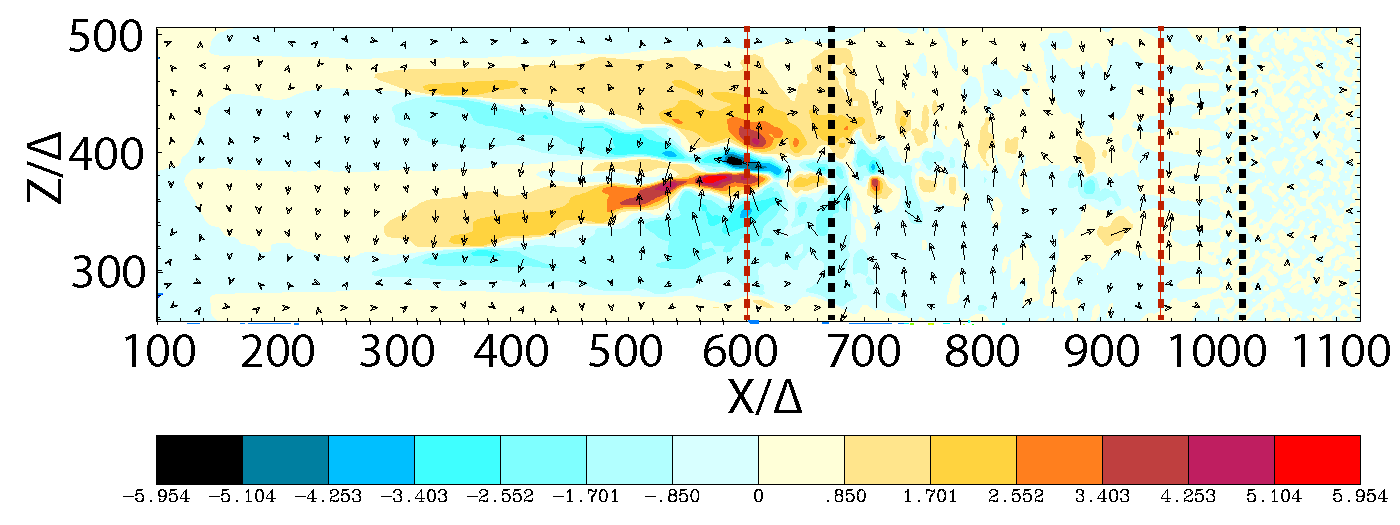}
\caption{Color maps of the $B_{\rm y}$ magnetic field  with arrows depicting the magnetic field components in the $x$ - $z$ plane, at  $t=900\, \omega_{\rm pe}^{-1}$. 
Upper panels (a) and (b) show unmagnetized jets and lower panels (c) and (d) jets with a toroidal magnetic field. See Fig.~\ref{ede4}.
The squares with dashed lines indicate the areas plotted in the 3D displays in Fig. \ref{3DBV} (red) and  Fig. \ref{ReconS} (blue).
The maximum and minimum are (a): $\pm 1.525$, (b): $\pm 4.073$, (c): $\pm 4.505$ and (d): $\pm 5.954$.}
\label{By4}
\end{figure*}

\begin{figure*}[h!]
\hspace*{2.6cm} {\bf unmagnetized e$^{\pm}$ jet} \hspace*{1.6cm} (a)  \hspace*{3.7cm} {\bf unmagnetized e$^{-}$ - p$^{+}$ jet}
\hspace*{1.7cm} (b) 

\includegraphics[scale=0.48,angle=0]{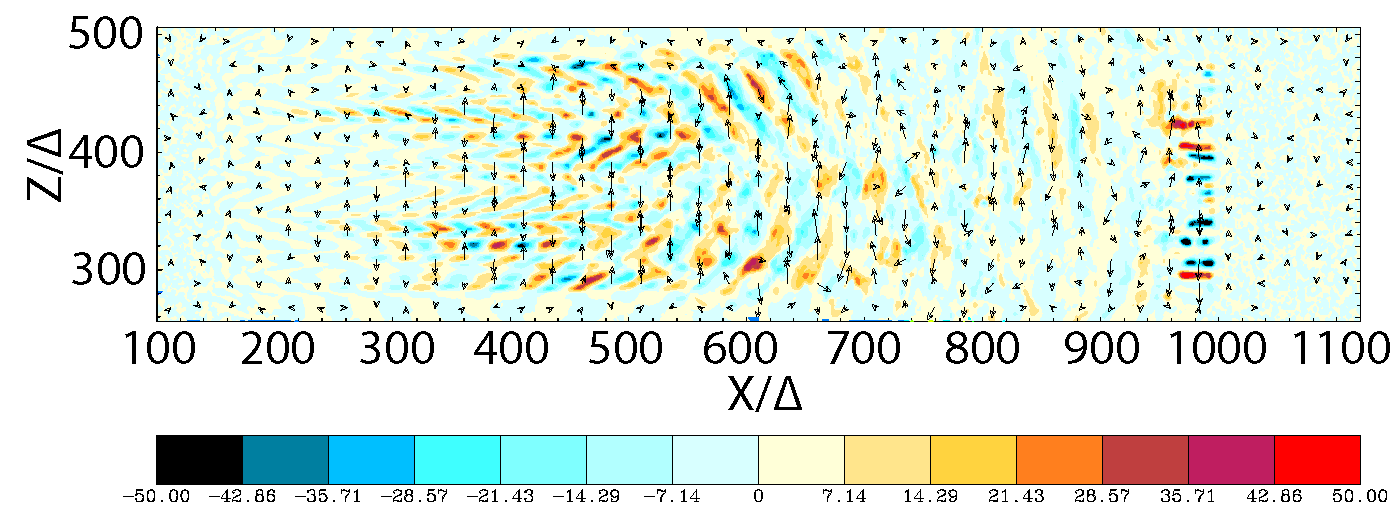}
\includegraphics[scale=0.48,angle=0]{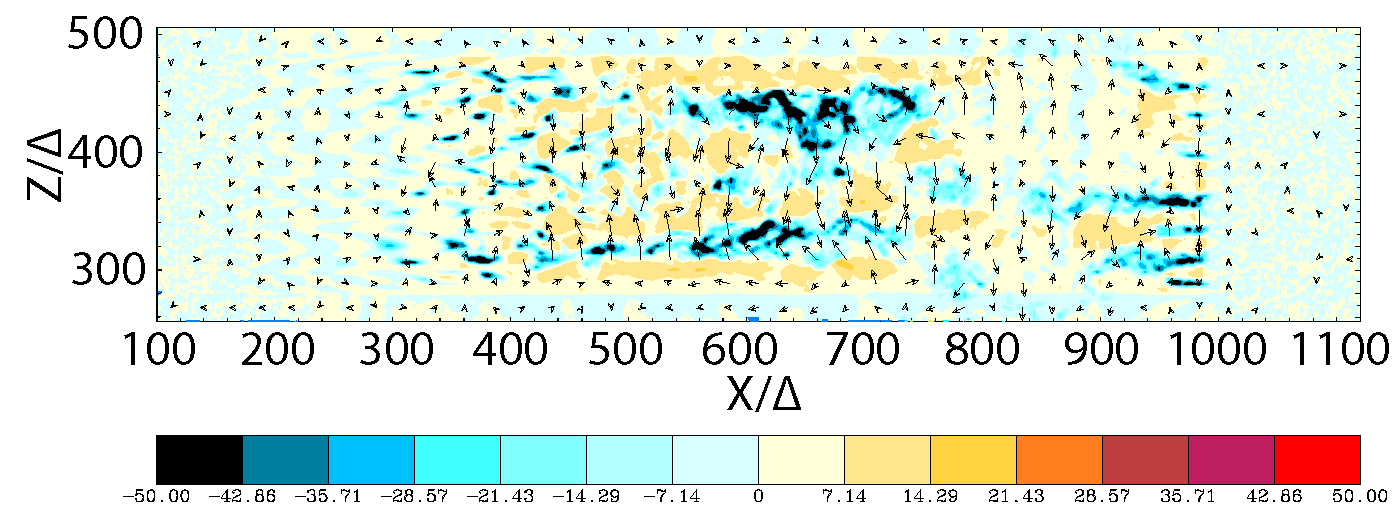}

\hspace*{2.6cm} {\bf magnetized e$^{\pm}$ jet} \hspace*{1.8cm} (c)  \hspace*{3.3cm} {\bf magnetized e$^{-}$ - i$^{+}$ jet}
\hspace*{1.7cm} (d) 
\includegraphics[scale=0.48,angle=0]{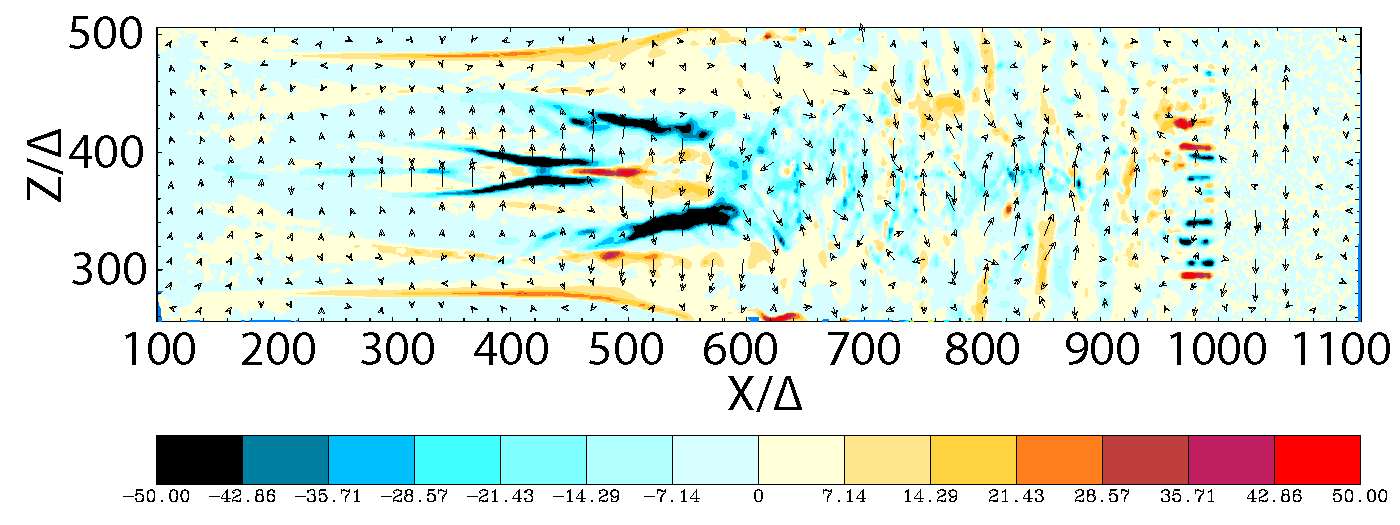}
\includegraphics[scale=0.48,angle=0]{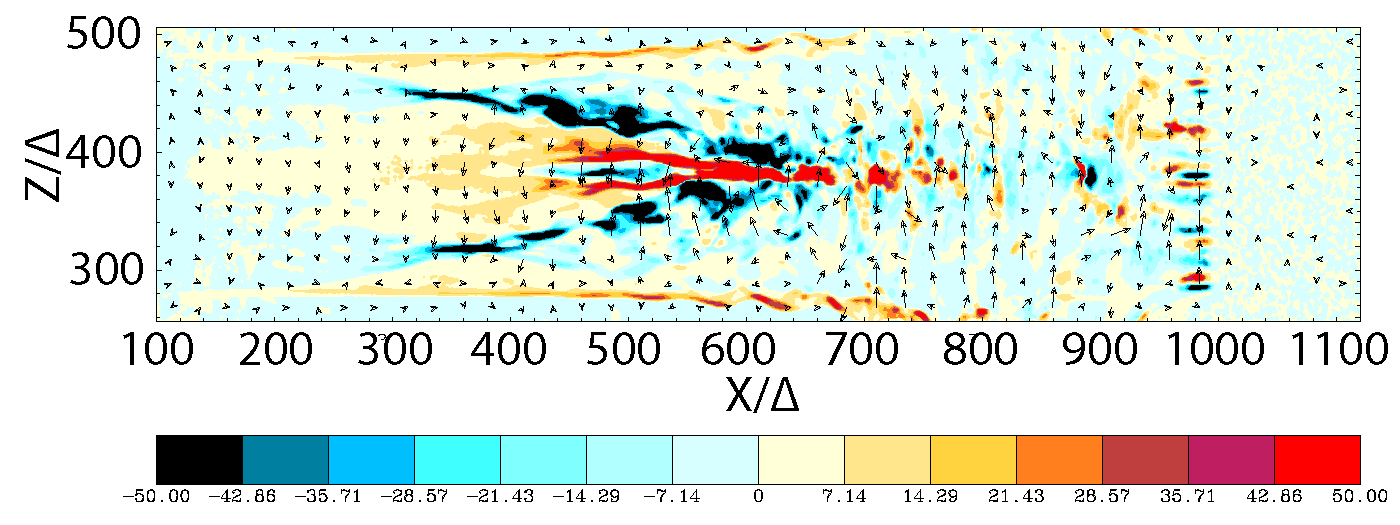}
\caption{Color maps of the 
$J_{\rm x}$ component of the electric current with arrows showing the magnetic field components in the 
$x$ - $z$ plane,  at   $ t=900\, \omega_{\rm pe}^{-1}$. See Fig.~\ref{ede4}. 
In order to view the weaker current, the maximum and minimum values are set to $\pm 50$.}. 
\label{Jx4}
\end{figure*}

We present simulation results for e$^{\pm}$ and e$^{-}$ -\, i$^{+}$ jets with a toroidal 
magnetic field, applying a new and improved jet injection scheme. We are in particular interested in the differences in the dynamical
behavior of the jets of different plasma compositions and in the way these jets interact with the surrounding  environment.
In order to give an overview of how the toroidal magnetic field affects the evolution of the jet, in Figs.~\ref{Lorentz}-\ref{Jx4} 
we present simulation results for jets with a toroidal field (shown in the lower panels of each figure) and compare them to the results obtained 
for an unmagnetized jet.

Figure \ref{Lorentz} presents the global jet structure and shows the Lorentz factor of the jet electrons for an e$^{\pm}$ (a, c),  
an e$^{-}$ - p$^{+}$ (b), and an e$^{-}$ - i$^{+}$ (d) jet. In all panels the magnetic field direction in the $x$ - $z$ plane is indicated by the black arrows.
It is noted that the applied toroidal magnetic field in the $x$ - $z$ plane should be zero but at the center of the jet (as shown the regions ($x/\Delta < 250$),
the $x$ and $z$ components are generated by the excited instabilities (WI, MI, and kKHI). 
For the e$^{\pm}$ jets, the structures seen at the boundary between the jet and the ambient plasma at 
$400< x/\Delta < 1000$ show the excitation and development of kKHI and MI, where the MI represents the transverse component/dynamics of the kKHI.  
It is interesting that at $600< x/\Delta< 700$ in panel c) one can see at an early stage, 
a central reversal of the magnetic field with an associated deceleration of the flow, and then  a tight ``bottle-neck'' form, 
featuring a strong jet collimation trend. This strong collimation occurs due to the slender jet and an 
excited strong MI growth. This collimation of jet electrons is  also seen in Fig. \ref{ede4}c.
After the dissipation of the magnetic field around $x/\Delta = 750$ (see Fig.~\ref{By4}), the disruption around the 
location, where the non-linear stage starts, is evident for both the unmagnetized and magnetized e$^{\pm}$ jets.  Especially for the magnetized 
e$^{-}$ - p$^{+}$ jet, a disruption and a reversal of the magnetic field occurs at earlier stages, which is not shown in this report,
since the drastic expansion of jet electrons outside the jet is caused by the extremely small jet radius.
We chose to re-ran the simulation with a mass ratio ($m_{\rm i}/m_{\rm e} = 4$) which shows a much weaker growth of MI, as we explained in Section 2.
In the magnetized jet cases we also see a radial expansion, forming spikes with a simultaneous stratification of the electric field
generated by the kKHI. 
It seems that the jet electrons are 
radially pushed as a main result of the excited instabilities (MI and kKHI) within the jet.

For the simulations of the unmagnetized e$^{-}$ - p$^{+}$ jet case, the jet electrons remain highly energetic for 
longer within the jet and before the development of the non-linear stages, where some deceleration with collimation is seen.
We further discern that at the non-linear stage 
the jet electrons are expanded outside and become 'diluted', which in turn deduces the average Lorentz factor. 
Longer simulations will provide further insights about the important non-linear stage of the jet evolution. 
We further discuss the particle Lorentz factor distributions in the jets in Section~3.2.
{\rm In typical hydrodynamic simulations, the jets are characterized by a channel with a relativistic outflow, which is separated from the surrounding materials by cocoons \citep{bromberg2011}. In our simulations, the jet particles are injected into the ambient medium. Nevertheless, the jet particles and the ambient ones mix with each other only at the edge of the jet where the sheath instabilities take place. When we represent the plasma parameters for the jet, e.g., the Lorentz factor, we pick up only the jet particles from simulations. In Fig.2, the 2D representation of the Lorentz factor of the jet particles shows the formation of a Mach-like cone, which, however, is not well developed as in the case of hydrodynamic simulations. One of the reasons for this shortcoming might be the fact that our simulations do not run long enough to generate a well-defined shock. In future work, we plan to use other setups for the jet density profile to be able to compare our results with hydrodynamic jet simulations. The PIC simulation results by \cite{Ardaneh16}, for example, show some similar structure of the jet head and contact discontinuity in the jet plasma as for the hydrodynamic jets.}
 
\begin{figure*}
\includegraphics[scale=0.48,angle=0]{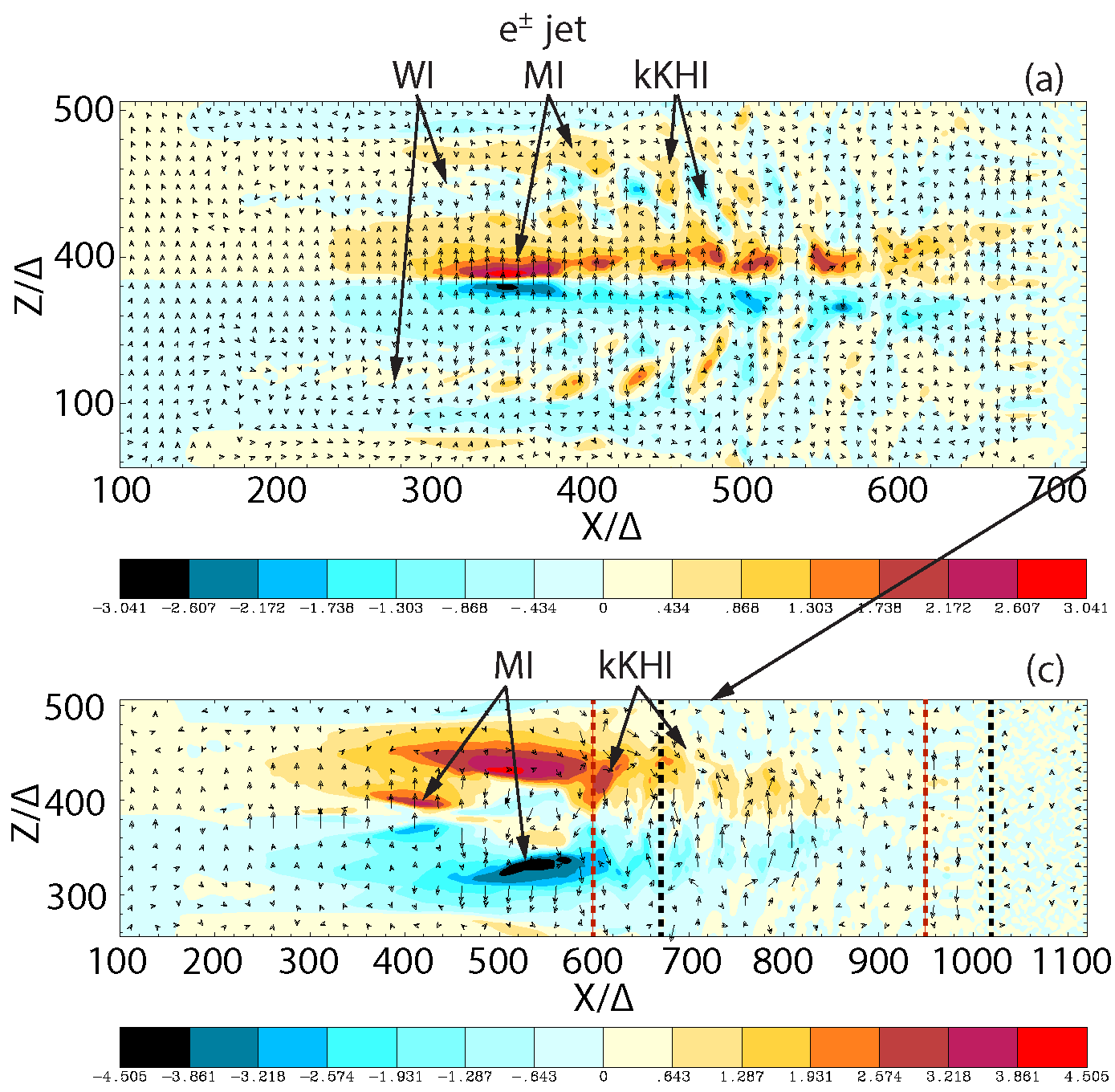}
\includegraphics[scale=0.48,angle=0]{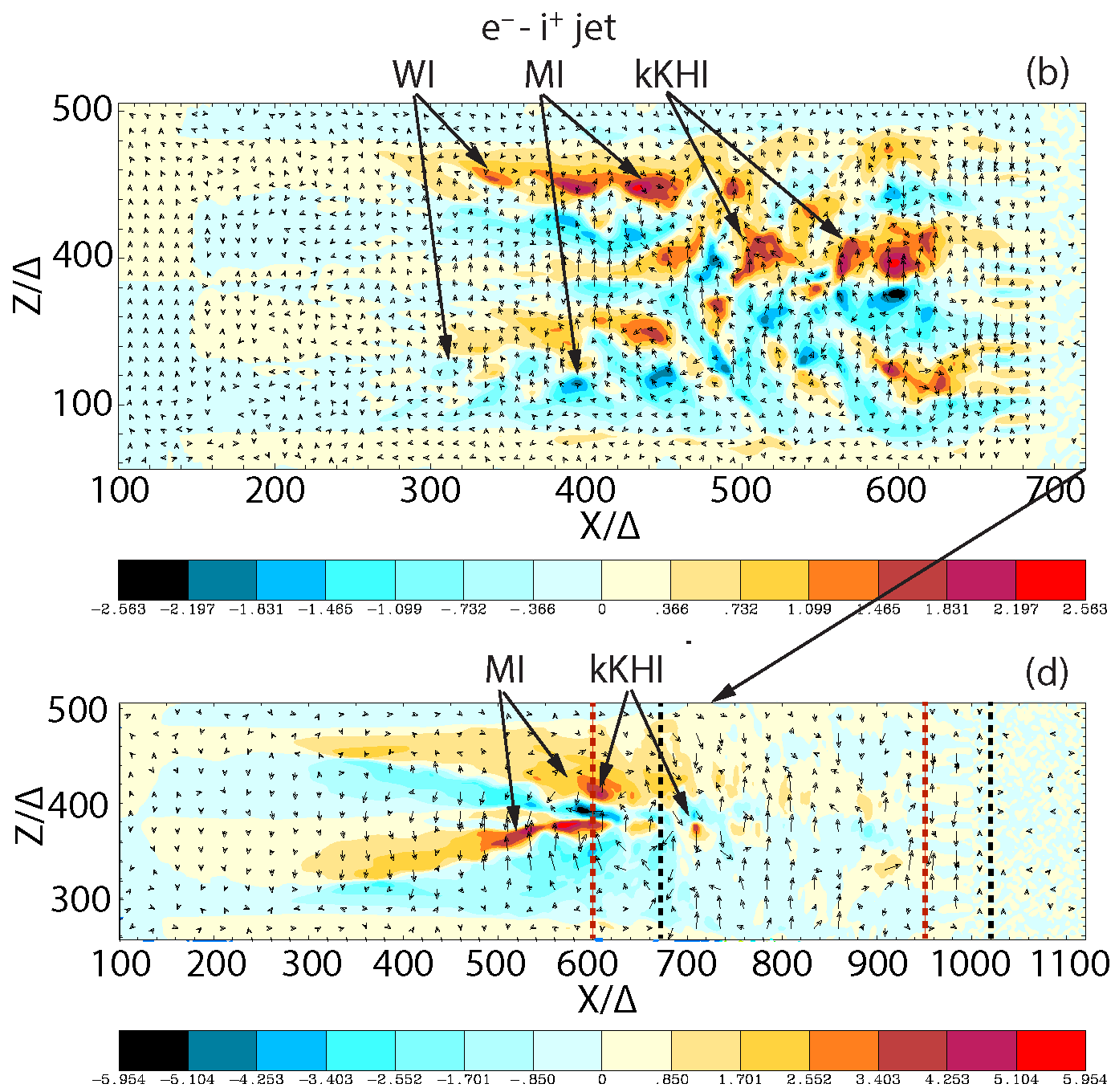}
\caption{Color maps of the magnetic field amplitude $B_{\rm y}$ and arrows depicting the magnetic field components in the $x$ - $z$ plane, 
both at $t=600\, \omega_{\rm pe}^{-1}$ (upper panels) and $900\, \omega_{\rm pe}^{-1}$ (lower panels), respectively. 
The jet is injected at $x=100\Delta$ in the middle of the $y$ - $z$ plane and propagates in $+x$-direction. 
Panels (a, c) are for an e$^{\pm}$ plasma while panels (b, d) are for an e$^{-}$ - i$^{+}$ composition. 
The peak amplitudes of $B_{\rm y}$ are (a) $\pm 3.041$, (b) $\pm 2.583$,  
(c) $\pm 4.505$ and (d) $\pm 5.954$.
\label{ByBxz}}
\end{figure*}

Figure \ref{ede4} depicts the density of the jet electrons (ambient electrons excluded) for e$^{\pm}$ jets (left column) and 
e$^{-}$ - p$^{+}$ (b) and 
e$^{-}$ - i$^{+}$  (d) jets (right column) at  $t=900\, \omega_{\rm pe}^{-1}$.
We observe that the jet electron density structures fluctuate significantly between the simulated cases. 
This is due to the expected excited instabilities, whose growth depends on the plasma conditions which are different for the two jet compositions and magnetizations. 
We note that in the jets with toroidal magnetic fields the jet electrons are strongly collimated towards the center of (i) the pair jet 
in a region $550\lesssim x/\Delta\lesssim 750$ (compare also with Fig. \ref{Lorentz}, and (ii) for the e$^{-}$ - i$^{+}$ jet in a region  
$550\lesssim x/\Delta\lesssim 850$,  which is coincident with a stronger MI and kKHI excited in the presence of the toroidal field, 
as we also show below (Fig. \ref{ByBxz}).

Figure \ref{By4} shows the amplitude of $B_{\rm y}$ component with the arrows indicating the magnetic field components in the $x$ - $z$ 
plane\footnote{Supplementary videos at \href{7017747}{doi: 10.5281/zenodo.7017747}}. 
For the unmagnetized cases with an initial magnetic field $B_{0} = 0$, an amplified magnetic field ($B_{\rm y}$) is generated 
by the excited instabilities, as shown blow. 
{\rm For the unmagnetized e$^{\pm}$ jet, panel (a) shows that a two-stream instability (WI) is dominantly excited.} 
{\rm For electron-proton (e$^{-}$-p$^{+}$) case,
first, the two-streaming instability grows, later kKHI and MI grow at the jet boundary and they propagate into the jet center due 
to $m_{\rm p}/m_{\rm e} =1836$.} {\rm Consequently,}
Panel (b) shows the dominantly excited MI {/rm (and kKHI)} in the unmagnetized 
e$^{-}$ - p$^{+}$ jet.
For the magnetized jets, we apply $B_{\rm 0} = 0.5$ initially at the jet orifice while $B_{\rm y}$ is measured in the same units. 
In the presence of the toroidal magnetic field for the e$^{\pm}$ jet,
we find a maximum value of $B_{\rm y} = 4.505 $ (Fig. \ref{By4}c), which means that $B_{\rm 0}$ is amplified by a factor of 9.01 over the initial value. 
It is a factor of 2.95 stronger than magnetic fields in the unmagnetized e$^{\pm}$ jet. The evident differences in the magnetic field structure in cases 
with and without the initial field, indicate a significant impact of the toroidal field on the development of the kinetic instabilities. The same is true for 
the  e$^{-}$ - i$^{+}$ jet, in which the magnetic field amplification is comparably stronger and the field amplitudes reach $B/B_0\approx 10.6$, 
a factor of 1.3 stronger compared to the unmagnetized e$^{-}$ - p$^{+}$ jet case.  
As aforementioned, for the unmagnetized e$^{-}$ - p$^{+}$ jet the large mass ratio contributes to a stronger growth of the
MI, that is the reason we needed to reduce the mass ratio to control the growth of MI (from $m_{\rm p}/m_{\rm e} =1836$ to $m_{\rm i}/m_{\rm e} =4$).
We note that for the e$^{\pm}$ jet the magnetic field dissipates, i.e., becomes considerably weakened, at the jet 
region $x/\Delta\gtrsim 680$ and similarly for the e$^{-}$ - i$^{+}$ jet the weakening occurs around $x/\Delta\gtrsim 700$. 
By comparing both magnetized jet species, at the non-linear stage close to $x/\Delta\gtrsim 950$, one discerns 
that for the electron-ion jet 
the $B_y$ field almost dissipates and becomes disorganized (turbulent).

Figure \ref{Jx4} shows the current amplitude $J_{\rm x}$ and the magnetic field components (by arrows) in the $x$ - $z$ plane.  
For the magnetized pair jet the strong negative component $-J_{\rm x}$ (blue)  near the center of the jet (Fig. \ref{Jx4}c), which 
is initially excited at the jet boundary moves into the center of the jet. The second outer MI mode is seen around $x/\Delta = 550$.
These MI modes are  modulated by the growth of a kKHI, in particular in the non-linear
stage $x/\Delta > 700$ (see also in  Fig. \ref{ByBxz}c). 
The thin elongated negative currents in Fig. \ref{Jx4}c away from the central region of the jet, correspond to 
the outer MI mode as indicated in Fig. \ref{ByBxz}c. Moreover, the strong merged concentric positive current at the center of the jet 
for the magnetized e$^{-}$ - i$^{+}$ jet (Fig. \ref{Jx4}d)  $x/\Delta > 600$, shows  more prominently the merged MI mode which is 
further modulated by the growth of  the kKHI  $x/\Delta > 750$ (Fig. \ref{ByBxz}d, correspondingly). Comparing these two cases 
(Figs. \ref{Jx4}c,d), the structures of excited MI modes are clearly different in the linear stage. 


\subsection{Kinetic instabilities in the linear and non-linear stage}

Figure \ref{ByBxz} shows the magnetic field component $B_{\rm y}$ in the $x$ - $z$ plane at $y/\Delta=381$ with an in-plane magnetic field depicted with black arrows. 
Results for jets with toroidal magnetic fields are shown and the upper panels present the field structures in the linear regime at time $t =600\, 
\omega_{\rm pe}^{-1}$ (Figs.~\ref{ByBxz}a,b), whereas the lower panels depict the non-linear stage at time $t=900 \, \omega_{\rm pe}^{-1}$ (compare Figs.~\ref{ByBxz}c,d).
The pair jet is shown on the left column (a, c), and the e$^{-}$ - i$^{+}$ jet on the right column (b, d).  

\begin{figure*} 
\hspace*{1.0cm} {\bf magnetized e$^{-}$ - i$^{+}$ jet} \\ 
\hspace*{5.5cm} (a) \hspace*{8.0cm} (b)   

\includegraphics[width=0.47\linewidth]{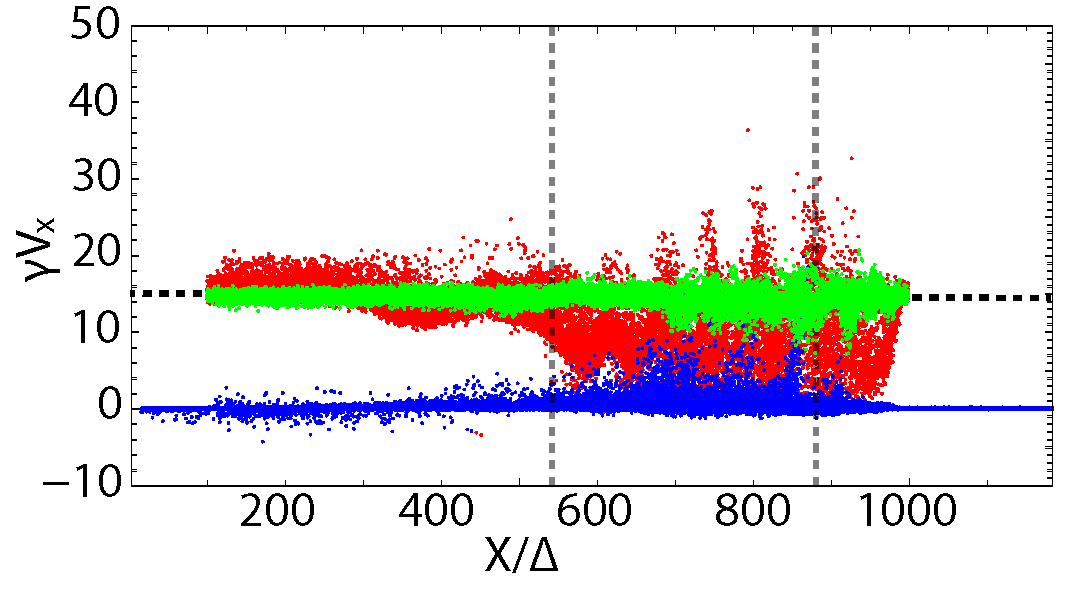}
\includegraphics[width=0.47\linewidth]{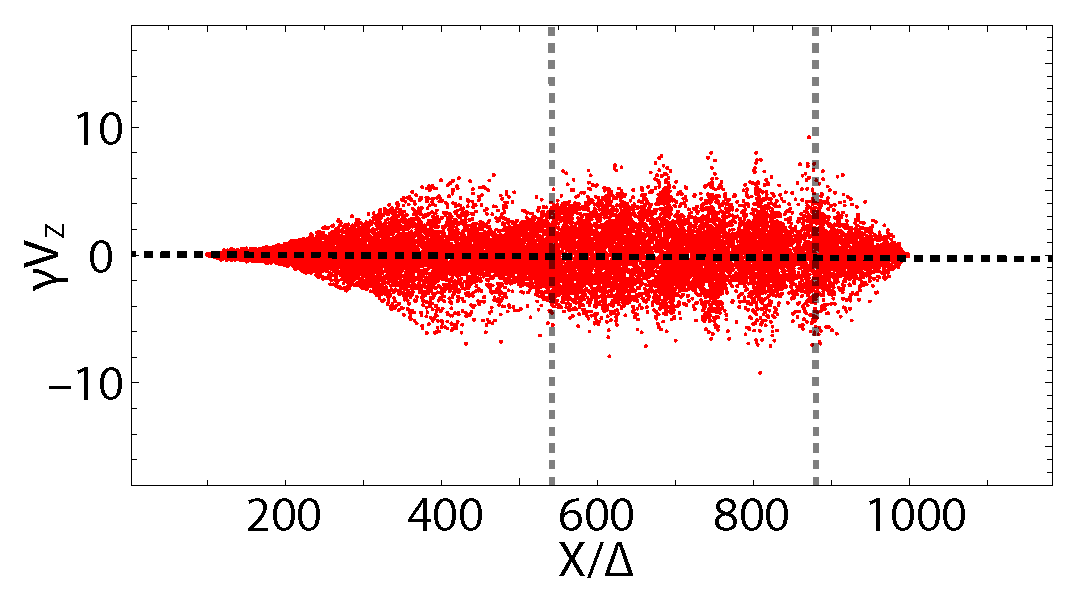}

\hspace*{5.5cm} (c) \hspace*{7.0cm} (d)   

\includegraphics[width=0.93\linewidth]{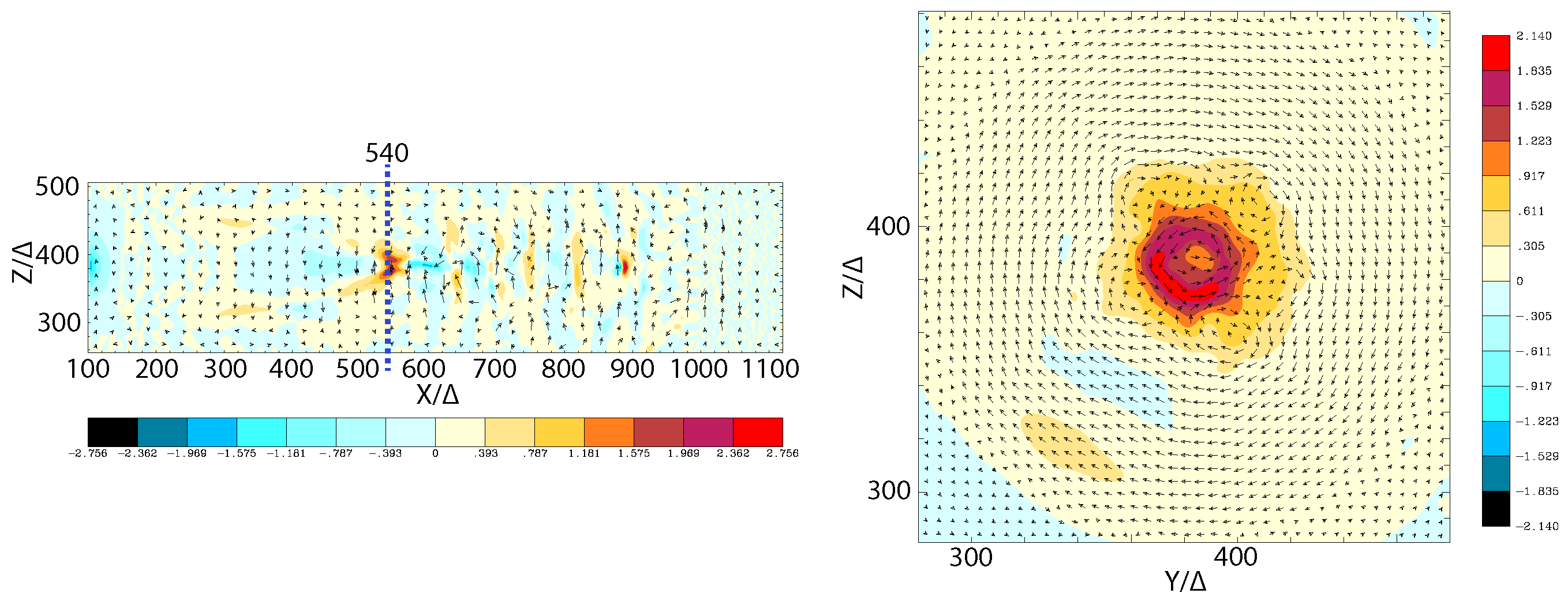}

\caption{a) $x$ - $\gamma v_{\rm x}$ distribution of jet (red), and  ambient (blue) electrons and jet ions (green)
at $t = 900\,\omega_{\rm pe}^{-1}$, for an electron-ion jet. The initial $\gamma v_{\rm x}  = 15$ is marked with the horizontal 
dashed black line. b) $x$ - $\gamma v_{\rm z}$ distribution of jet 
electrons at $t = 900\,\omega_{\rm pe}^{-1}$. 
The initial $\gamma v_{\rm z}  = 0$ is marked with the horizontal dashed black line.
c)  The color-map of  $E_x$ in the $x$ - $z$ plane 
 at $y/\Delta = 381$, the dashed blue line is located at  $x/\Delta = 540$.
d) Color-map of $E_x$ in the $y$-$z$ plane at $x/\Delta = 540$, marked by the vertical lines in panels (a), (b) and (c), 
arrows indicating $B_{y,z}$. The maxima and minima of $E_x$ are (c): $\pm 2.754$ and (d): $\pm 2.140$.} 
\label{bcontours}
\end{figure*}

\begin{figure} 
\hspace*{2.7cm} {\bf  magnetized e$^{\pm}$ jet} \hspace*{1.1cm} (a)
\vspace{-0.1cm}

\hspace*{.40cm}
\includegraphics[width=0.77\linewidth]{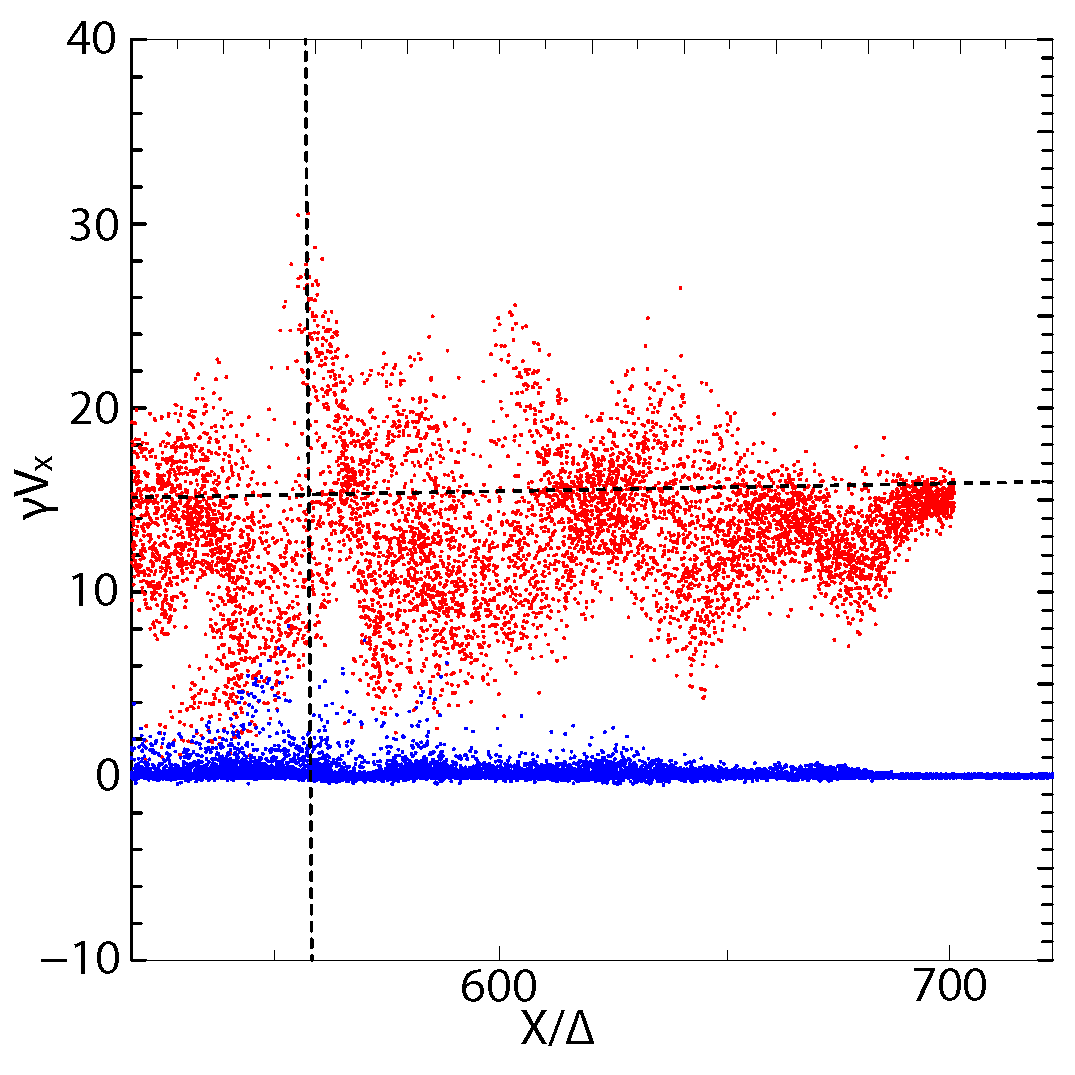}

\hspace*{6.cm} (b)

\hspace*{0.3cm}       
\includegraphics[width=0.90\linewidth]{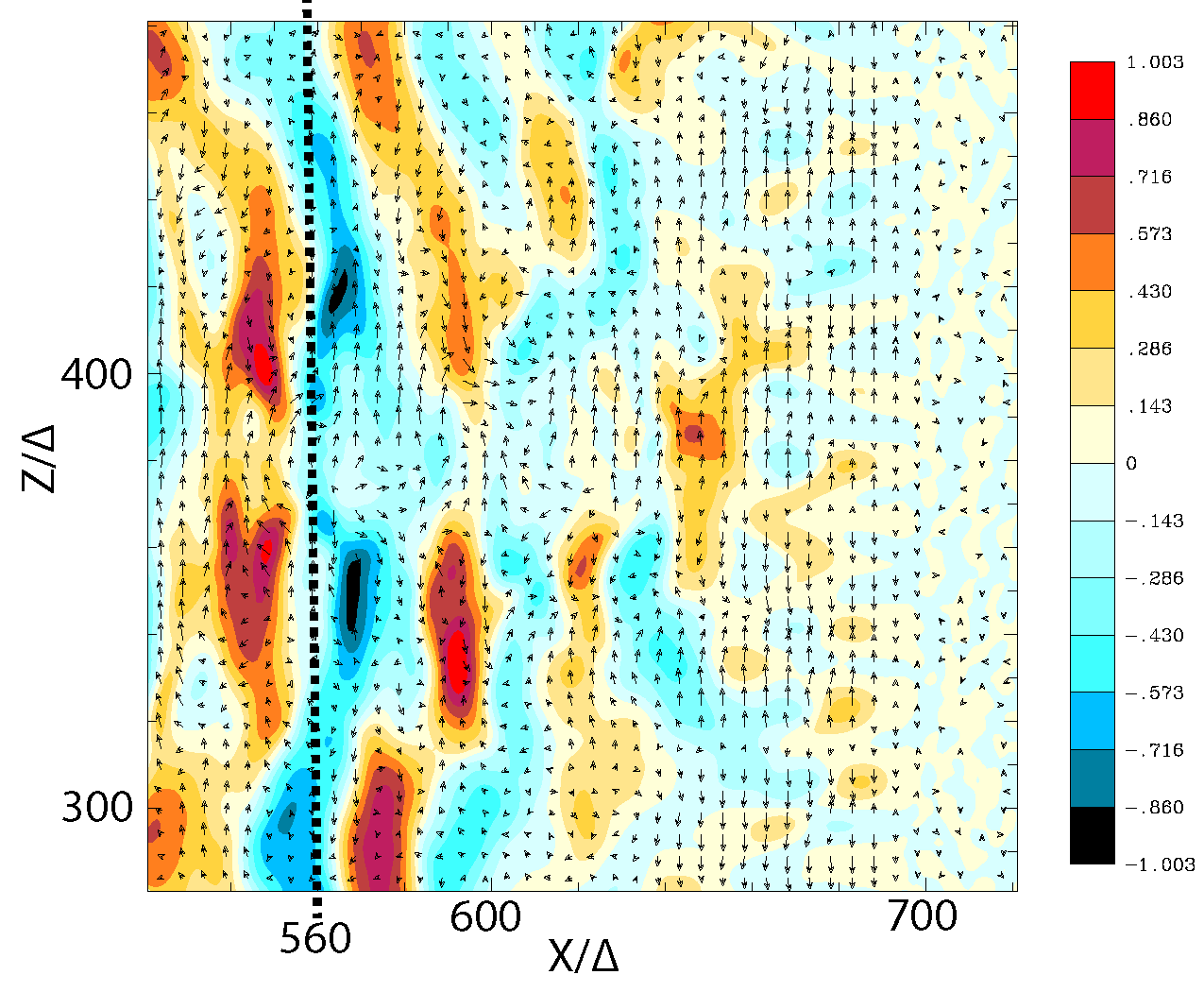}

\hspace*{6.0cm} (c)

\hspace*{0.3cm}    
\includegraphics[width=0.90\linewidth]{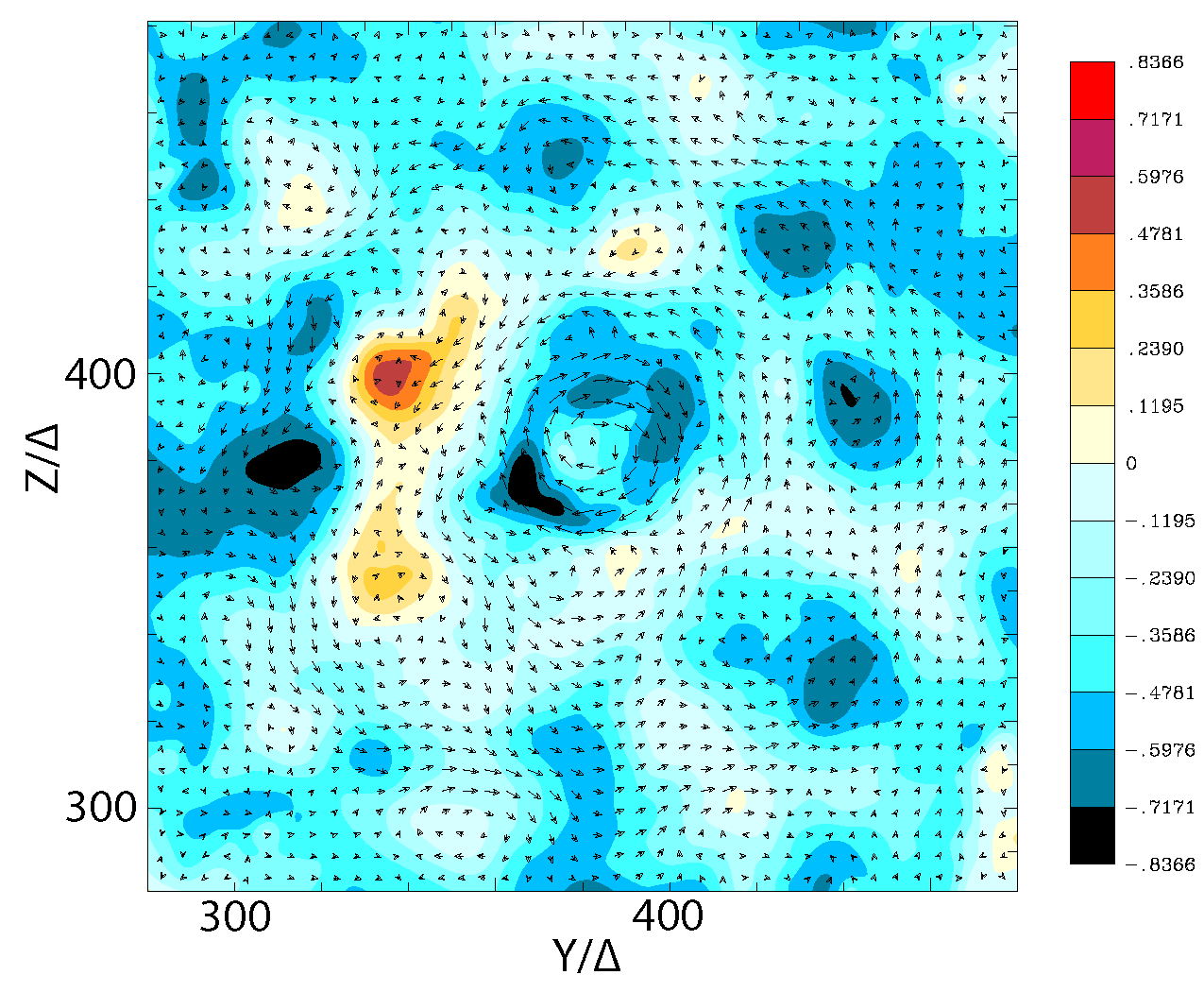}
\vspace*{-0.2cm}
\caption{a) $x$ - $\gamma v_{\rm x}$ distribution of jet (red) and ambient (blue)
electrons at $t = 600\,\omega_{\rm pe}^{-1}$, for an e$^{\pm}$ jet. The initial $\gamma v_{\rm x}  = 15$ is marked with the horizontal dashed black line. 
b) Color map of $E_{\rm x}$ with arrows showing $B_{\rm x, z}$ at  $y/\Delta = 381$. 
The cross section at $x/\Delta = 560$ is marked by vertical dashed lines. 
c)Color map of $E_x$ in the $y$ - $z$ plane at $x/\Delta = 560$, marked by the vertical line in 
panels a and b, with arrows indicating $B_{y,z}$. The maxima and minima of $E_x$  is (b) $\pm 1.003$ and (c) $\pm 0.8366$.}
\label{econtours}
\end{figure}

\begin{figure} 
\hspace*{4.1cm} {\bf e$^{\pm}$ jet} \hspace*{2.5cm} (a)

\hspace*{-0.05cm}
\includegraphics[width=0.99\linewidth]{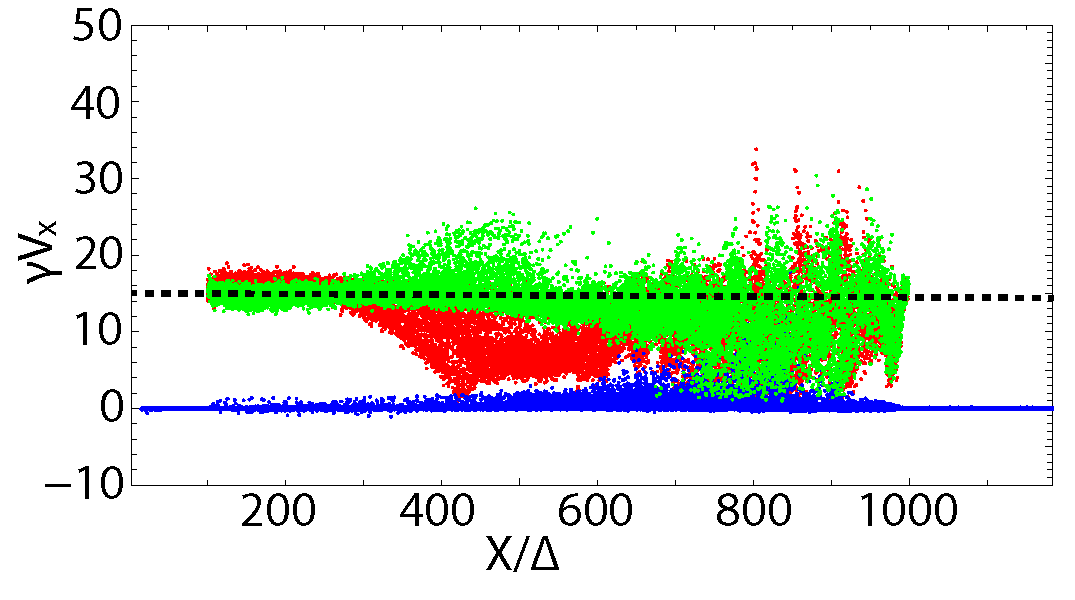}

\hspace*{7.2cm} (b)

\includegraphics[width=0.99\linewidth]{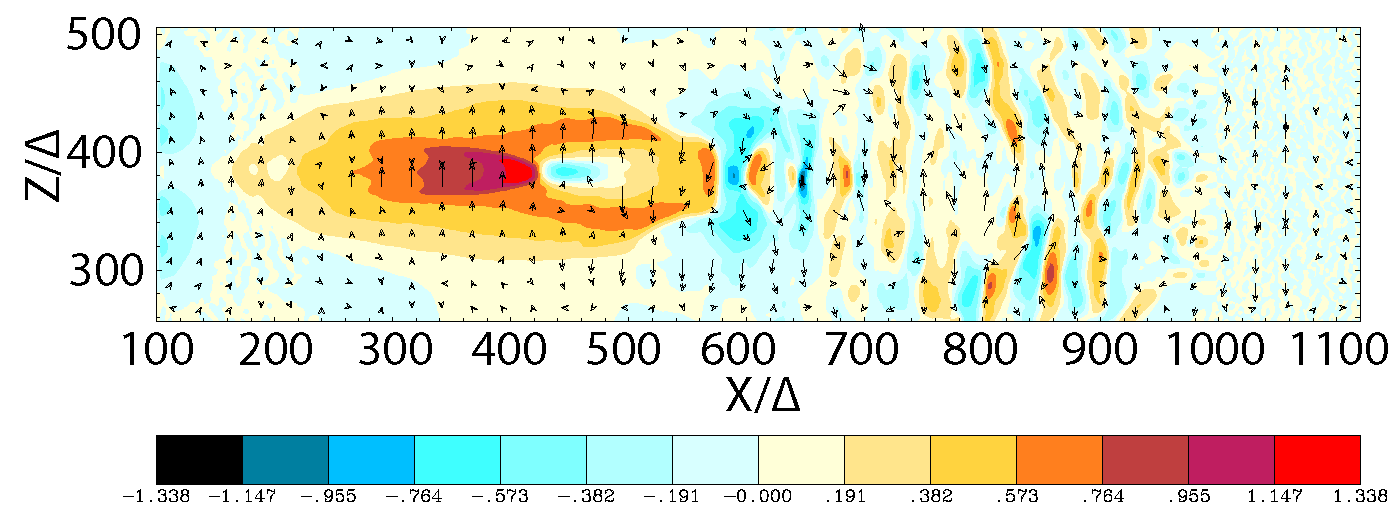}
\vspace*{-0.2cm}
\caption{a) The $x$ - $\gamma v_{\rm x}$ distribution of jet electrons (red), jet positrons (green) and 
ambient electrons (blue) at $t = 900\,\omega_{\rm pe}^{-1}$. 
b) Color map of $E_{\rm x}$ in the $x$-$z$ plane at $y/\Delta = 381$, with arrows indicating $B_{x,z}$. 
The dislocation of jet electrons and positrons generates the strips of the positive and negative $E_{\rm x}$
The maximum and minimum are $\pm 1.338$.}
\label{outofface}
\end{figure}

\begin{figure*}  

\hspace*{2.6cm} {\bf magnetized e$^{\pm}$ jet} \hspace*{1.8cm} (a)  \hspace*{3.3cm} {\bf magnetized e$^{-}$ - i$^{+}$ jet}
\hspace*{1.7cm} (b) 
\includegraphics[scale=0.6,angle=0]{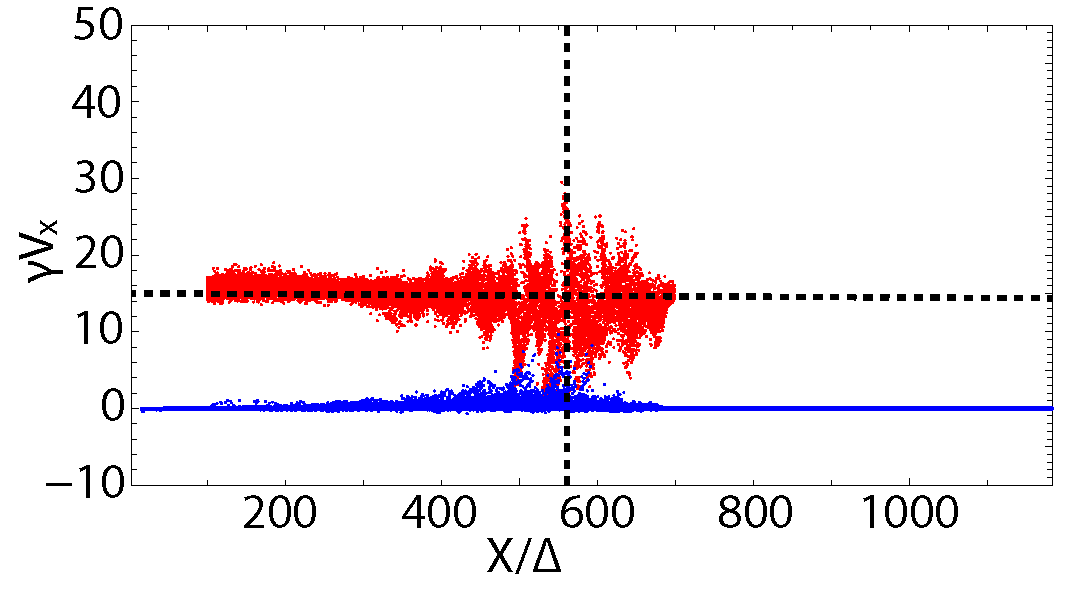}
\includegraphics[scale=0.6,angle=0]{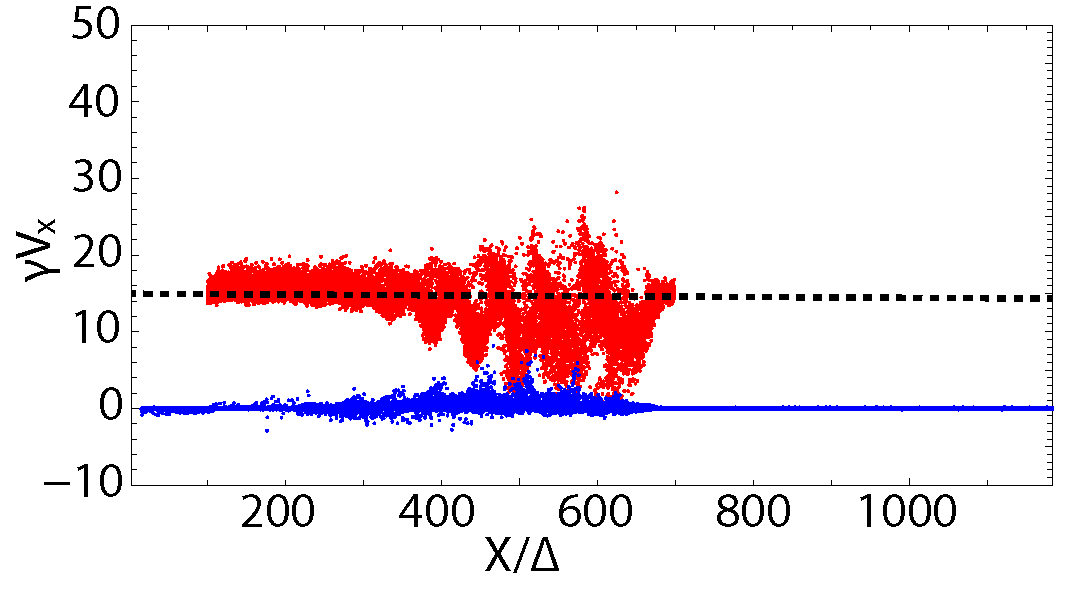}

\hspace*{6.3cm} (c) \hspace*{8.cm} (d)
\includegraphics[scale=0.6,angle=0]{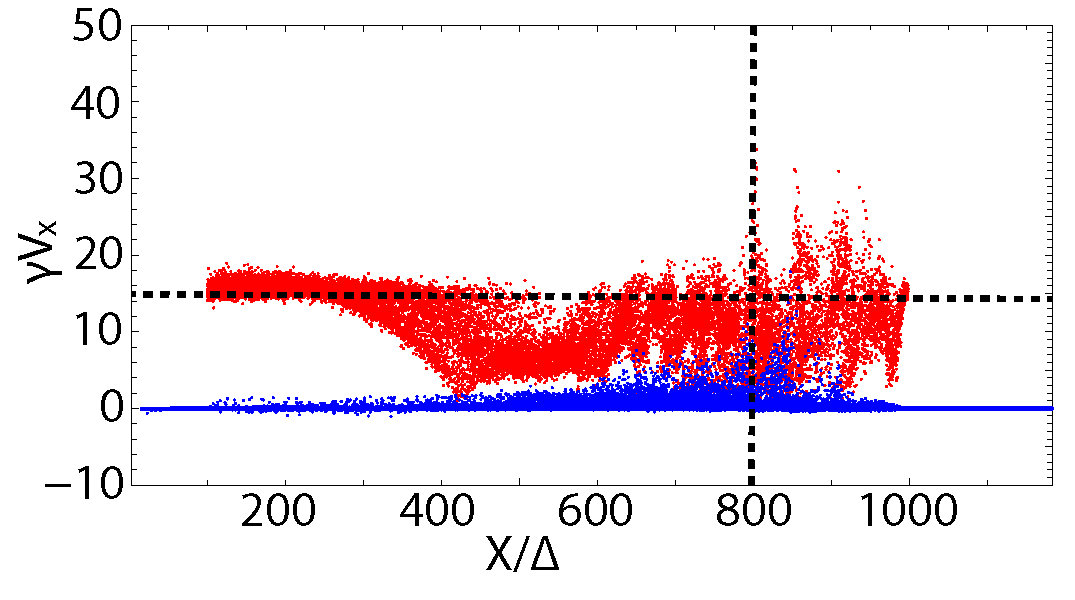}
\includegraphics[scale=0.6,angle=0]{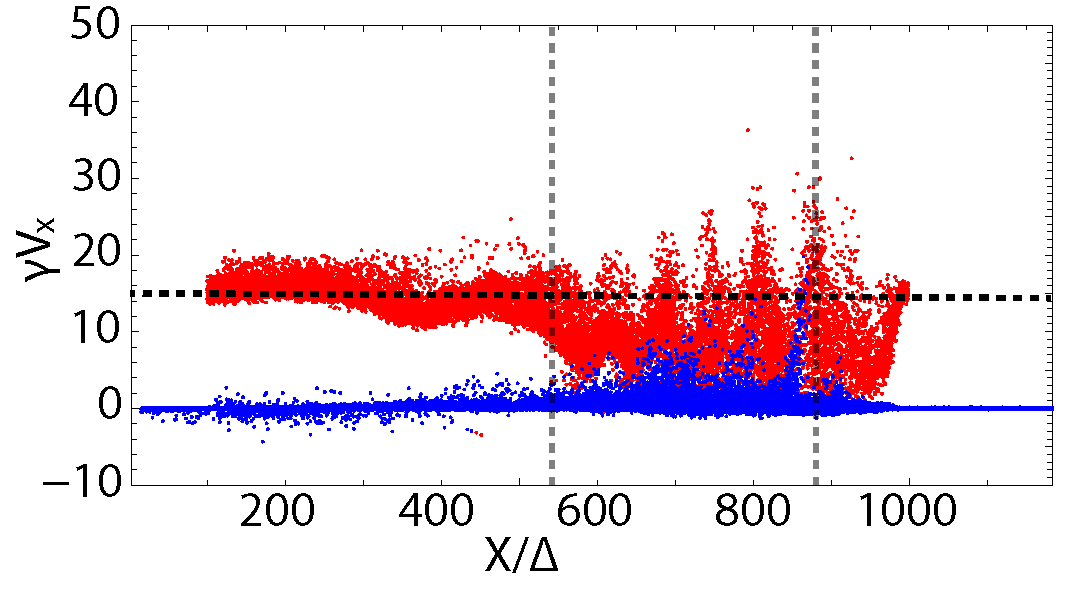}

\hspace*{6.3cm} (e) \hspace*{8.cm} (f)
\includegraphics[scale=0.6,angle=0]{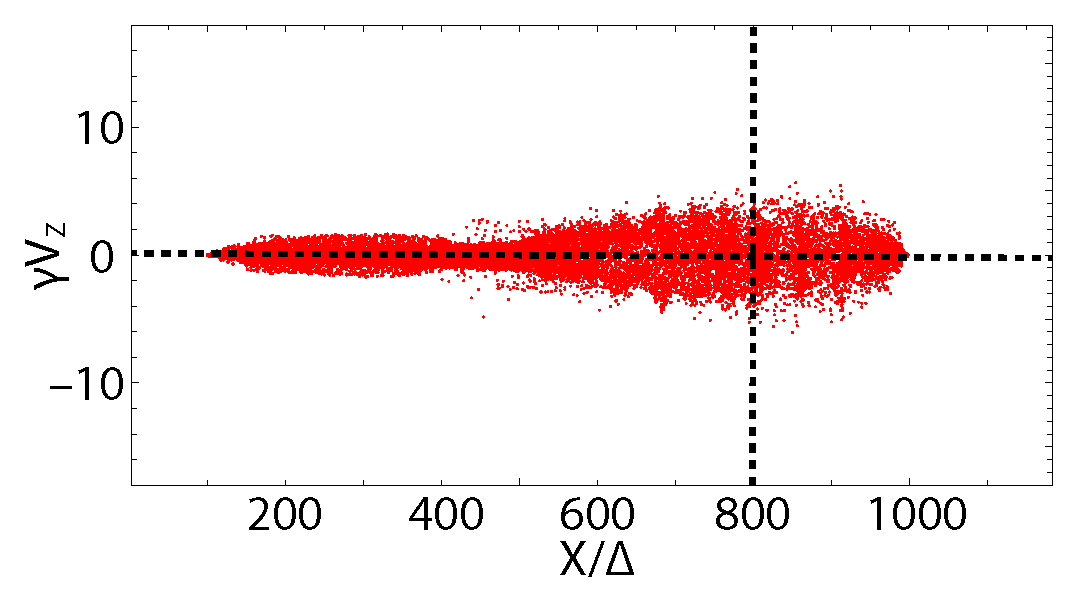}
\includegraphics[scale=0.6,angle=0]{dpx_vz16raS_010Ln}
\caption{Phase-space $x$ - $\gamma V_{\rm x}$ distributions, for the e$^{\pm}$ (a, c) jet and e$^{-}$ - i$^{+}$ (b, d) jet,
respectively, for $t =600\, \omega_{\rm pe}^{-1}$ (upper row) and $t =900\, \omega_{\rm pe}^{-1}$ (middle row). 
Panels (e) and (f) show the phase-space $x$ - $\gamma V_{\rm z}$ distributions, for the e$^{\pm}$  jet and e$^{-}$ - i$^{+}$ 
jet, respectively, at $t= 900\omega_{\rm pe}^{-1}$. 
In the e$^{\pm}$ jet the jet electrons are slightly decelerated and
later develop oscillations caused by kKHI and MI. In the e$^{-}$ - i$^{+}$ jet, the jet electrons decelerate in bulk 
before the oscillatory pattern is established. The red colour indicates jet electrons, the blue colour, ambient electrons.}
\label{dphas}
\end{figure*}

For both magnetized jets (Figs. \ref{ByBxz}a,b),  as expected, an WI is initially generated inside the jet.  
The wavelength of the WI is about $4\lambda_{\rm se}$.
One can see how residues of the WI remain at the jet head after $x/\Delta > 600$. This phenomenon was investigated (without the toroidal magnetic field) by \cite{Ardaneh16}.
Specifically, at the jet head of the e$^{\pm}$ jet, at $x/\Delta \approx 700$ an instability generates magnetic field filaments 
aligned with the jet propagation direction. Downstream along the jet an oblique mode  (slanted stripes) of the WI dominates, 
visible as consecutive concentrated magnetic fields between $x/\Delta \approx 280 - 630$  (Fig. \ref{ByBxz}a,b). 
Then, the MI and kKHI start to grow simultaneously in the same region of the jet at the jet-ambient medium boundary and across the jet, as seen in panel a). 
The excitation of the MI and the kKHI is merged with the WI, which results in slanted striped structures of the magnetic field in the e$^{\pm}$ jet.  
Because the growth rates of MI and kKHI are similar, the excited modes propagate towards the 
jet center. For the e$^{-}$ - i$^{+}$ jet the instabilities are stronger and more prominent, nevertheless they look rather similar
to the e$^{\pm}$ jet. Note that the wavelength of the kKHI mode is about $6\lambda_{\rm se}$, whilst the wavelength of the MI mode is about $5\lambda_{\rm se}$ 
along the jet radius  (i.e., perpendicular to the jet axis).  

Figure \ref{ByBxz}c shows the grown MI and kKHI in the non-linear stage at $x/\Delta > 700$, as indicated with the black arrow. 
We see here how the MI grows stronger and generates two dominant modes along the jet radius, the inner mode having larger amplitude. Simultaneously, the longitudinal 
kKHI wave modes modulate the magnetic field along the jet. 
For the e$^{-}$ - i$^{+}$ jet, Figure \ref{ByBxz}b shows that the MI grows around $x/\Delta = 450$ dominantly, it propagates toward
the jet center modulated by the kKHI modes already at the linear stage. 
The supplemental movie \footnote{MovieBy$\_$pairJet.mp4 at \href{https://zenodo.org/record/7017747}{doi: 10.5281/zenodo.7017747}} shows clearly the MI mode growth. 

In the non-linear stage, in Figure \ref{ByBxz}d we find a strong $B_{\rm y}$ component excited by  MI  at $t=900 \omega_{\rm pe}^{-1}$ (at approximately $580\lesssim x/\Delta \lesssim 620$). 
There is a magnetic field amplification in the non-linear stage, which can be attributed to the kKHI and MI which was 
similarly observed in the  unmagnetized case of \cite{nishikawa2016a} (see also Fig.~\ref{By4}b).
Here the outer MI mode merges with the inner mode (Fig. \ref{ByBxz}d) closer to the jet center, where one sees
comparatively the highly concentrated negative and then positive $J_x$ current in Fig. \ref{Jx4}d.
That means that the magnetic field structure is pinched by the MI which is strongly modulated 
due to the growth of kKHI along the jet, and  demonstrates that the field collimation is caused primary by the pinching of the jet electrons 
by the MI of the strong toroidal magnetic field present, towards the center of the jet. 
At the non-linear stage both MI and kKHI are saturated, and they 
subsequently become weakened and quasi-steady strips of $E_x$ are generated (see Fig. \ref{Bacc2} below) 
with a kKHI modulation along the jet propagation \footnote{Supplementary movies MovieBy$\_$eiJet.mp4, MovieBy$\_$pairJet.mp4 at \href{https://zenodo.org/record/7017747}{doi: 10.5281/zenodo.7017747}}.

\subsection{Electromagnetic fields and particle acceleration}

Figure \ref{bcontours} shows the pattern of the electron and ion  acceleration and deceleration in 
comparison with the structure of the electromagnetic field at $t = 900\,\omega_{\rm pe}^{-1}$ for the e$^{-}$ - i$^{+}$ jet.
The top panel (Fig.~\ref{bcontours}a) shows the $x$ - $\gamma v_{\rm x}$ distribution of the jet electrons (red), jet 
ions (green) and the 
ambient electrons (blue) electrons whilst the middle panel shows the $x-\gamma v_{\rm z}$ distribution of the jet 
electrons (Fig.~\ref{bcontours}b). The jet ions are propagated out of phase with jet electrons as electron-positron jet as shown in
\ref{econtours}a, however, jet ions are not accelerated as jet positrons are, due to their heavier mass. 
The initial $\gamma v_{\rm x}  = 15$ and $\gamma v_{\rm z}  = 0$ are marked 
with horizontal dashed black lines in Figures~\ref{bcontours}a and \ref{bcontours}b, respectively. The cross section 
at $x/\Delta = 540$ and it is marked with vertical dashed lines. In panel a) we see how electrons 
are accelerated-up in bunches between  $x/\Delta \approx 550 - 900$, reaching $\gamma v_{\rm x}\approx 35$.
Ambient electrons are also accelerated but slightly delayed (shifted), and consequently accelerated to
$\gamma v_{\rm x}\approx 20$. Figure \ref{bcontours}b showing the $x$ - $\gamma v_{\rm z}$ distribution for the jet
electrons, indicates as well bunches of jet electrons accelerating mainly between $x/\Delta \approx 550 - 900$, reaching values
of $\approx 10$. 

Figure \ref{bcontours}c shows the $E_x$ component of the electric field at the cross section 
at the center of the jet ($z/\Delta = 381$), the strong positive electric field is located at $x/\Delta = 540$ indicated by the blue dashed line.
Figure \ref{bcontours}d shows the $E_x$ component of the electric field
in the $y-z$ plane at $x/\Delta = 540$, which is marked by the vertical lines in panels (a), (b) and (c).
Black arrows indicate $B_{y,z}$. 
A strong electric field 
region visible in  Fig. \ref{bcontours}d is located near the center of the jet ($x/\Delta = 540$), which is confirmed with
the collimation of jet electrons in Fig. \ref{ede4}d, and it corresponds to the MI mode that 
is dominant at this same $x$-location of the jet in Fig.~\ref{ByBxz}d. The concentric pattern around the jet center ($m = 0$) is excited.
Outside this mode, another MI is excited with $m=5$.

Figure~\ref{econtours}a shows the $x-\gamma v_{\rm x}$ distribution of the jet (red) 
and the ambient (blue) electrons for the e$^{\pm}$ jet
in the linear stage at time $t=600 \omega_{\rm pe}^{-1}$.  
It illustrates the electron acceleration and 
deceleration along the jet, with some jet electrons reaching $\gamma v_{\rm x} \approx 30$ at $x = 560\Delta$. 

To understand how the jet electrons are accelerated we will examine the $E_{\rm x}$ in the $x - z$ (Fig.~\ref{econtours}b) 
and $y - z$ (Fig.~\ref{econtours}c) planes. 
The cross section is $x/\Delta = 560$ and it is marked with the vertical dashed line for comparison needs.
Figures \ref{econtours}b and \ref{econtours}c show two MI modes in the transverse plane that form a symmetric, concentric 
pattern ($m > 5$) around the jet center ($z/\Delta =381$) and have a wavy, complicated structure along the $z$ direction. 
The jet electrons that are accelerated most significantly up to $\gamma v_x\approx 30$ are located  
around $x/\Delta \approx 560$ where a negative quasi-steady$E_{\rm x}$ is found at the outer layer of the jet.
It is further discerned that excited MI and kKHI accelerate electrons (at the linear stage) at $x/\Delta = 560$. 
The kKHI modulates the jet structure along the jet propagation direction which leads to the complex electromagnetic 
field pattern along the jet (see also Fig. \ref{outofface}b).  This is responsible for the stratified jet electrons (positrons),
with the out-of-phase mode as will be shown in Fig. \ref{outofface}a below.
\begin{figure*} 
\hspace*{3.1cm} {\bf e$^{\pm}$ jet} \hspace*{2.7cm} (a)  \hspace*{3.8cm} {\bf e$^{-}$ - i$^{+}$ jet}
\hspace*{2.2cm} (b) 
\includegraphics[scale=0.35,angle=0]{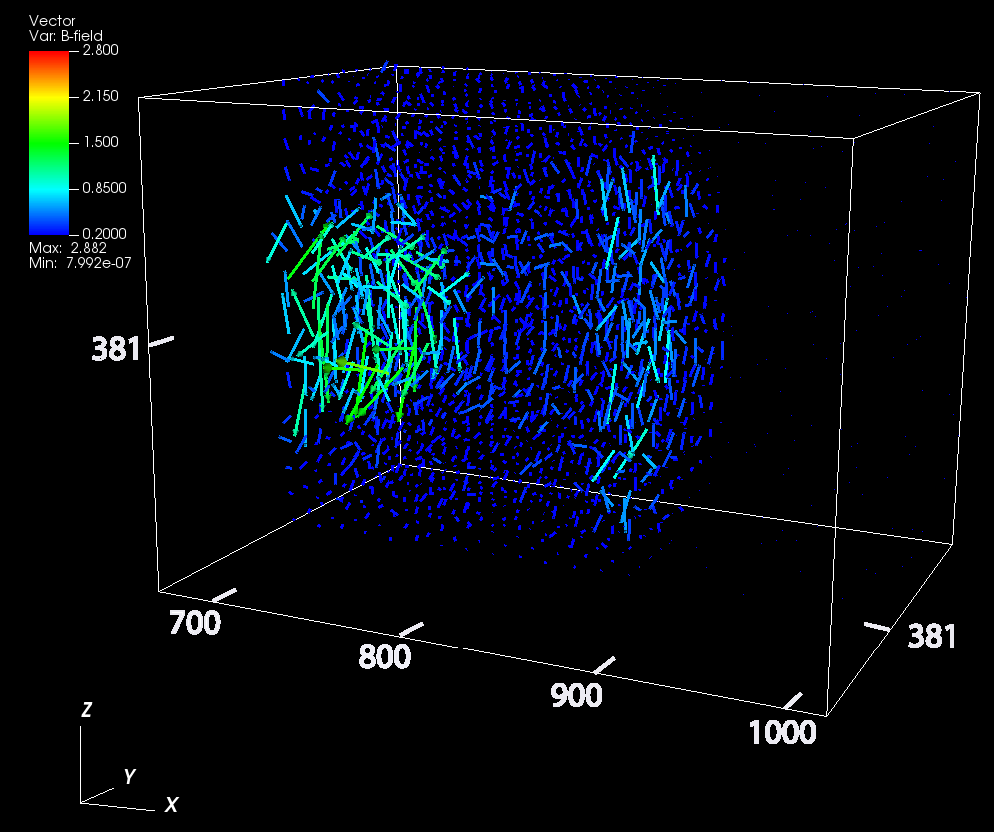}
\includegraphics[scale=0.35,angle=0]{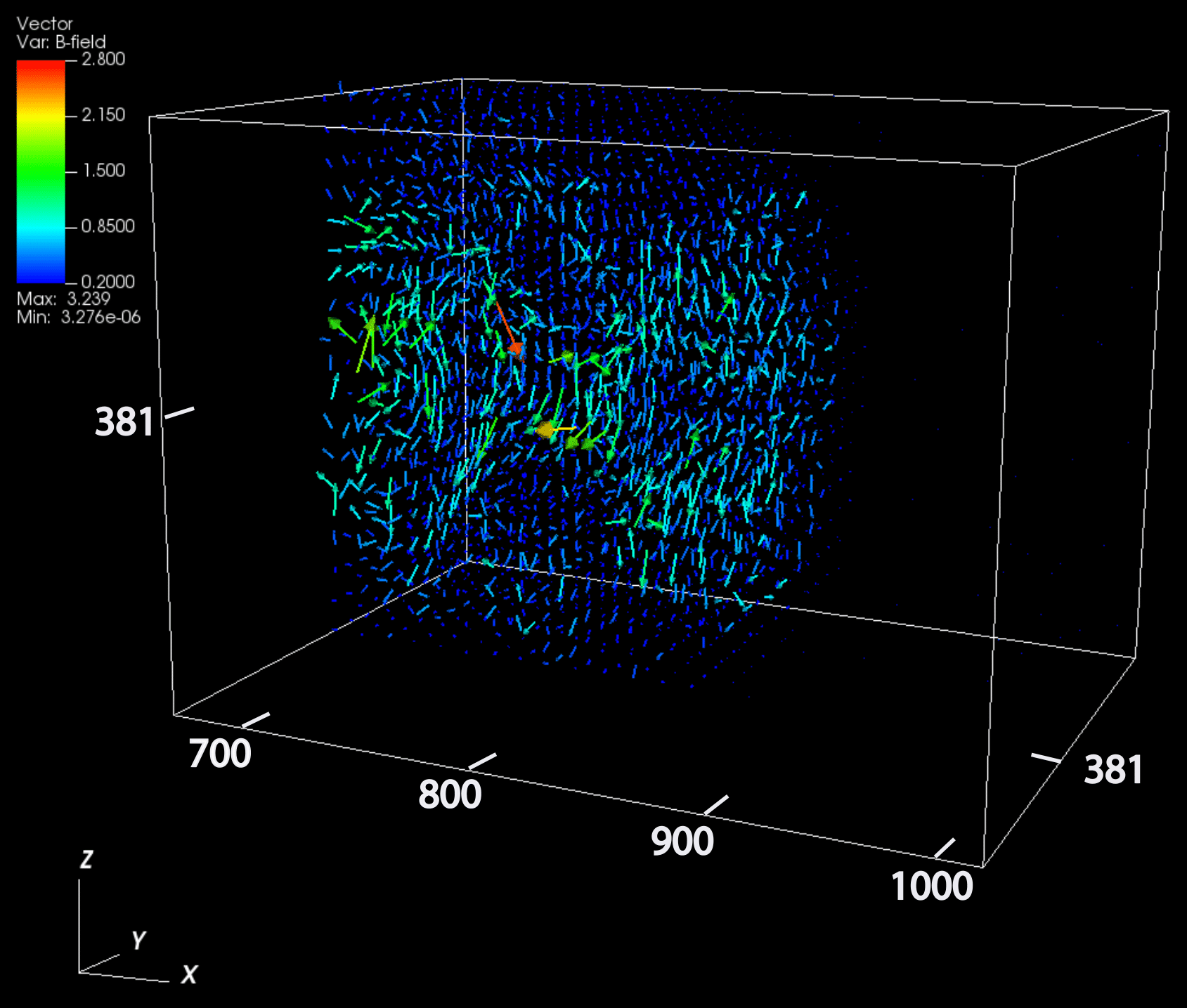}
\hspace*{6.6cm} (c) \hspace*{7.3cm} (d)
\includegraphics[scale=0.35,angle=0]{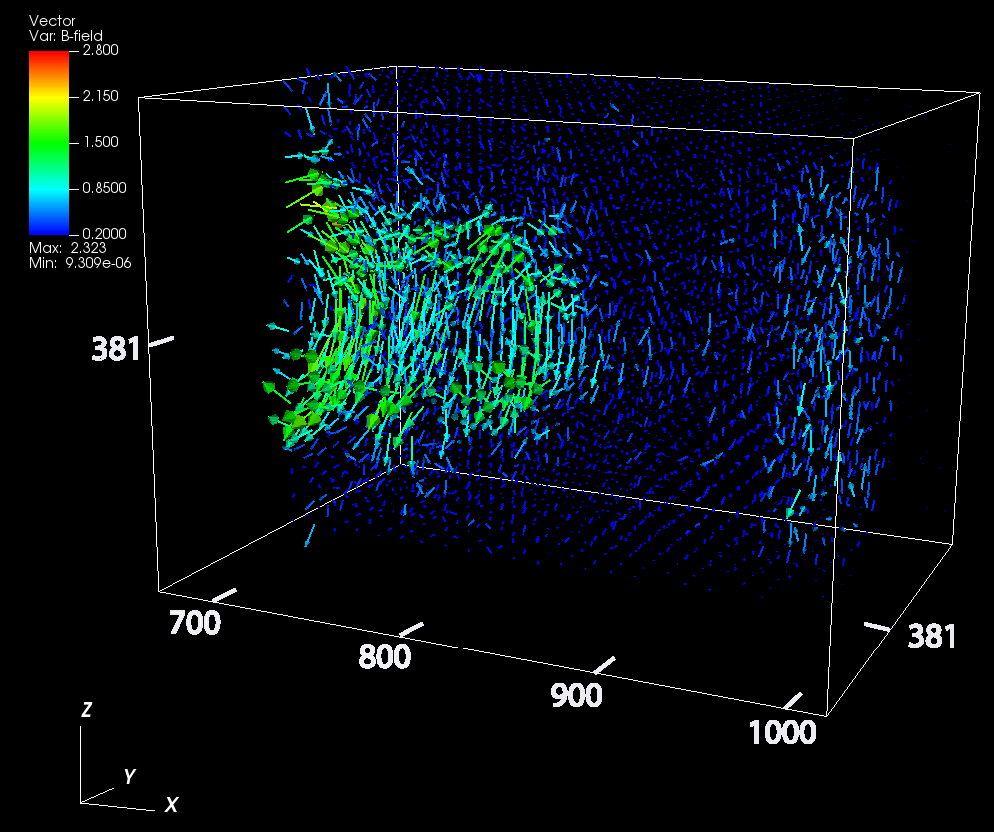}
\includegraphics[scale=0.35,angle=0]{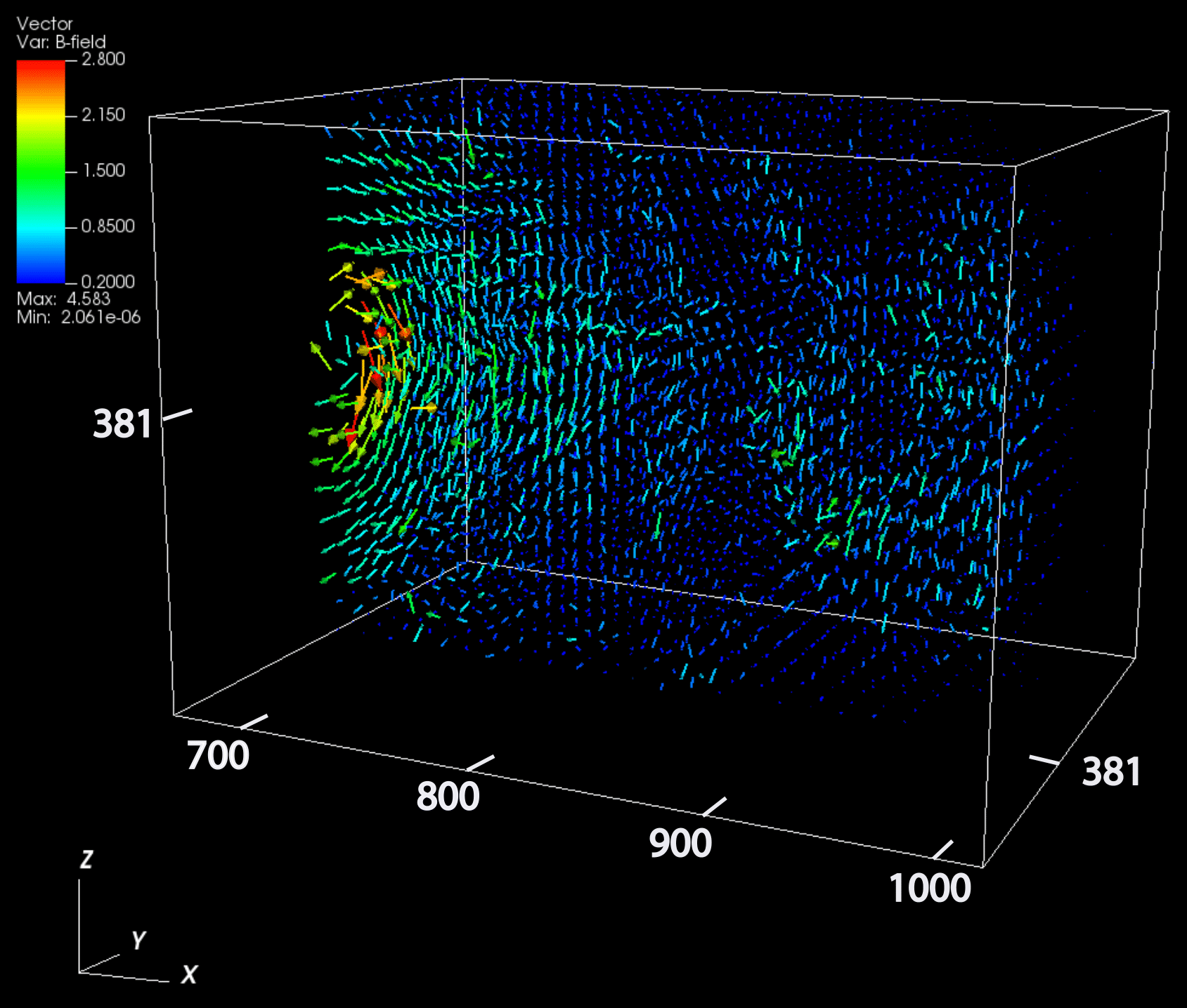}
\caption{The magnetic field vectors within the cuboid  
($670 < x/\Delta < 1020; 256<y/\Delta, z/\Delta < 506$) at $t =800\, 
\omega_{\rm pe}^{-1}$ (first row) and $t =900\, 
\omega_{\rm pe}^{-1}$ (second row) for e$^{\pm}$ (panel a, c) and e$^{-}$ - i$^{+}$ (panel b, d). The center 
of the jet is at $y/\Delta =z/\Delta=381$. For the magnetic field inside the jet,  
the plots show half of the regions clipped at the center of jet in the $x$ - $z$ plane 
($381 < y/\Delta < 506$). The red dashed squares in Fig. \ref{By4}c and \ref{By4}d show the volume of these plots.
The maximum and minimum of the legend of the magnetic field strength are 2.8 and 0.2.}
\label{3DBV}
\end{figure*}
Note that the energetic jet electrons around $x/\Delta\approx 560$ at $t = 600\,\omega_{\rm pe}^{-1}$ are accelerated further around $x/\Delta\approx 960$ at $t = 900\,\omega_{\rm pe}^{-1}$, as shown in Figs. \ref{outofface}a and \ref{dphas}c below.
This indicates that the quasi-steady negative electric field propagates with the jet and accelerates 
the jet electrons up to $\gamma v_{\rm x}  \sim 35$ as shown above.

Figure \ref{outofface}a shows the $x$ - $\gamma v_{\rm x}$ distribution of jet electrons (red) with the out-of-phase jet positrons (green) 
and ambient  electrons (blue), 
for a pair jet at $t = 900\,\omega_{\rm pe}^{-1}$.
The temporal analysis of phase-velocity distributions ($x$ - $\gamma v_{\rm x}$) reveals that the bunched jet electrons 
(positrons) propagate with the jet velocity, which indicates that the generated patterns of $E_{\rm x}$ are quasi-steady 
in time as shown in Fig. \ref{outofface}b. To note that it is the dislocation of jet electrons and positrons which generates 
the strips of the positive and negative $E_{\rm x}$. Such structures of $E_{\rm x}$ are in other words formed because
the combined modes of MI and kKHI first propagate obliquely at the linear stage as shown in 
Figs. \ref{econtours}b and \ref{econtours}c, and later in the non-linear stage 
they become more vertical ($550 < x/\Delta <900$), see Fig.~\ref{outofface}b. 

It should be noted that Figure \ref{By4}c shows a rather weak $B_{\rm y}$ ($650 < x/\Delta <1000$); however 
Figure \ref{outofface}b shows striped patterns of $E_{\rm x}$ in the $x$ - $z$ plane at $y/\Delta = 381$, 
which are generated by the out-of-phase distributions of jet electrons and positrons as shown in Fig. \ref{outofface}a.  
We understand that the slight dislocations of the jet electrons and positrons constitute a response to the magnetic field structures, 
as the corresponding variations in $v_z$ suggest (cf. Fig.\ref{dphas}e below), generating the strips of the positive and negative $E_{\rm x}$ through 
$\dot{\mathbf{E}}\propto \mathbf{j}$. 
The strips of the positive and negative $E_{\rm x}$ seen in the color-map of the same figure.  
\textit {In other words this means that the quasi-steady $E_x$ accelerates jet electrons and positrons out of phase}.

Figure \ref{dphas} shows the phase-space $x-\gamma v_{\rm x}$ distributions for the e$^{\pm}$ (a, c) 
jet and e$^{-}$ - i$^{+}$ (b, d) jet respectively, for $t =600\, \omega_{\rm pe}^{-1}$ (upper row) and $t =900\, \omega_{\rm pe}^{-1}$ (middle row). 
Red colour indicates the jet electrons and blue the ambient electrons.
At first glance, the phase-space distributions indicate electron acceleration at different locations. 
Initially the jet electrons have a Lorentz factor of $\gamma \simeq 15$.
Figure \ref{dphas}a shows that these get accelerated and decelerated by the excited instabilities as shown in Fig. \ref{econtours}. 
Very similar behavior we see for the magnetized e$^{-}$ - i$^{+}$ jet as well (Fig. \ref{dphas}a).

In figure \ref{dphas}c (e$^{\pm}$  jet) we can observe two different stages of acceleration at $t =900\, \omega_{\rm pe}^{-1}$ : 
The linear stage at $300 <x/\Delta <800$  where jet electrons are weakly accelerated and decelerated (by the MI and kKHI),  
and the non-linear stage at $800 <x/\Delta <1000$,  where they are considerably accelerated on three bunches. 
At the non-linear stage electrons reach maximum energies of $\gamma \approx 35$.

\begin{figure*} 
\hspace*{3.5cm} (a)  \hspace*{1.5cm} {\bf magnetized e$^{\pm}$ jet} \hspace*{5.0cm} (b) 
\hspace*{0.1cm}
\includegraphics[width=0.41\linewidth]{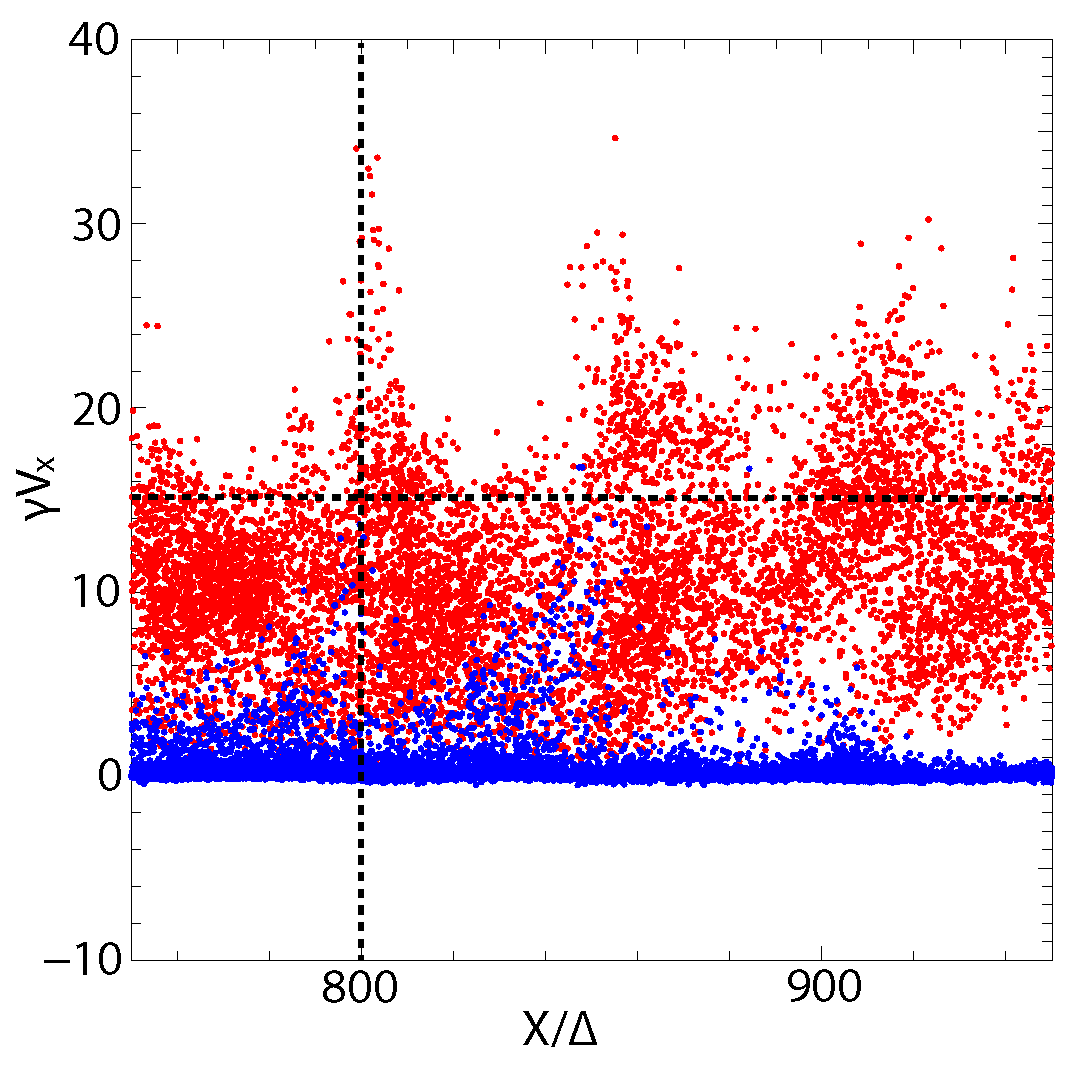}
\hspace*{1.1cm}
\includegraphics[width=0.49\linewidth]{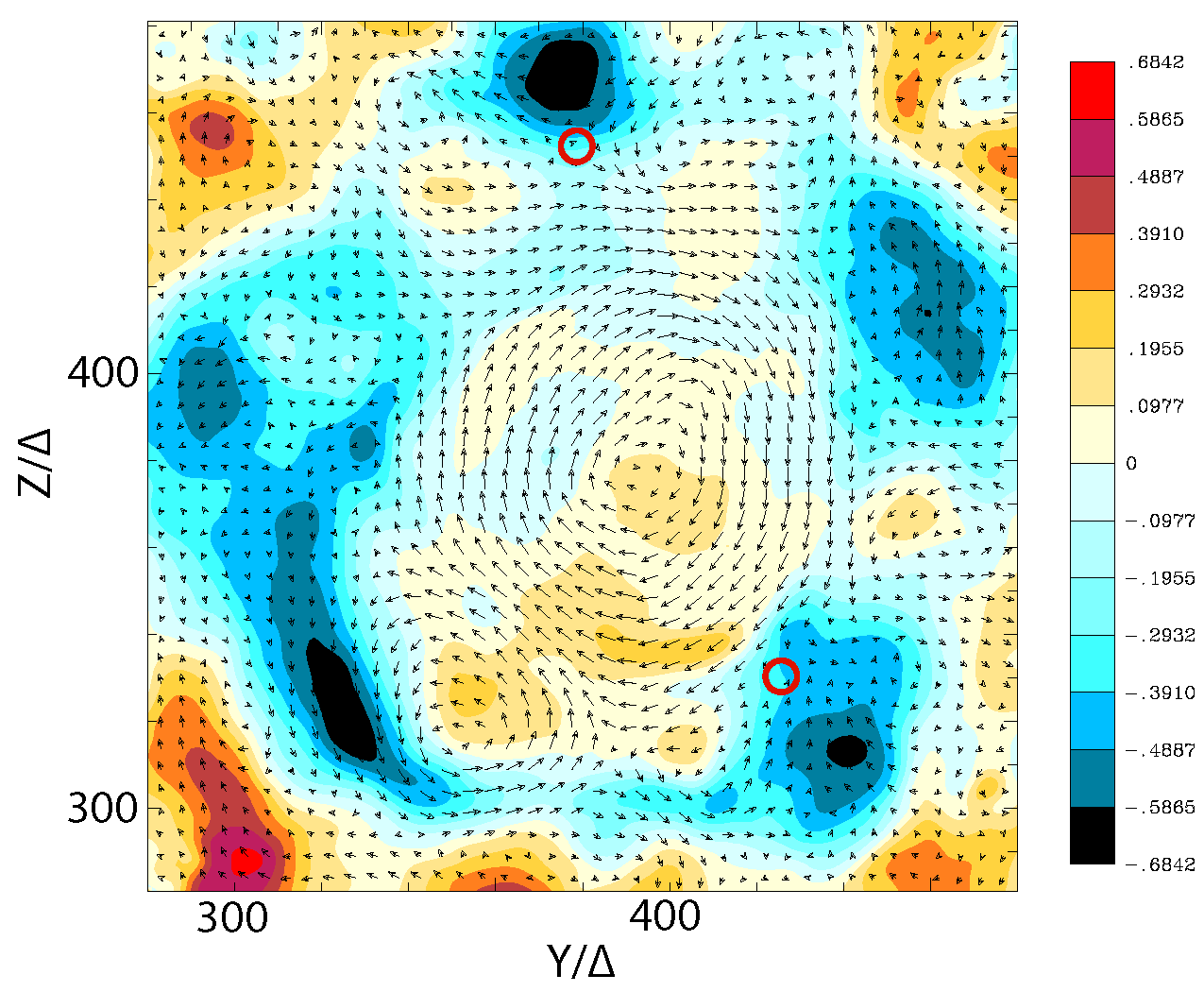}

\hspace*{3.2cm} (c) \hspace*{8.3cm} (d)

\includegraphics[width=0.49\linewidth]{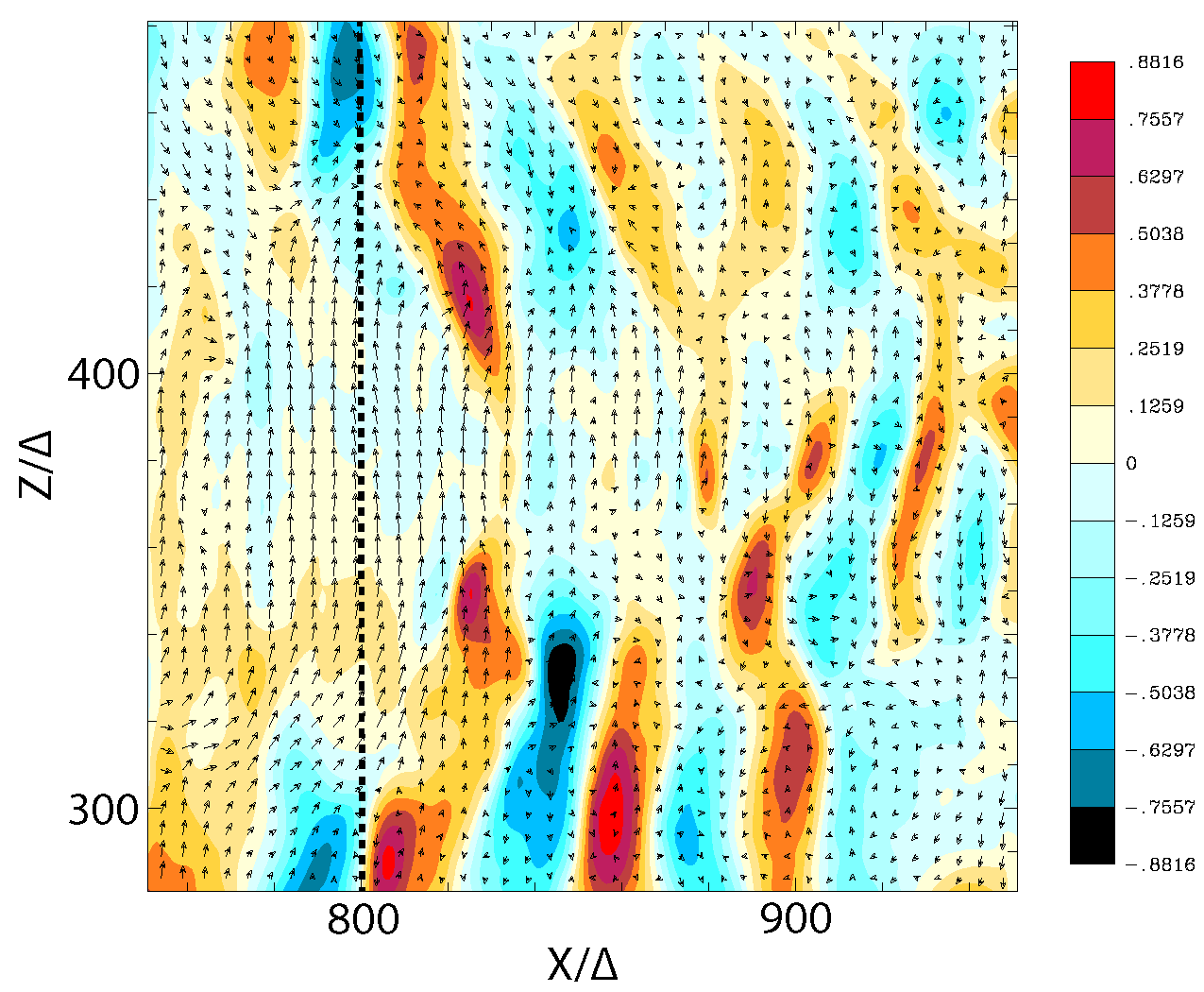}
\includegraphics[width=0.49\linewidth]{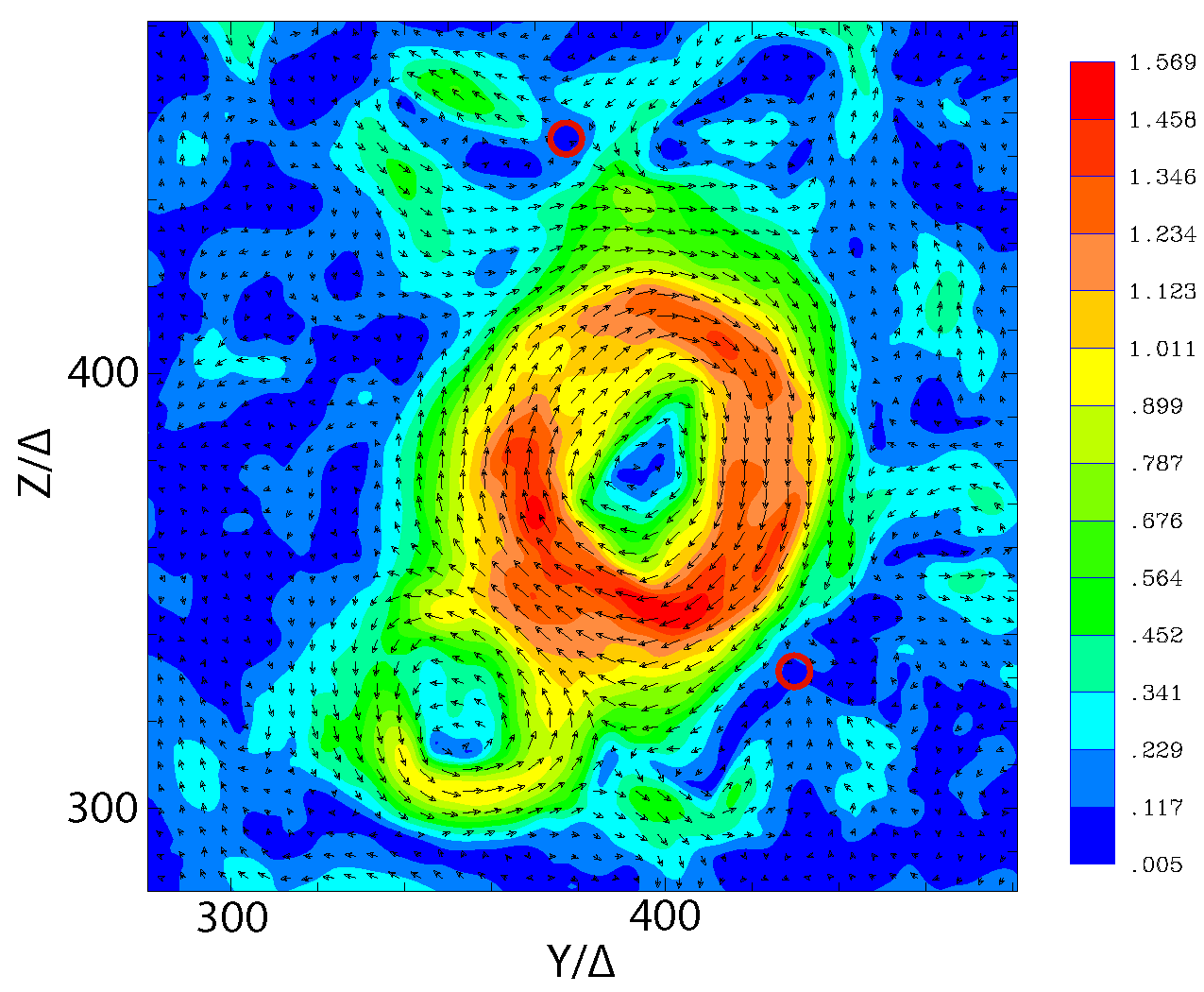}
\caption{Possible acceleration mechanism of jet electrons in the $e^{\pm}$ jet at time $t =900\, \omega_{\rm pe}^{-1}$.  
a) The phase space $x- \gamma V_{\rm x}$, the peak of the jet electrons (red dots) is located at $x = 800\Delta$. b) Color map of $E_{\rm x}$ 
in the $y - z$ plane at $x/\Delta = 800$, marked by the vertical line in panels a and c with the arrows of $B_{\rm y, z}$.  
c) Color map of $E_{\rm x}$ in the $x - z$ plane at $y/\Delta = 381$, with arrows indicating $(B_{\rm x},B_{\rm z})$.  d)
the total magnetic field strength  at $x/\Delta=800$, for $281<y/\Delta, z/\Delta < 481$. The arrows indicate 
the magnetic field $(B_{\rm y},B_{\rm z})$; red circles indicate possible reconnection sites.
The maximum and minimum vales of $E_{\rm x}$ are (b): $\pm 0.6842$ and (c): $\pm .08816$.  The maximum and minimum values in panel (d) are 1.569 and 0.005, respectively.}
\label{Bacc}
\end{figure*}

Respectively, in figure \ref{dphas}d for the e$^{-}$ - i$^{+}$ jet, we can discern that the most significant acceleration
occurs between $550 <x/\Delta <900$ (linear and non-linear stage) where as we discussed above,
a quasi-steady $E_x$ prominently appears, and  the jet electrons reach values of $\gamma \approx 35$. 
So in this case the acceleration is stronger because of the instabilities present but also because of the $E_x$ presence (as shown
in Fig.\ref{bcontours}). Interestingly, the maximum electron energy is similar for both jet species. Also, the ambient electrons (blue) 
in the both jets are accelerated up to  $\gamma \approx 15$ within multiple bunches, in the range $650\lesssim x/\Delta \lesssim 850$, 
but in the  e$^{-}$ - i$^{+}$ jet ambient electrons are participating earlier in the acceleration comparing to the e$^{\pm}$ jet case.
There are correlations of the energy gains and losses between the ambient and the jet
electrons, because the jet electrons propagate with the saturated instabilities and the peaks of the accelerated
jet electrons are rather sharp. However, the ambient electrons do not move with the excited waves, and thus instead the peaks of the accelerated ambient electrons are mostly slanted toward the jet propagation. It should be noted that the acceleration of
the ambient electrons stops at approximately $x/\Delta\approx 950$, which is explained by the fact that the
electromagnetic fields dissipate around $x/\Delta\approx 900$.

At $t=900\ \omega_{\rm pe}^{-1}$ the acceleration region of the ambient electrons around the  $e^{\pm}$ jet (Fig. \ref{dphas}c) approximately coincides with the jet region at $800 < x/\Delta < 950$, in which a strong magnetic turbulence after the dissipation is observed, see Fig. \ref{3DBV}.  
The non-linear saturation of the instabilities ends, hence the magnetic fields dissipate 
and acceleration of jet electrons (not the ambient electrons) occurs.  

Figures \ref{dphas}e and \ref{dphas}f, show the $x-\gamma v_{\rm z}$ as a function of $x/\Delta$ for the 
jet electrons in e$^{\pm}$ and e$^{-}$ - i$^{+}$ jets.
In the e$^{\pm}$ jet, the jet electrons are strongly pinched around  $400 < x/\Delta < 500$ being decelerated, while gradually 
start to accelerate at the later stages as stratifications develop, caused by the kKHI and MI as shown in Fig. \ref{dphas}e. 
In contrast, Fig. \ref{dphas}f shows that in the e$^{-}$ - i$^{+}$ jet, electrons are accelerated perpendicularly and some of them are reflected 
(in the case with $m_{\rm p}/m_{\rm e} = 1836$)- this is due to the large mass ratio and the slim jet used in this case - and then
 decelerated along the $x$-direction 
due to the MI (or kKHI) (Fig. \ref{dphas}d), which clearly indicates magnetic deflection, as also shown in Fig. \ref{bcontours} and discussed
above (Figs. \ref{bcontours}, \ref{econtours}).
Some of the ambient electrons are also strongly accelerated as they are swept up into the relativistic jet plasma.

\begin{figure*} 
\hspace*{3.5cm} (a)  \hspace*{1.5cm} {\bf magnetized e$^{-}$ - i$^{+}$ jet} \hspace*{5.0cm} (b) 
\hspace*{-0.2cm}
\includegraphics[width=0.41\linewidth]{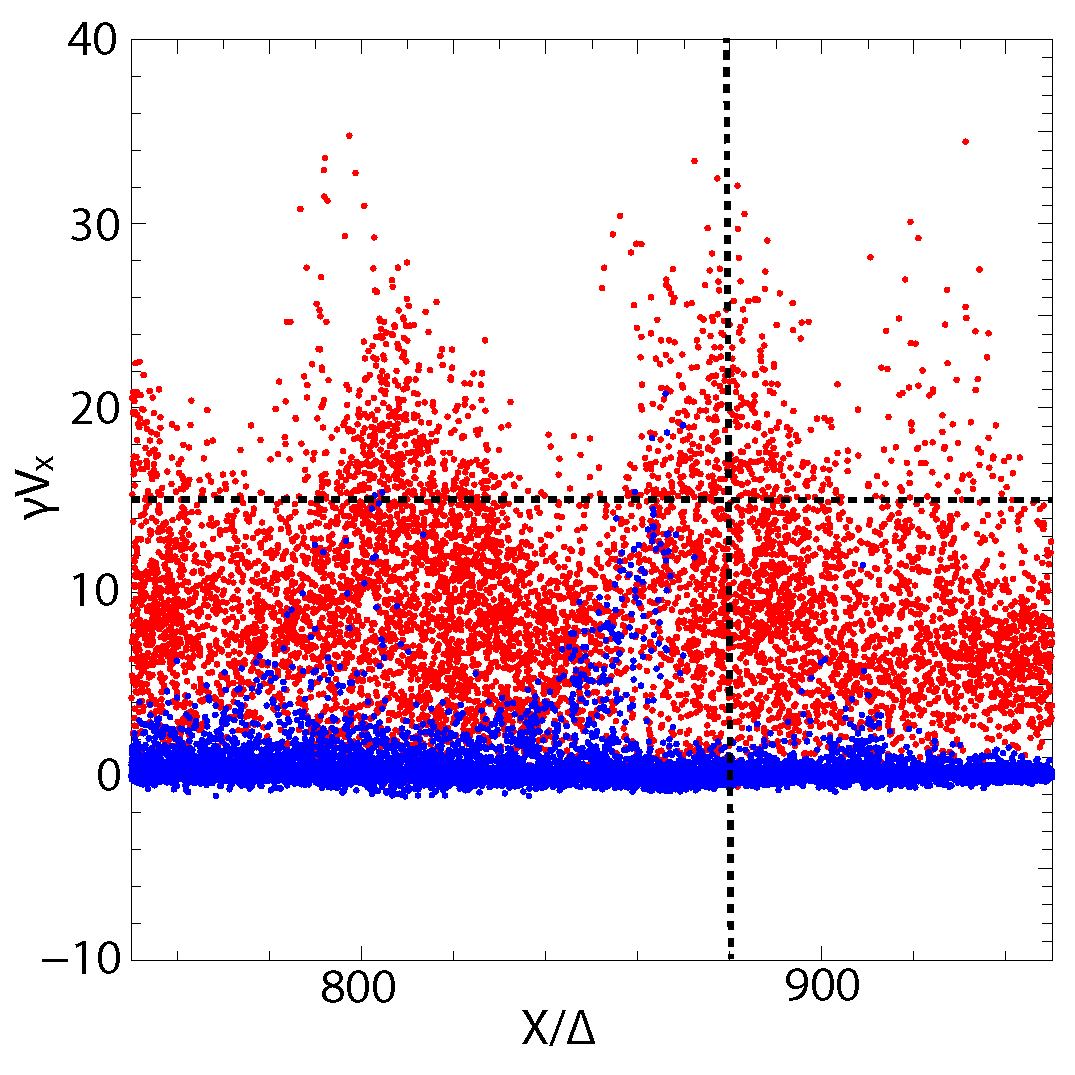}
\hspace*{1.2cm}
\includegraphics[width=0.49\linewidth]{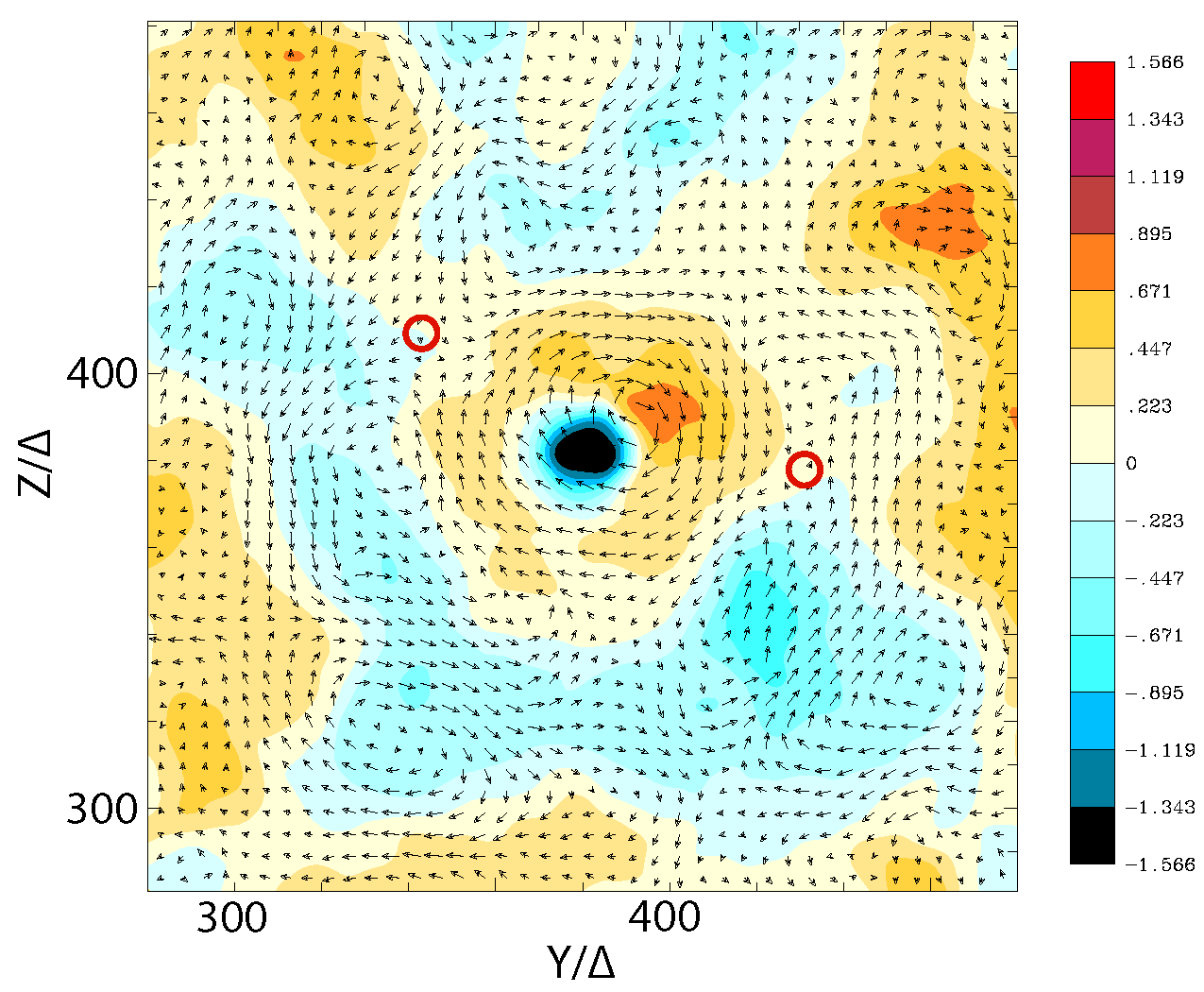}

\hspace*{3.2cm} (c) \hspace*{8.3cm} (d)

\includegraphics[width=0.49\linewidth]{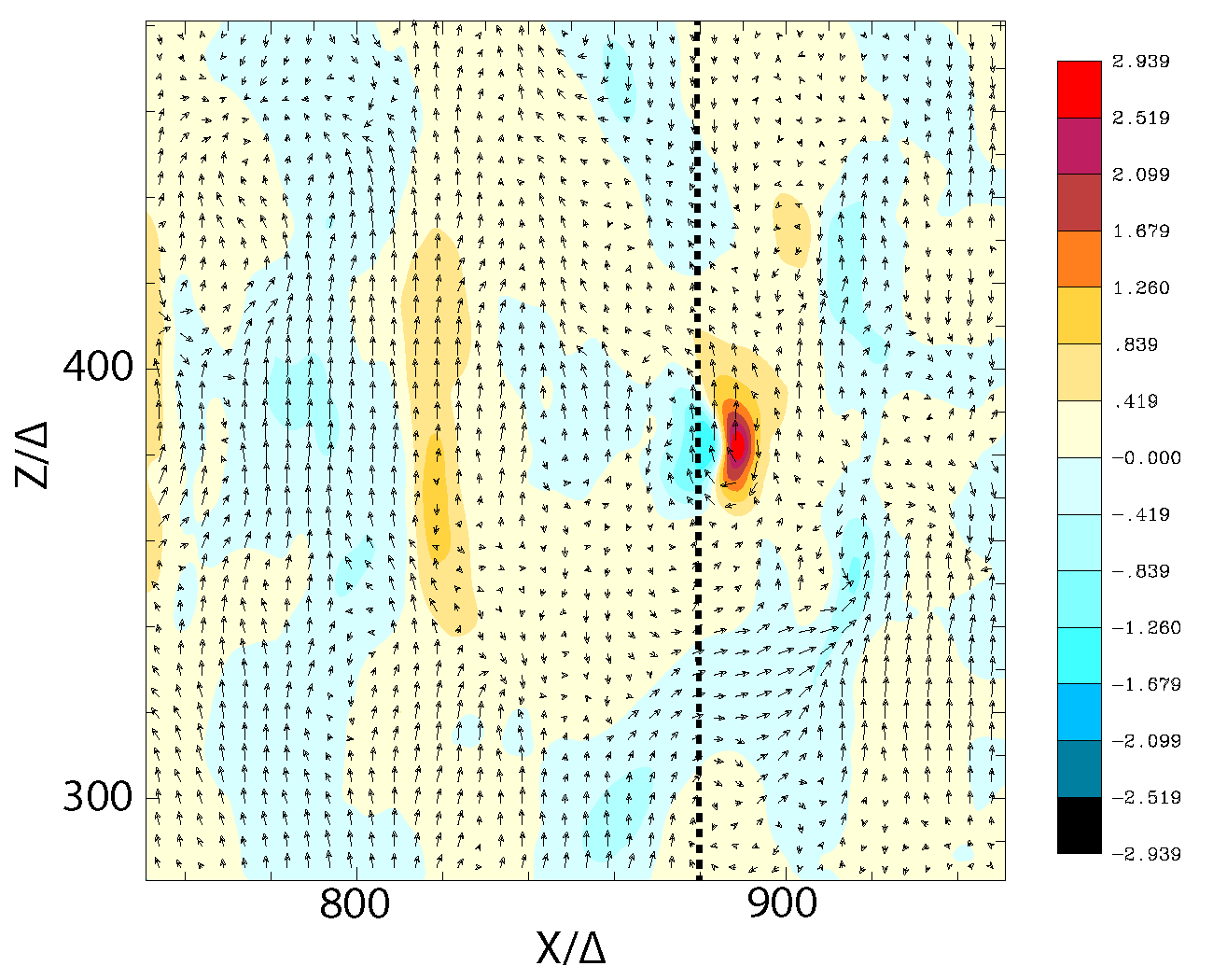}
\includegraphics[width=0.49\linewidth]{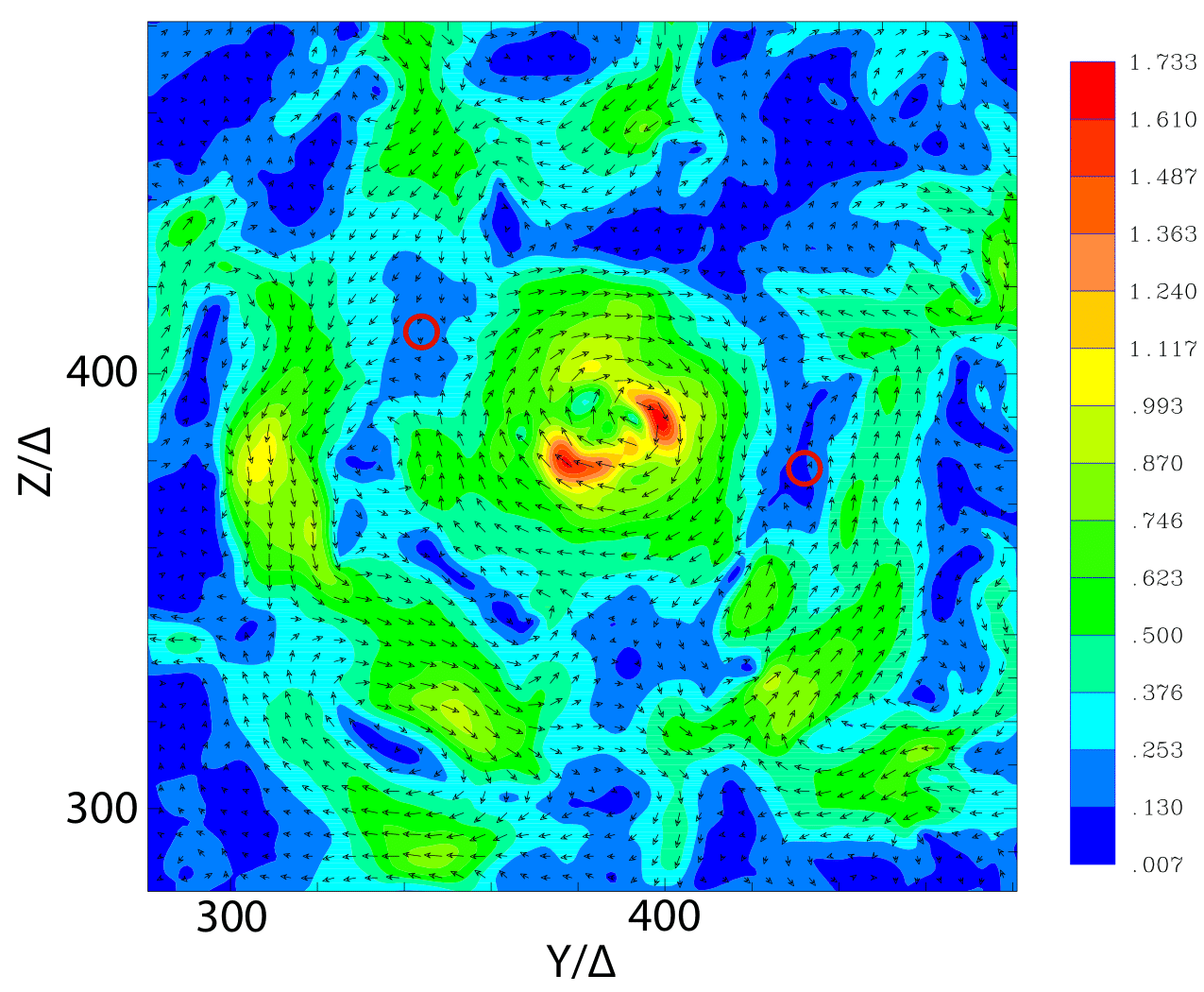}
\caption{Possible acceleration mechanism of jet electrons in the $e^{-}-i^{+}$ jet at time $t =900\, \omega_{\rm pe}^{-1}$.  
a) The phase space $x- \gamma V_{\rm x}$, the peak of the jet electrons (red dots) is located at $x = 880\Delta$. b) Color map of $E_{\rm x}$ in the $y - z$
plane at $x/\Delta = 880$ marked by the vertical line in panel a with the arrows of $B_{\rm y, z}$.  c) Color map of $E_{\rm x}$ in the $x - z$ plane 
at $y/\Delta = 381$, with arrows indicating $(B_{\rm x},B_{\rm z})$.  d)
the total magnetic field strength  at $x/\Delta=880$, for $281<y/\Delta, z/\Delta < 481$. The arrows indicate 
the magnetic field $(B_{\rm y},B_{\rm z})$; red circles indicate possible reconnection sites.}
\label{Bacc2}
\end{figure*}


\subsection{Three-dimensional magnetic field evolution - reconnection}

Figure \ref{3DBV} shows the 3D evolution of the magnetic field near the jet head in the region indicated by the red rectangle in Fig. \ref{ByBxz}.
The figure shows the magnetic-field vectors within a cuboid ($670 < x/\Delta < 1020;\,  256<y/\Delta, z/\Delta < 506$) at $t =800\, \omega_{\rm pe}^{-1}$ and $t =900\, \omega_{\rm pe}^{-1}$ for the e$^{\pm}$ (a, c) and the e$^{-}$ - i$^{+}$ (b, d) jets with their center at $y/\Delta =z/\Delta=381$. The plots show a half-section of the jet's center 
in the $x-z$ plane with $381 < y/\Delta < 506$ in order to view the interior of the jets. 
For the e$^{\pm}$ jet, the jet-head is located at $x/\Delta = 900$ at time $t=800\ \omega_{pe}^{-1}$ (panel a), and it moves to $x/\Delta = 1000$ 
later at $t=900\ \omega_{pe}^{-1}$ (c). Comparing panels a) and c) one observes that the magnetic fields between $800 < x/\Delta < 1000$ first get twisted and then dissipate. 
Around $x/\Delta\approx 850$, at both time-steps - at the end of the linear jet evolution stage - 
the magnetic fields (generated by the outer MI mode) have dissipated. 
The magnetic fields in the inner MI mode get weak, but reappear after $x/\Delta \approx 980$, as shown in Fig. \ref{3DBV}c. 
Fig. \ref{3DBV} indicates that a similar form of electron acceleration occurs during the dissipation of magnetic 
fields around $x/\Delta =800$ for the e$^{\pm}$ jet species.  It is important to note that the acceleration of
the electrons in the non linear stage, between $800 <x/\Delta < 1000$ is not only due to the electric field of the generated instabilities, 
but also it is correlated to the dissipation of the magnetic fields. 
At the non-linear stage the magnetic field generated by instabilities dissipates, and although the cause of this is not clear presently, 
it might be due to the termination of the non-linear saturation as described by \cite{blandford2017}.

The e$^{-}$ - i$^{+}$ jet, in Figure \ref{3DBV}b, shows a weakened magnetic field 
at the jet centre, surrounded by swirling magnetic fields. For this jet the front edge of the toroidal magnetic field at the centre is peeled-off  during the jet propagation, as seen in Fig. \ref{3DBV}d. 
This indicates that the toroidal magnetic field might have dissipated at the non-linear stage and as a consequence moved towards the 
jet boundary where flux ropes are formed, as discussed in \cite{blandford2017}, see Figs. \ref{Bacc} and \ref{Bacc2}.

\begin{figure*} 
\hspace*{2.1cm} {\bf e$^{\pm}$ jet} \hspace*{2.7cm} (a) \hspace*{3.7cm} {\bf e$^{-}$ - i$^{+}$ jet}
\hspace*{2.5cm} (b) 
\hspace*{-0.3cm}
\includegraphics[scale=0.36,angle=0]{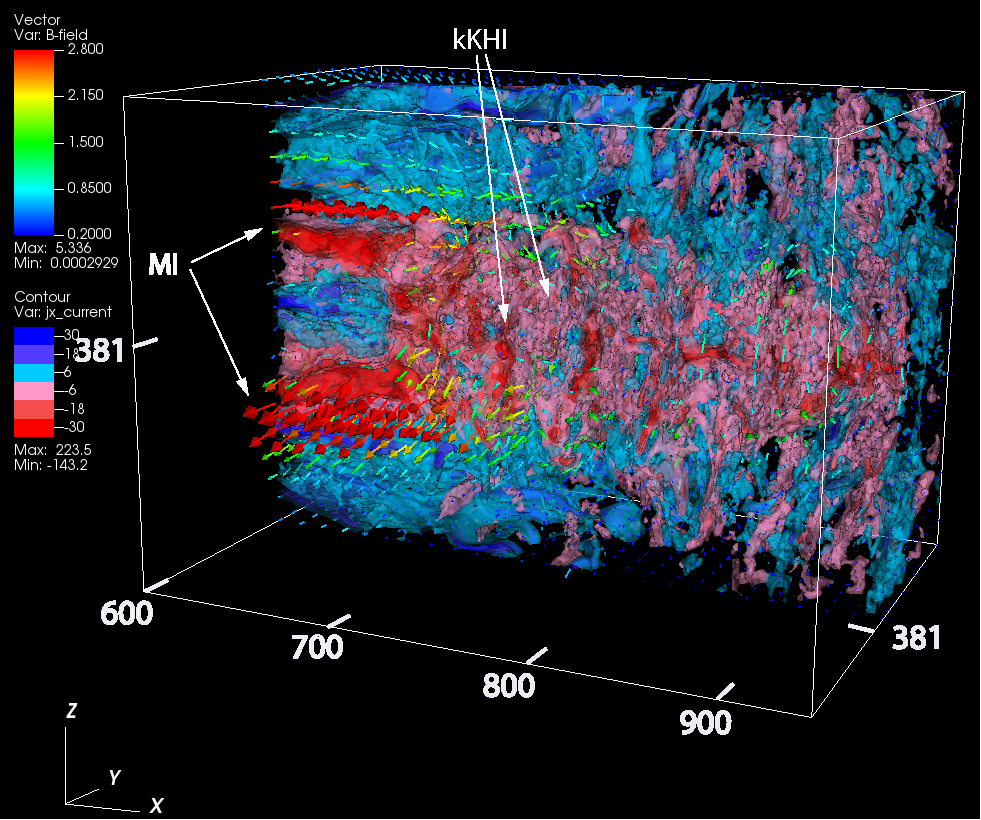}
\includegraphics[scale=0.36,angle=0]{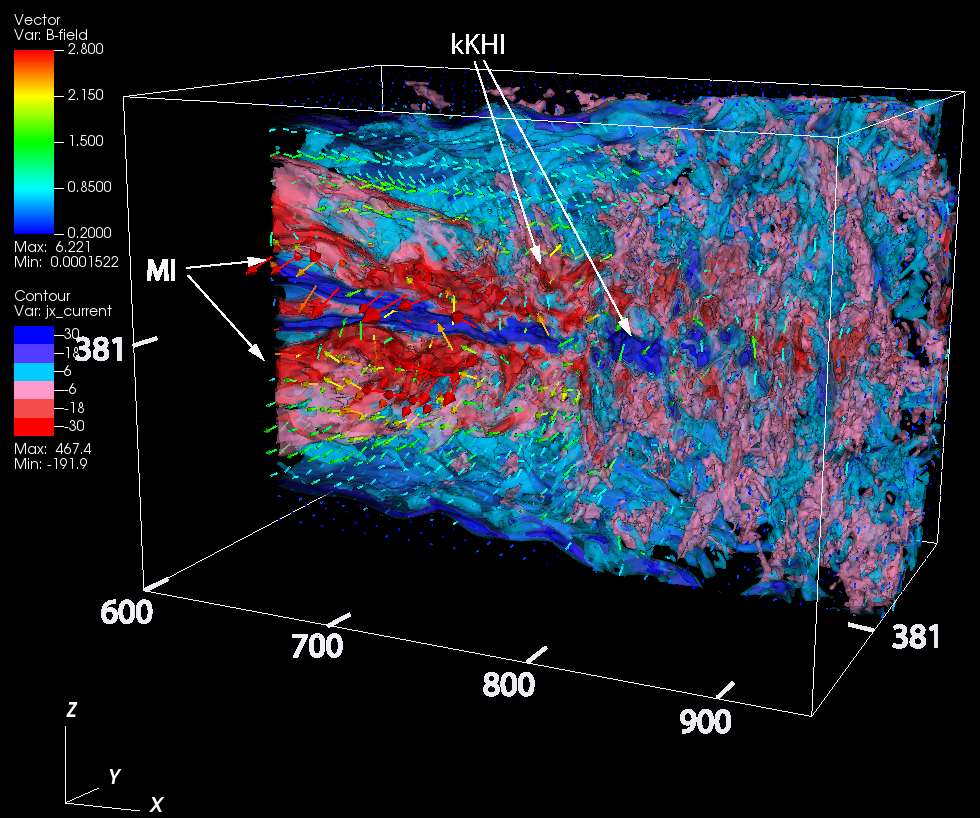}
\caption{The $x$-component $J_{\rm x}$ of the current within a cuboid of $600 < x/\Delta < 950$; $256<y/\Delta,
z/\Delta < 506$) at $t =1000\, \omega_{\rm pe}^{-1}$ for an e$^{\pm}$ (a) and e$^{-}$ - i$^{+}$ (b) jet. 
Both plots show the cross section at the centre of the jet ($y/\Delta =z/\Delta=381$). 
The blue dashed squares in Fig. \ref{By4}c and \ref{By4}d show the volume of these plots. 
There is  an overlap with Figs. \ref{3DBV}c and \ref{3DBV}d including the range $770 < x/\Delta < 1120$.} 
\label{ReconS}
\end{figure*}

Figure \ref{Bacc} shows four different plots indicating a possible acceleration mechanism of jet electrons for an $e^{\pm}$ jet, at 
$t =900\, \omega_{\rm pe}^{-1}$.  Figure \ref{Bacc}a shows the phase space $x- \gamma V_{\rm x}$.  
The peak of the jet electrons (red dots) is located at $x = 800\Delta$. 
A color map of $E_{\rm x}$ is shown in Fig. \ref{Bacc}b with the arrows of $B_{\rm y, z}$ marked by the vertical line
in Fig. \ref{Bacc}a,at $x/\Delta = 800$. The strong negative $E{\rm x}$ near the jet boundary is responsible for the acceleration of electrons. Figure \ref{Bacc}c correspondingly shows a colour map of $E_{\rm x}$ in the $x - z$ plane at $y/\Delta = 381$,  with the arrows indicating 
a $(B_{\rm x},B_{\rm z})$.  Figure \ref{Bacc}d shows
the total magnetic field strength  at $x/\Delta=800$, for $281<y/\Delta, z/\Delta < 481$. The arrows indicate 
the magnetic field components $(B_{\rm y},B_{\rm z})$; red circles indicate possible reconnection sites.
As a necessary condition, the reconnection location should coincide with the regions of minimum magnetic field strength, indicated by the dark blue color in Fig. \ref{Bacc}d.   The red circles indicate possible reconnection sites where the magnetic fields are oriented on the opposite direction.

For an $e^{-}$ - i$^{+}$ jet we correspondingly show in Fig. \ref{Bacc2} four panels, indicating a possible 
acceleration mechanism for the jet electrons, at time $t =900\, \omega_{\rm pe}^{-1}$ comparable to Fig. \ref{Bacc}.
Panel a) shows the phase space $x- \gamma V_{\rm x}$ where the peak of the jet electrons (red dots) is located at $x = 880\Delta$.
A color map of the electric field component $E_{\rm x}$ with the arrows of the magnetic field component $B_{\rm y, z}$
is shown in panel b). The strong negative $E{\rm x}$ at the center of the jet is responsible for the electron acceleration. 
Panel c) depicts a color map of $E_{\rm x}$ in the $y - z$ plane at $x/\Delta = 880$, marked by the vertical line in panel a), 
with arrows indicating the $(B_{\rm y},B_{\rm z})$.  Panel d) shows the total magnetic field strength  
at $x/\Delta=$880, for $281<y/\Delta, z/\Delta < 481$. The arrows indicate the magnetic field $(B_{\rm y},B_{\rm z})$ and the red circles 
on panels b and d, indicate possible reconnection sites. 

What we observe in this study is that the inner MI mode becomes dominant due to the collimation of the jet electrons, which generate 
the clockwise circular magnetic field. At the same time, the outer MI mode in the jet starts to split, resulting in a number of magnetic structures through the residues of the MIs. These structures are found in the areas of weak magnetic fields (blue) where swirling and/or 
oppositely directing magnetic fields exist, thus a possible reconnection site is marked by a red circle where the direction of the magnetic fields changes. 
In Fig. \ref{Bacc}d specifically, we see typical morphologies of a magnetic reconnection site. In our simulation results we have observed 
that possible reconnection sites are surrounded by oppositely swirling magnetic field patterns with a minimum magnetic field strength (dark blue). 
Note that the magnetic fields are produced by the jet current modulated by the excited kKHI and MI. At this time the outer MI mode is 
dissipated as shown in Fig. \ref{3DBV}c as well where the magnetic field is weak near the jet boundary.  The toroidal field structure gets 
distorted and dispersed, as also seen in the movies\footnote{The total magnetic field in the $y-z$ plane at $289 < y/\Delta,\, z/\Delta <480$; 
for e$^{\pm}$ and e$^{-}$ - i$^{+}$ jets respectively, 
see supplementary videos Movie Btot$\_$pairJet.mp4, MovieBtot$\_$eiJet.mp4, at \href{https://doi.org/10.5281/zenodo.7017747}
{doi:10.5281/zenodo.7017747}}  
provided as supplementary material for the two jet compositions (electron-positron and electron-ion), showing the spatial evolution 
of the magnetic field and their differences. 
Subsequently comparing panels d) of Figs. \ref{Bacc}, \ref{Bacc2} while inspecting the provided movies, one discerns how  
the magnetic field gets reorganized and forms multiple magnetic flux ropes. 
The supplemental movies for both magnetized jet compositions\footnote{see MovieBtot$\_$pairJet.mp4, MovieBtot$\_$eiJet.mp4 at \href{https://doi.org/10.5281/zenodo.7017747}{doi: 10.5281/zenodo.7017747}} further show that the magnetic structures interact with each other and with the surrounding environment, generating the right conditions for magnetic reconnection.

\subsection{Non-linear instabilities growth and acceleration}

To investigate the transition from the late linear stage to the non-linear stage, Figure \ref{ReconS} shows
the 3D magnetic field vectors within a cuboid of $600 < x/\Delta < 950$, $256<y/\Delta, z/\Delta < 506$ 
(indicated by the red dashed square in Fig. \ref{By4}c and \ref{By4}d)
at t =$1000\, \omega_{\rm pe}^{-1}$ for e$^{\pm}$ and e$^{-}$ - i$^{+}$ jets. 
For the magnetic field to be shown inside the jet, the plots depict half of the jet regions clipped at the center of jet in the $x$ - $z$ plane
($381 < y/\Delta < 506$). Note that these plots overlap with Fig. \ref{3DBV} at $x/\Delta = 770$.  
Beyond $x/\Delta=750$ for the e$^{\pm}$ jet, a magnetic field disruption of the outer mode of the MI occurs, 
resulting  in a disordering via a non-linear saturation of the kKHI \& MI, that is seen up to $x/\Delta \approx 950$. 
Around $x/\Delta = 780$ as the magnetic field near the jet boundary dissipates, the magnetic field near the centre 
of the jet gets dissipated around $x/\Delta = 820$. 

\begin{figure*} 
\hspace*{2.9cm} {\bf magnetized e$^{\pm}$ jet} \hspace*{1.8cm} (a)  \hspace*{3.9cm} {\bf magnetized e$^{-}$ - i$^{+}$ jet}
\hspace*{1.5cm} (b)
\includegraphics[scale=0.60,angle=0]{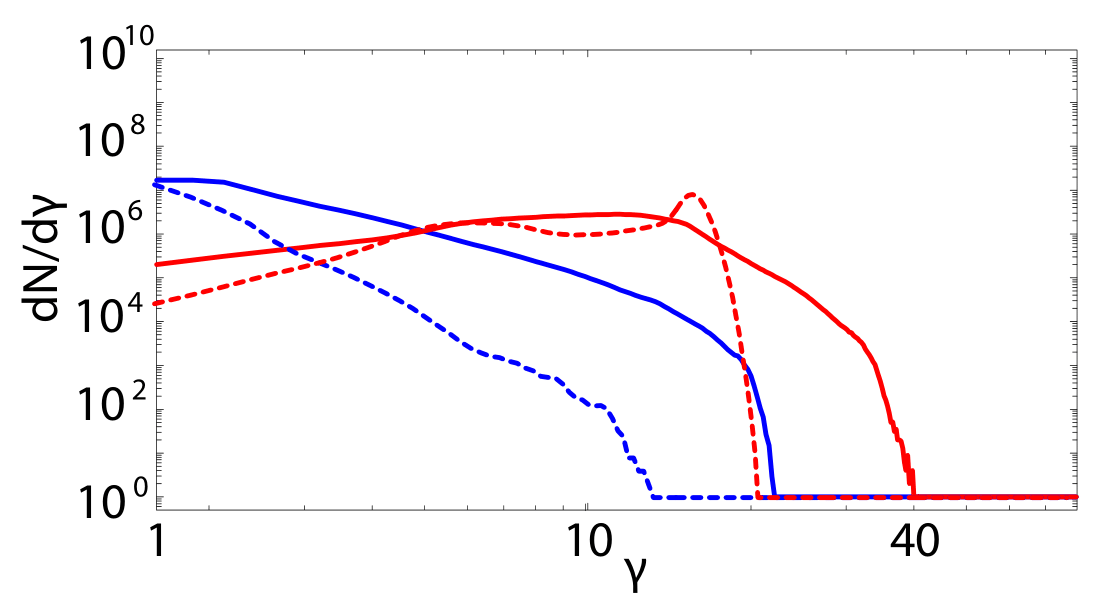}
\includegraphics[scale=0.60,angle=0]{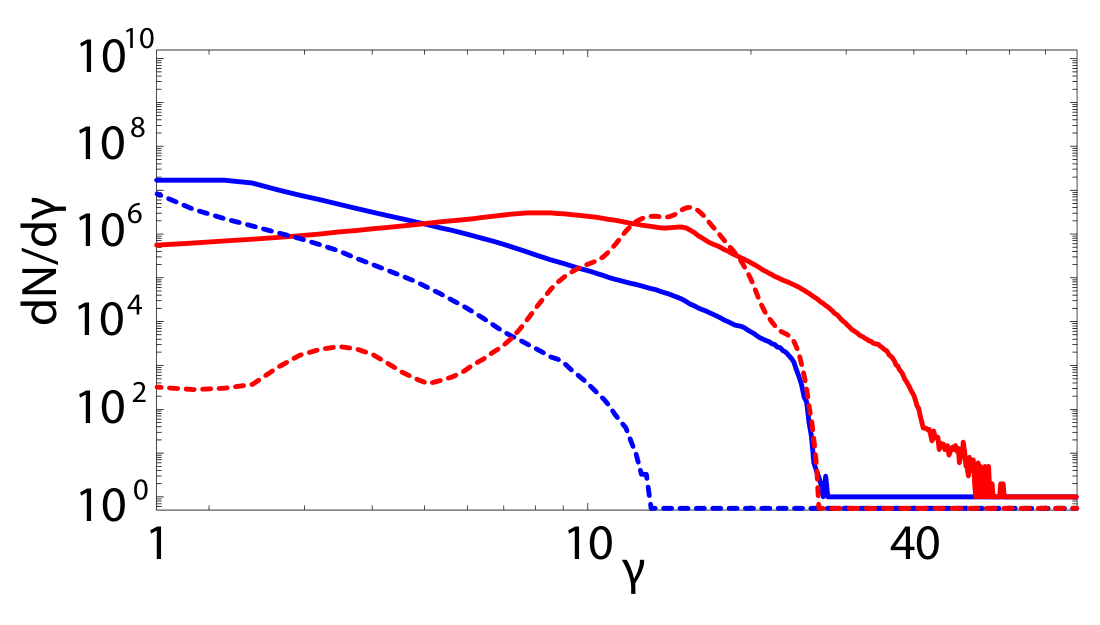}  
\caption{Particle energy distributions of the jet (red) and
 ambient (blue) electrons in and around 
the e$^{\pm}$ jet (a) and e$^{-}$ - i$^{+}$ jet (b) in the two regions $x/\Delta < 550$
(dashed lines) and $x/\Delta > 550$ (solid lines) at $t = 900\, \omega_{\rm pe}^{-1}$.} 
\label{veldis2}
\end{figure*}

Figure \ref{ReconS}a shows the growth of the kKHI and MI (and of the WI)
as well as the generation of two modes of MIs (indicated by the two red dotted lines in Fig. \ref{ByBxz}c), along the jet 
radius ($z$) up to $x/\Delta = 750$. 
These correspond to the groups of accelerated and decelerated bunches of electrons in the calculated phase-space 
distributions shown in, e.g., Fig. \ref{dphas}c. 

Note that without the magnetic field the jets propagate collimated, but instabilities
push the jet particles out of the original jet boundary, as shown in Figs. \ref{Lorentz}a,b, in particular at the non-linear
stage. Moreover, in Figs. \ref{By4}c,d all jets are collimated most probably due to the MI. On the other hand with a toroidal 
magnetic field the MI grows stronger, therefore the jet electrons are stronger collimated.

At around $x/\Delta=700$ the collimated jet structure becomes weakened and stratified, as shown
in Fig. \ref{ReconS}a.  
This indicates the existence of a late non-linear stage of the MI and kKHI. The MI mode near the jet boundary dissipates 
around $x/\Delta = 750$ first, but the inner mode (MI and kKHI) stays longer and dissipates at approximately $x/\Delta = 800$,
which are more easily recognized in Figs. \ref{3DBV}a,c.
It is important to note that by comparing the previous panels of Fig. \ref{3DBV} with Figs. \ref{ByBxz} and \ref{Bacc}, the 3D perspectives indicate distinctive differences in the location of the magnetic flux ropes and vector directions, contrary to the information 
extracted from the 2D projections onto the $y$ -$z$ and $x$ - $z$ planes. 
The latter indicates that the viewing-angle sensitivity should be taken into account for future studies combined with 
observational jet polarization maps, as we will discuss in the following section. In a future work we will use larger set-ups to 
investigate comparable cases with observed astronomical sources.
Figure \ref{ReconS}b shows the strong negative current in the center of the jet and the toroidal magnetic field which is 
opposite to the original direction.

Figure \ref{veldis2} shows the energy distribution of the jet (red) and ambient (blue) electrons in and around 
the e$^{\pm}$ and e$^{-}$ - i$^{+}$  jet in the two  
regions $x/\Delta < 550$ and $x/\Delta > 550$ at  $t = 900\, \omega_{\rm pe}^{-1}$.
We see that the electron acceleration is mostly significant in the non-linear stage ($x/\Delta > 550$).
We observe significant acceleration for both cases reaching Lorentz factors of $\gamma \sim 35-40$. 
Specifically, for the mangetized e$^{-}$ - i$^{+}$  jet, the jet electrons even in the non-linear stage are more efficiently 
accelerated compared to the non-linear stage of the e$^{\pm}$ jet case, reaching a maximum of Lorentz factor $\sim 45$. 
Nevertheless for the e$^{\pm}$ jet, the jet electrons  are mostly accelerated in the non-linear stage reaching a maximum 
Lorentz factor of 40.

Comparing Figs. \ref{veldis2} and \ref{dphas} we deduct that the jet electrons are further 
accelerated once a dissipation of the magnetic fields occurs in the non-linear stage,
as it is also similarly seen in the kinetic simulations of driven magnetized turbulence \citep[e.g.,][]{Zhdankin18}.  
Note that in the latter simulation studies, the turbulent magnetic fluctuations were externally forced into the simulation system 
and therefore were not self-consistent. On the contrary, the turbulent magnetic field in our simulations 
(where we see multiple magnetic flux ropes (e.g. in Figs. \ref{ByBxz}-\ref{3DBV})) is self-consistently created in the relativistic 
jets, through the self-consistent dissipation of the toroidal magnetic field. Previous studies \citep[][]{kowal11,kowal12,lazarian2016} discuss the particle acceleration process in turbulent magnetic reconnection. 
\cite{Comisso18} have investigated particle acceleration at reconnecting current sheets due to stochastic interactions 
with turbulent fluctuations (plasmoids and vertexes).  In our simulations at the non-linear stage, it is comprehensible to claim that 
plasmoids (vertexes) are generated as shown above which may accelerate jet electrons in a similar way.


\section{Summary and discussion} 

We have conducted extensive 3D PIC simulations to study the spatio-temporal evolution of magnetized relativistic electron-positron and electron-ion 
(electron-proton) jets, examining their kinetic instabilities and the associated particle acceleration.  
We investigated the excited kinetic instabilities and the associated  
magnetic fields at the linear and non-linear stage, as they may occur in astrophysical relativistic jets. The dissipation of the magnetic
fields was observed to generate electric fields that are sufficiently strong to further accelerate particles to Lorentz factors of up to around 35.

In this work we used a new jet injection scheme. We injected both e$^{\pm}$
and e$^{-}$ - i$^{+}$ jets, with a co-moving toroidal magnetic field while we used a top-hat jet density profile. 
The current was self-consistently carried by the jet 
particles , $\mathbf{J}=\nabla \times \mathbf{B}$. In order to sustain the 
toroidal magnetic field carried by the jet, the current
was applied at the jet orifice and a motional electric field was applied in order to compensate the bending by the applied toroidal
magnetic field. Both jets were initially moderately magnetized while the ambient medium remained unmagnetized. In the new jet injection scheme we applied the induced electric field $E_{\rm x} =\nabla B_{\phi}$ in order to avoid a non-linear growth. We have run 
the simulations sufficiently long in order to examine the non-linear effects of the jet evolution.

We found that the dominant excited modes of instabilities depend on the different jet compositions. Three different instabilities (WI, MI, kKHI) 
grow in similar timescales, however, depending on the plasma conditions (unmagnetized, different species) some instabilities grow faster
and stronger. Particularly the MI grows faster and stronger with larger mass ratio (4 and 1826) 
and an excitation of the MI or kKHI is weaker for the unmagnetized case. In general,  stronger MI and kKHI grow
in the jets with toroidal magnetic fields. The MI and kKHI are associated with a
quasi-steady electric field ($E_{\rm x}$) for both jet species.
These accompanied electric fields accelerate and decelerate electrons and positrons. Additionally we found that the 
electrons can be further accelerated by the development of twisted and turbulent magnetic fields which 
are generated by the dissipation of the toroidal magnetic fields possibly accompanying reconnection. 
For an e$^{\pm}$ jet, the MI is excited combined with a kKHI, while the produced quasi-steady $E_{\rm x}$ 
modulates the jet particles. For the electron-ion jet, 
the jet electrons are pinched dominantly by the MI at the later linear stage.
Further simulations will be important to decisively confirm possible supplemental acceleration mechanisms with
varying simulation parameters such as, jet radius, magnetization factor, jet density profile, etc.
From the present extensive simulation studies we conclude, that a moderate initial magnetic field can change 
the development of the kinetic plasma instabilities, for different jet species, even if the instabilities significantly amplify 
the magnetic field, here by factors reaching 50.

Figure \ref{econtours} shows that the quasi-steady electric fields accelerate and decelerate jet electrons
at the linear stage and that the consequent magnetic field dissipation accelerates jet electrons further at the non-linear stage (Fig. \ref{Bacc})
for the e$^{\pm}$ jet. 
At the non-linear stage we also observed that jet electrons propagate outside the
jet and that the jet boundaries seem to get distorted by the kKHI. Near the jet head of
the e$^{\pm}$ jet we witness a re-arrangement of  the magnetic field and a general weakening of the currents 
with some vortices (shown in Fig. \ref{Bacc}d). We found that at the initial stages along 
the jet a strong toroidal magnetic field is maintained by a strong $-J_{\rm x}$ current associated with
collimated jet (ambient) electrons.

The magnetic field generated and amplified by different instabilities, in the non-linear stage dissipates and reorganizes into a new topology. 
The 3D magnetic field topology (Fig. \ref{3DBV}) indicates possible reconnection sites and the electrons accelerated in the linear stage
are significantly re-accelerated in the non-linear stage up to a Lorentz factor of 35, accompanied by the dissipation 
of the magnetic field possibly associated with the reconnection, which requires further investigation.

We have identified potential sites of magnetic reconnection in our simulations;
however, an unambiguous determination in 3D is not trivial. The magnetic field structure of a 
reconnection site in 2D simulations consists of $X$ and $O$ shapes which can be recognized rather
easily by the changes of the magnetic field direction and the position of null (very weak) magnetic fields
in the 2D projections. The complex structures of 3D reconnections have been investigated  
in e.g., \cite{parnell10, lalescu15, lazarian2016, bentkamp19, lazarian2020, borissov20}. In order to determine the reconnection 
locations analytically, we would need to investigate the eigenvalues of Jacobian matrix, which is beyond the scope of this work,
for more details, see \cite{Cai07}. 

It is evident that the currents and magnetic structures are very different for e$^{-}$ - i$^{+}$ and e$^{\pm}$ jets. 
These differences arise from the different mobilities of ions and positrons manifested in the polarization signatures of radiation.
It follows that the resulting magnetic field structures are different enough 
to yield distinctive polarizations in VLBI (Very Long Baseline Interferometry) observations of AGN jets at the highest angular resolutions 
\cite[e.g.,][]{gomez2016}.
For example, toroidal magnetic fields (like in our 
simulations) inside and outside of an e$^{-}$ - i$^{+}$ jet contribute to circular polarization.
This may help us to distinguish an e$^{-}$ - i$^{+}$ jet clearly from an e$^{\pm}$ jet, 
at least partially, and in accordance with the present and recent studies, also to establish if and when a possible dissipation of the 
swirling magnetic fields occur in accordance with the present and recent studies \citep[e.g.][]{nishikawa2020}. 

The present work is an important first step from a future follow-up simulation studies with 
larger jet radii that will cover more modes of kinetic instabilities. Therefore, although our present results 
offer important insights, we expect that larger simulations will shed further light onto understanding 
the evolution of different jet species, the dissipation of twisted magnetic fields,
the consequent development of electric and magnetic field instabilities, 
organized magnetic patterns, reconnection events, and particle acceleration 
which are critical to observational astronomy.

VLBI observations of AGN jets are not sufficient to distinguish signatures of
different jet plasma compositions.
Circular polarization (CP) has been detected so far in a limited number
of AGN jets \cite[e.g.,][]{wardle98, homan99, homan01, gabuzda08, thum18}, 
in all cases at very low percentage levels that do not exceed $1-2\%$, in contrast to the relatively high values of linear polarization that usually can reach $30-40\%$. 
CP in AGN jets can be produced either as an intrinsic component of the synchrotron radiation
(for electron-ion jets), or through Faraday conversion of linear polarization \citep{jonesodell77},
the latter appearing to be the prevailing mechanism  \cite[e.g.,][]{wardle98, gabuzda08}. 
Constraints on the jet composition requires not only extremely accurate CP measurements, but also a robust determination of the three-dimensional 
magnetic field structure, in order to quantify the level of Faraday conversion from linear to circular polarization \citep{wardle03}, 
as well as stringent constraints on the energy distribution of radiating particles, see \cite{wardle98}. 

There are currently a lot of efforts, both within the MOJAVE and BU Blazar programs 
\citep[e.g.,][]{jorstad05, pushkarev12}, to generate additional
VLBI images of CP which will help advance our knowledge. In addition, the participation of phased ALMA in VLBI arrays, such 
as the Event Horizon Telescope and the GMVA \cite[e.g.,][]{EHT19, issaoun21, goddi}
provides a significant boost in sensitivity that will allow a better characterization of the CP in AGN jets, and most importantly 
the CP spectra which may hold the key to determine the jet composition.

We propose that several of the complicated magnetic field structures we report in this work could be observed and verified in the near future
with polarimetric VLBI observations at extremely high angular resolutions, if they can resolve the transverse structure of the jet, as with 
space VLBI \cite[e.g.,][]{gomez2016,RA3C84} and with the Event Horizon Telescope \cite[e.g.,][]{EHT19,EHT-3C279}. For example, 
flares may be found associated with reconnection where the dissipation of a significant fraction of the magnetic energy occurs.
Particularly, this may happen when an accelerated particle beam is directed along the 
line of sight \citep[see][]{komissarov2012,mcKinney2012,sironi2015}. Our study might have 
important implications in this context such as for example, the prompt GRB 
emission which could be due to reconnection events\cite[e.g.,][]{Metzger2017,Burns2020,Zhang2018b}. 
Our immediate future work aims to investigate this phenomenon
by combining observations with the temporal and spectral properties of simulation studies by applying
simulations with a systematic parameter survey in order to understand the jet evolution with toroidal magnetic fields 
and with Gaussian (not top-hat) jet density profiles.

\section*{Data availability}
The data underlying this article will be shared on reasonable request to the corresponding author.
The supplementary videos for this publication can be obtained from \href{https://doi.org/10.5281/zenodo.7017747}{doi: 10.5281/zenodo.7017747}

\section*{Acknowledgements}
The authors would like to thank the collaborators Martin Pohl and Jacek Niemiec for the insightful discussions during the development of this work.   
This work was supported by the NASA-NNX12AH06G, NNX13AP-21G, and NNX13AP14G grants.
Recent work was also provided by the NASA through Chandra Award Number GO7-18118X 
(PI: Ming Sun at UAH) issued by the Chandra X-ray Center, which is operated by the SAO for 
and on behalf of the NASA under contract NAS8-03060. The simulations presented in this report have been performed 
by Frontera supercomputer at the Texas Advanced Computing Center under the AST21038: Computational Study of Astrophysical Plasmas,
and also provided by  the NASA through by the grant:  
Nature Of Hard X-rays From A TeV-detected RadioGalaxy
(PI: Ka Wah Wong at SUNY Brockport) issued by the  NuSTAR Guest Observer Cycle 6 2019.
Y.M. is supported by the ERC Synergy Grant ``BlackHoleCam: Imaging the 
Event Horizon of Black Holes'' (Grant No. 610058). 
The work of I.D. has been supported by the NUCLEU project. 
Simulations were performed using Pleiades and Endeavor facilities at NASA Advanced Supercomputing 
(NAS: s2004), using Comet at The San Diego Supercomputer Center (SDSC), and Bridges at the 
Pittsburgh Supercomputing Center, 
which are supported by the NSF. JLG acknowledges the support of the Spanish Ministerio de Econom\'{\i}a y 
Competitividad (grants AYA2016-80889-P, PID2019-108995GB-C21), the Consejer\'{\i}a de Econom\'{\i}a, 
Conocimiento, Empresas y Universidad of the Junta de Andaluc\'{\i}a (grant P18-FR-1769), 
the Consejo Superior de Investigaciones Cient\'{\i}ficas (grant 2019AEP112), and the State Agency 
for Research of the Spanish MCIU through the Center of Excellence Severo Ochoa award for the Instituto de
Astrof\'{\i}sica de Andaluc\'{\i}a (SEV-2017-0709).



\begin{thebibliography}{99}

\bibitem[{{Alves} {et~al.}(2012)}]{alves2012}
Alves, E.P., Grismayer, T., Martins, S.F., Fiuza, F., Fonseca, R.A., \& Silva, L.O.,  \href{https://doi.org/10.1088/2041-8205/746/2/L14} 
{2012,  ApJ, 746, L14}

\bibitem[{{Alves} {et~al.}(2015)}]{alves2015}
Alves, E.P., Grismayer, T., Fonseca, R.A., Silva, L.O., 
\href{https://doi.org/10.1103/PhysRevE.92.021101}{2015, Phys. Rev. E, 92, 021101} 

\bibitem[{{Ardaneh} {et~al.}(2016)}]{Ardaneh16}
Ardaneh, K., Cai, D., Nishikawa, K.I., \href{https://doi.org/10.3847/0004-637X/827/2/124}{2016, ApJ,  827, 124}

\bibitem[{{Ascenzi} {et~al.}(2021)}]{Ascenzi2021}
Ascenzi, S., Oganesyan, G., Branchesi, M., Ciolfi, R.,  \href{https://doi:10.1017/S0022377820001646}
{2021, Electromagnetic counterparts of compact binary mergers, 
Journal of Plasma Physics  87, 845870102}

\bibitem[{{Bentkamp} {et~al.} (2019)}]{bentkamp19}
Bentkamp, L., Lalescu, C. C.,  Wilczek, M., 
\href{https://doi.org/10.1038/s41467-019-11060-9}{2019, Nature communications, Volume 10, id. 3550}

\bibitem[{{Blandford} {et al.} (2019)}] {blandford2019}
Blandford, R., Meier, D., Readhead, A.,  \href{https://www.annualreviews.org/doi/pdf/10.1146/annurev-astro-081817-051948}{2019, ARA \& A, 57, 467}

\bibitem[{{Blandford} {et al.} (2017)}] {blandford2017}
 Blandford, R., Yuan, Y, Hoshino, M.,  Sironi, L., \href{https://doi.org/10.1007/s11214-017-0376-2}{2017, Space Sci Rev, 207, 291}

\bibitem[{{Borissov} {et~al.} (2020)}]{borissov20}
Borissov, A.,  Neukirch, T.,   Kontar, E. P.,   Threlfall, J., and  Parnell, C. E.,
 \href{https://doi.org/10.1051/0004-6361/201936977}{2020, A\&A. 635, A63}

\bibitem[{{Bret} (2009)}]{Bret2009}
Bret, A., 
\href{http://dx.doi.org/10.1088/0004-637X/699/2/990}{2009, ApJ, 699,990}

\bibitem[{{Bret} {et al.} (2010)}]{Bret2010}
Bret, A., Dieckmann, M. E., Gremillet, L.,
\href{https://angeo.copernicus.org/articles/28/2127/2010/}{2010, Ann. Geophys., 28, 2127} 


\bibitem[Bret \& Dieckmann (2010)]{bretdieckmann10} 
Bret, A. and Dieckmann, M. E., 
\href{https://doi.org/10.1063/1.3357336}{2010, Physics of Plasmas 17, 032109}

\bibitem[{{Bromberg} {et~al.} (2011)}]{bromberg2011}
Bromberg, O.,  Nakar, E  Piran, T., and  Sari, R., 
\href{http://dx.doi.org/10.1088/0004-637X/740/2/100}{2011, ApJ, 740, 100}

\bibitem[Buneman(1993)]{tristan} Buneman, O.\ 1993, in {\it Computer Space Plasma Physics:
Simulation Techniques and Software}, Eds.:  Matsumoto \& Omura, \href{https://doi.org/10.1109/27.199533}{Tokyo: Terra, p.67}
 
\bibitem[{{Burns} (2020)}]{Burns2020} Burns, E., 
\href{https://doi.org/10.1007/s41114-020-00028-7}{2020, Living Reviews in Relativity, 23, 4}

\bibitem[{{Cai} {et~al.} (2007)}]{Cai07}
Cai, D.S., Nishikawa, K.-I., Lembege, B.,
2007, in {\it 
Advanced Methods for Space Simulations}, edited by H. Usui and Y. Omura, 
\href{https://www.terrapub.co.jp/e-library/amss/}{TERRAPUB, Tokyo, pp. 145-166}

\bibitem[{{Christie} {et~al.} (2019)}]{Christie19}
Christie, I. M., Lalakos, A., Tchekhovskoy, A., Fernandez, R., Foucart, F., Quataert, E., 
Kasen, D., \href{https://doi.org/10.1093/mnras/stz2552}{2019, MNRAS,  490, 4811} 

\bibitem[{{Comisso} \& {Sironi} (2018)}]{Comisso18}
Comisso, L., Sironi, L.,  \href{https://doi.org/10.3847/1538-4357/ab4c33}{2018, \prl, 121, 255101}

\bibitem[{{Daughton} (2011)}]{daughton2011} Daughton, W., Roytershteyn, V., Karimabadi, H., 
et al. \href{https://doi.org/10.1038/nphys1965}{2011, Physics Nature, 7, 539}

\bibitem[{{de Gouveia Dal Pino} \& {Lazarian} (2005)}]{Pino05} de Gouveia Dal Pino, E. M., 
\& Lazarian, A., \href{https://ui.adsabs.harvard.edu/link_gateway/2005A&A...441..845D/doi:10.1051/0004-6361:20042590}{ 2005,  A\&A 441, 845} 

\bibitem[{{de Gouveia Dal Pino} {et~al.} (2010)}]{Pino10}
de Gouveia Dal Pino, E. M., Piovezan, P. P., Kadowaki, L. H. S.
 \href{http://dx.doi.org/10.1051/0004-6361/200913462}{2010, A\&A, 518, 5}

\bibitem[{{de Gouveia Dal Pino} {et~al.} (2018)}]{Pino18}
de Gouveia Dal Pino, E. M., Alves Batista, R.,  Kowal, G., Medina-Torrejn, T., 
Ramirez-Rodriguez, J. C. 2018,
\href{https://pos.sissa.it/cgi-bin/reader/conf.cgi?confid=329}{BHCB2018}   
\bibitem[Dieckmann et al. (2008)]{dieckmann2008}
Dieckmann, M.E., Shukla, P.K., Drury, L.O.C., 
 \href{https://doi.org/10.1086/525516}{2008, ApJ, 675, 586}

\bibitem[{{Drenkhahn} \& {Spruit} (2002)}]{Drenkhahn2002}
  Drenkhahn, G., \& Spruit, H. C., 
\href{https://ui.adsabs.harvard.edu/link_gateway/2002A&A...391.1141D/doi:10.1051/0004-6361:20020839}{2002, A\&A 391, 1141}


 \bibitem[{{EHT} {Collaboration}
 (2019)}]{EHT19}
EHT Collaboration,  \href{https://doi.org/10.3847/2041-8213/ab0ec7}{2019, ApJL, 875, L1}

\bibitem[{{Fowler} {et~al.} (2019)}]{Fowler19} Fowler, T. K., Li, H., Anantua, R., \href{https://ui.adsabs.harvard.edu/link_gateway/2019ApJ...885....4F/doi:10.3847/1538-4357/ab44bc}{2019, ApJ, 
885, 4}



\bibitem[{{Gabuzda} (2019)}]{gabuzda2019} Gabuzda, D., \href{http://dx.doi.org/10.3390/galaxies7010005}{2019, Galaxies,  7, 5}

\bibitem[{{Gabuzda} {et~al.} (2008)}]{gabuzda08}
Gabuzda, D., et al. \href{https://ui.adsabs.harvard.edu/link_gateway/2008MNRAS.384.1003G/doi:10.1111/j.1365-2966.2007.12773.x}{2008, MNRAS, 384, 1003}

\bibitem[{{Giannios} {et~al.} (2009)}]{giannios2009} Giannios, D., Uzdensky, D. A., \& Begelman,
M. C., \href{https://doi.org/10.1111/j.1745-3933.2009.00635.x}{2009, MNRAS, 395, L29}

\bibitem[{{Giannios} (2010)}]{giannios2010} Giannios, D., \href{https://doi.org/10.1111/j.1745-3933.2010.00925.x}{2010, MNRAS, 408, L46}


\bibitem[{{Giannios} (2013)}]{giannios2013} Giannios, D., 
\href{https://academic.oup.com/mnras/article/431/1/355/1043962}{2013, MNRAS, 431, 355}

\bibitem[{{Giovannini} {et~al.}} (2018)]{RA3C84}
Giovannini, G., et~al. \href{https://ui.adsabs.harvard.edu/link_gateway/2018NatAs...2..472G/doi:10.1038/s41550-018-0431-2}{2018, Nature Astronomy, 2, 472}

\bibitem[{{Goddi} {et~al.} (2021)}]{goddi}
Goddi, C., Martí-Vidal, I., Messias, H., et al.,
 \href{https://doi.org/10.3847/2041-8213/abee6a}{2021, ApJL,910,L14}

\bibitem[{{Granot} (2012)}]{granot2012} Granot, J., \href{https://doi.org/10.1111/j.1365-2966.2012.20489.x}{2012, MNRAS, 421, 2442}

\bibitem[{{Grismayer} {et al.} (2013)}]{grismayer2013}
Grismayer, T.,  Alves, E. P., Fonseca, R. A. and  Silva,  L. O., 
\href{http://dx.doi.org/10.1103/PhysRevLett.111.015005}{Phys. Rev. Lett.  2013, 111, 015005}

\bibitem[{{Gomez} {et al.} (2016)}]{gomez2016} G\'omez, Jos\'e L., Lobanov, Andrei P., Bruni,
Gabriele, Kovalev, Yuri Y., Marscher, Alan P., Jorstad, Svetlana G., Mizuno, Yosuke, Bach, Uwe,
Sokolovsky, Kirill V., Anderson, James M., Galindo, Pablo, Kardashev, Nikolay S., Lisakov, 
Mikhail M., \href{http://dx.doi.org/10.3847/0004-637X/817/2/96}{2016, ApJ, 817, 96}

\bibitem[{{Guo} {et~al.} (2015)}]{guo2015} Guo, F., Liu, Y.-H., Daughton, W., \& Li, H.,  
\href{https://doi.org/10.1088/0004-637x/806/2/167}{2015, ApJ, 806, 167}

\bibitem[{{Guo} {et~al.} (2016a)}]{guo2016a} Guo, F., Li, H., Daughton, W., et al., \href{https://doi.org/10.3847/2041-8205/818/1/l9}{2016a, 
ApJL, 818, L9}

\bibitem[{{Guo} {et~al.} (2016b)}]{guo2016b} Guo, F., Li, H., Daughton, W., Li, X., \& Liu, 
Y.-H., \href{https://doi.org/10.1063/1.4948284}{2016b, PhPl, 23, 0055708}

\bibitem[{{Hawley} {et~al.} (2015)}]{hawley2015} Hawley, J. F., Fendt, C., Hardcastle, M., 
et al.,  \href{https://doi.org/10.1007/s11214-015-0174-7}{2015, Space Sci. Rev., 191, 441}

\bibitem[{{Homan} \& {Wardle} (1999)}]{homan99}
Homan, D.C. \& Wardle. J.F.C.,  \href{http://dx.doi.org/10.1086/301108}{1999, AJ, 118, 1942}

\bibitem[{{Homan} {et~al.} (2001)}]{homan01}
Homan, D.C. et al. \href{http://dx.doi.org/10.1086/321568}{2001, ApJ, 556, 113}

\bibitem[{{Homan} {et~al.} (2006)}]{homan06}
Homan, D.C. et al. \href{http://dx.doi.org/10.1086/500256}{2006, AJ, 131, 1262}

\bibitem[{{Homan} {et~al.} (2009)}]{homan09}
 Homan, D.C. et al. \href{https://ui.adsabs.harvard.edu/link_gateway/2009ApJ...696..328H/doi:10.1088/0004-637X/696/1/328}{2009, ApJ, 696, 328} 

\bibitem[{{Issaoun} {et~al.} (2021)}]{issaoun21}
Issaoun, S., Johnson, M. D., Blackburn, L.,  et al. 2021, \href{https://ui.adsabs.harvard.edu/link_gateway/2021ApJ...915...99I/doi:10.3847/1538-4357/ac00b0}{2021, ApJ, 915,99}

\bibitem[Jaroschek et al. (2005)]{jaroschek2005}
Jaroschek, C.H., Lesch, H., Treumann, R.A., 
 \href{https://doi.org/10.1086/426066}{2005, ApJ, 618, 822}

\bibitem[{{Jones} \& {O'Dell} (1977)}]{jonesodell77}
Jones \& O'Dell, \href{https://ui.adsabs.harvard.edu/link_gateway/1977A%26A....61..291J/ADS_PDF}{1977  A\&A, 61, 291}

\bibitem[{{Jorstad} {et al.}(2005)}]{jorstad05}
Jorstad, S. G., Marscher, A. P., Lister, M. L., et al. \href{https://ui.adsabs.harvard.edu/link_gateway/2005AJ....130.1418J/doi:10.1086/444593}{2005, AJ, 130, 1418}




\bibitem[{{Kadowaki} {et~al.} (2018)}]{Kadowaki18}
Kadowaki, L. H. S., De Gouveia Dal Pino, E. M., Stone, J. M., \href{https://ui.adsabs.harvard.edu/link_gateway/2018ApJ...864...52K/doi:10.3847/1538-4357/aad4ff}{2018, ApJ, 864, 52}

\bibitem[{{Kadowaki} {et~al.} (2019)}]{Kadowaki19} Kadowaki, L. H. S., de Gouveia Dal Pino, 
E. M., Medina-Torrejn, T. E.
\href{https://pos.sissa.it/cgi-bin/reader/conf.cgi?confid=329}{PoS}

\bibitem[{{Kagan} {et~al.} (2013)}]{kagan2013} Kagan, D., Milosavljevic, M., \& Spitkovsky, 
A., \href{https://doi.org/10.1088/0004-637X/774/1/41}{2013, ApJ, 774, 41}

\bibitem[{{Karimabadi} {et~al.} (2014)}]{karimabadi14} Karimabadi, H., Roytershteyn, V., Vu, 
H. X., et al., \href{https://doi.org/10.1063/1.4882875}{2014, PhPl, 21, 2308}

\bibitem[{{Kato} {et~al.} (2010)}]{Kato10}
Kato, T. N.  \&  Takabe, H.,  \href{https://doi.org/10.1063/1.3372138}{2010, ApJ, 721, 828} 
 
 \bibitem[{{Kim} {et~al.} (2020)}]{EHT-3C279}
Kim, J-Y, et al. \href{https://ui.adsabs.harvard.edu/link_gateway/2020A&A...640A..69K/doi:10.1051/0004-6361/202037493}{2020, A\&A, 640, A69}
 
\bibitem[{{Komissarov} (2012)}]{komissarov2012} Komissarov, S. S.,  \href{https://doi.org/10.1111/j.1365-2966.2012.20609.x}{2012, MNRAS, 422, 326}

\bibitem[{{Kowal} {et al.} (2011)}]{kowal11} Kowal, G.,  de Gouveia Dal Pino, E.M.,  Lazarian, A.,
\href{https://doi.org/10.1088/0004-637X/735/2/102}{2011, ApJ. 735, 102} 

\bibitem[{{Kowal} {et al.} (2012)}]{kowal12} Kowal, G.,   de Gouveia Dal Pino, E. M., Lazarian,
A., \href{http://dx.doi.org/10.1103/PhysRevLett.108.241102}{2012, \prl, 108, 241102}

\bibitem[{{Lalescu} {et~al.} (2015)}]{lalescu15}
 Lalescu, C. C.,   Shi, Yi-K.,  Eyink, G. L.,  Drivas, T. D.,   Vishniac, E. T. 
and  Lazarian, A.,  \href{https://doi.org/10.1103/PhysRevLett.115.025001}{2015, \prl 115, 025001} 

\bibitem[{{Lazarian} {et al.} (2016)}]{lazarian2016}
Lazarian A., Kowal G., Takamoto M., de Gouveia Dal Pino E.M., Cho J. 2016, Theory and 
Applications of Non-relativistic and Relativistic Turbulent Reconnection. In: Gonzalez W., 
Parker E. (eds) Magnetic Reconnection. \href{https://link.springer.com/chapter/10.1007/978-3-319-26432-5_11}{Astrophysics and Space Science Library, vol 427. 
Springer, Cham}. \href{https://ui.adsabs.harvard.edu/link_gateway/2016ASSL..427..409L/arxiv:1512.03066}{arxiv:1512.03066}

\bibitem[{{Lazarian} {et al.} (2020)}]{lazarian2020}
 Lazarian, A., Eyink, G. L.,  Jafari, A.,  Kowal, G.,   Li, H.,  Siyao Xu, S.,  and 
Vishniac, E. T., \href{https://doi.org/10.1103/PhysRevLett.115.025001}{2020, Phys. Plasmas, 27, 012305} 



\bibitem[{{Liang} {et al.} (2013a)}]{liang2013a}
Liang,  E., Boettcher, M. and  Smith,  I., 
\href{http://dx.doi.org/10.1088/2041-8205/766/2/L19}{2013a, ApJL, 766, L19}
 
 
\bibitem[{{Liang} {et al.} (2013b)}]{liang2013b}
Liang, E.,  Fu, W.,  Boettcher,  M.,  Smith, I. and  Roustazadeh, P., 
\href{http://dx.doi.org/10.1088/2041-8205/779/2/L27}{2013b, ApJL, 779, L27}


\bibitem[{{MacDonald \& Nishikawa} (2021)}]{MacDonald2021}
 MacDonald, N.R. \& Nishikawa, K-I., 
 \href{https://doi.org/10.1051/0004-6361/201937241}{2021, A\&A, A10, 653} 

\bibitem[{{Matsumoto et al.} (2017)}]{Matsumo17}
Matsumoto, Y., Amano, T., Kato, T. N., Hoshino, M., \href{https://doi.org/10.1103/PhysRevLett.119.105101}{2017, \prl, 119, 105101}

\bibitem[{{McKinney} \& {Uzdensky} (2012)}]{mcKinney2012} McKinney, J. C., \& Uzdensky, 
D. A., \href{https://doi.org/10.1111/j.1365-2966.2011.19721.x}{2012, MNRAS, 419, 573}

\bibitem[{{Meli} \& {Nishikawa} (2021)}]{meli21} Meli, A. \& Nishikawa, K.,
   \href{https://doi.org/10.3390/universe7110450}{2021, Universe, 7, 450}

\bibitem[{{Metzger} (2027)}]{Metzger2017}
Metzger, B. D.,
 \href{https://doi.org/10.1007/s41114-017-0006-z}{2017, Living Reviews in Relativity, 20, 3}

\bibitem[{{Mizuno} {et~al.} (2014)}]{mizuno2014} Mizuno, Y., Hardee, P. E., \& Nishikawa, 
K.-I., \href{https://doi.org/10.1088/0004-637X/784/2/167}{2014, ApJ, 784, 167}

\bibitem[Niemiec et al.(2008)]{niemiec_2008} Niemiec, J., Pohl, M., Stroman, T., \& Nishikawa,
K.-I., \href{https://doi.org/10.1086/590054}{2008, \apj, 684, 1174-1189} 

\bibitem[{{Nishikawa} {et~al.} (2003)}]{nishikawa2003} Nishikawa, K.-I., Hardee, P., Richardson, G., Preece, R., 
Sol, H., Fishman, G.J., 
 \href{https://doi.org/10.1086/377260}{2003, ApJ 595, 555}

\bibitem[{{Nishikawa} {et~al.} (2009)}]{nishikawa2009} Nishikawa, K. -I., J. Niemiec, 
M. Medvedev, H. Sol, P. Hardee,Y. Mizuno, B. Zhang, M. Pohl, M. Oka, D. H. Hartmann,  
\href{https://doi.org/10.1088/0004-637X/698/1/L10}{2009, ApJ, 698, L10}

\bibitem[{{Nishikawa} {et~al.} (2013)}]{nishikawa2013}
Nishikawa, K.-I., Hardee, P., Zhang, B.,, Dutan, I., Medvedev, M., Choi, E.J., Min, K.W., Niemiec, J., Mizuno, Y.,
Nordlund, \AA., Frederiksen, J.T., Sol, H., Pohl, M., Hartmann, D.H., 
\href{https://doi.org/10.5194/angeo-31-1535-2013}{2013, Ann Geophys 31, 1535} 

\bibitem[{Nishikawa et~al.(2014)}]{nishikawa2014}
Nishikawa, K.I.., Hardee, P.E., Du\c{t}an, I., Niemiec, J., Medvedev, M., Mizuno, Y., Meli, A., 
Sol, H., Zhang, B., Pohl, M., et al., \href{https://doi.org/10.1088/0004-637X/793/1/60}{2014, ApJ, 793, 60}

\bibitem[{{Nishikawa} {et~al.} (2016a)}]{nishikawa2016a}
Nishikawa, K.-I., Frederiksen, J. T., Nordlund, \r{A}., et al., \href{https://doi.org/10.3847/0004-637X/820/2/94}{2016a, ApJ, 820, 94}

\bibitem[{{Nishikawa} {et~al.} (2016b)}]{nishikawa2016b} 
Nishikawa, K.-I., Mizuno, Y., Niemiec, J., et al., \href{https://doi.org/10.3390/galaxies4040038}{2016b, Galaxies, 4, 38}

\bibitem[{{Nishikawa} {et~al.} (2017)}]{nishikawa2017} Nishikawa, K.-I., Mizuno, Y., 
G\'omez, J. L., et al., \href{https://doi.org/10.3390/galaxies5040058}{2017, Galaxies, 5, 58}

\bibitem[{{Nishikawa} {et~al.} (2019)}]{nishikawa2019} 
Nishikawa, K.-I., Mizuno, Y., G\'{o}mez, J. L., et al., \href{https://doi.org/10.3390/galaxies7010029}{2019, Galaxies, 7, 29}

\bibitem[{{Nishikawa} {et al.} (2020)}]{nishikawa2020} Nishikawa, K.-I., Mizuno, Y., 
G\'omez, J. L., et al., \href{https://academic.oup.com/mnras/article-abstract/493/2/2652/5734521?redirectedFrom=fulltext}{2020,  MNRAS, 493, 
2652} 

\bibitem[{{Nishikawa} {et al.} (2021)}]{nishikawa2021} Nishikawa, K.-I., Dutan, I, K\"ohn, C., Mizuno, Y., 
\href{https://doi.org/10.1007/s41115-021-00012-0}{Living Reviews in Computational Astrophysics,   2021, 7:1}

\bibitem[{{Oka} {et~al.} (2008)}]{oka2008} Oka, M., Fujimoto, T. K., Nakamura, M., et al. 
\href{http://dx.doi.org/10.1103/PhysRevLett.101.205004}{2008, \prl, 101, 205004}

\bibitem[{{Ruiz} {et~al.} (2018)}]{ruiz2018} Ruiz, M., Shapiro, S. L., Tsokaros, A., \href{https://journals.aps.org/prd/abstract/10.1103/PhysRevD.97.021501}{2018, Phys. Rev. D. 97 (2): 021501}

\bibitem[{{O'Sullivan} {et~al.} (2013)}]{osu2013} O'Sullivan, S.P., McClure-Griffiths, 
N.M., Feain, I.J., Gaensler, B.M., Saul, R.J. \href{https://ui.adsabs.harvard.edu/link_gateway/2013MNRAS.435..311O/doi:10.1093/mnras/stt1298}
{2013, MNRAS, 435, 311}

\bibitem[{{Parnell} {et~al.} (2010)}]{parnell10}
Parnell, C. E., Rhona C. Maclean, R. C.,  Haynes, A. L.  and Galsgaard, K., \href{https://doi.org/10.1017/S1743921311017650}
{Astrophysical Dynamics: From Galaxies to Stars
Proceedings IAU Symposium No. 271, 227, 2010}

\bibitem[{{Pushkarev} {et~al.} (2012)}]{pushkarev12}
Pushkarev, A. B., Hovatta, T., Kovalev, Y. Y., et al., \href{https://ui.adsabs.harvard.edu/link_gateway/2012A&A...545A.113P/doi:10.1051/0004-6361/201219173}{2012, A\&A, 545, AA113}

\bibitem[Rieger \& Duffy(2021)]{rieger2021} Rieger, F.~M. \& Duffy, P.\ \href{https://iopscience.iop.org/article/10.3847/2041-8213/ab563f}{2021, ApJL, 886,, L26} 

\bibitem[Silva et al. (2013)]{silva2003}
Silva, L. O., Fonseca, R. A., Tonge, J. W., Dawson, J. M., Mori, W. B., Medvedev, M. V., 
 \href{https://doi.org/10.1086/379156}{2003, ApJL, 596, L121}

\bibitem[{{Sironi} {et~al.} (2013)}]{sironi2013} Sironi, L.,  Spitkovsky, A., Arons, J., 
\href{https://ui.adsabs.harvard.edu/link_gateway/2013ApJ...771...54S/doi:10.1088/0004-637X/771/1/54}{2013, ApJ, 771, 54}

\bibitem[{{Sironi} \& {Spitkovsky} (2014)}]{sironi2014} Sironi, L. \& Spitkovsky, A., \href{https://doi.org/10.1088/2041-8205/783/1/L21}{2014, 
ApJ, 783, L21}

\bibitem[{{Sironi} {et~al.} (2015)}]{sironi2015} Sironi, L., Petropoulou, M., \& Giannios, D.,
\href{https://doi.org/10.1093/mnras/stv641}{2015, MNRAS, 450, 183}

\bibitem[Spitkovsky (2008a)]{spitkovsky2008a}
Spitkovsky, A., 
 \href{https://doi.org/10.1086/527374}{2008a, ApJL 673, L39}

\bibitem[Spitkovsky (2008b)]{spitkovsky2008b}
Spitkovsky, A., 
 \href{https://doi.org/10.1086/590248}{2008b, ApJL, 682, L5}

\bibitem[{{Tchekhovskoy} (2015)}]{tchekhovskoy2015} Tchekhovskoy, A., \href{https://ui.adsabs.harvard.edu/link_gateway/2015ASSL..414...45T/doi:10.1007/978-3-319-10356-3_3}{2015, ASSL, 414, 45}

\bibitem[{{Thum} {et~al.} (2018)}]{thum18}
Thum, C., et al. \href{https://ui.adsabs.harvard.edu/link_gateway/2018MNRAS.473.2506T/doi:10.1093/mnras/stx2436}{2018, MNRAS, 473, 2506}

\bibitem[{{Uzdensky} (2011)}]{uzdensky2011} Uzdensky, D. A., \href{https://doi.org/10.1007/s11214-011-9744-5}{2011, Space Sci. Rev., 160, 45}

\bibitem[{{Wardle} {et~al.} (1998)}]{wardle98}
Wardle, J.F.C., et al. \href{https://ui.adsabs.harvard.edu/link_gateway/1998Natur.395..457W/doi:10.1038/26675}{1998, Nat, 395, 457}

\bibitem[{{Wardle} \& {Homan} (2003)}]{wardle03}
Wardle, J.F.C. \& Homan, D.C., \href{https://ui.adsabs.harvard.edu/link_gateway/2003Ap&SS.288..143W/doi:10.1023/B:ASTR.0000005001.80514.0c}{2003, Ap\&SS, 288, 143}

\bibitem[Weibel (1959)]{weibel59}
Weibel, E.S.,  (1959) 
\href{https://doi.org/10.1103/PhysRevLett.2.83}{1959, \prl, 2, 83}

\bibitem[{{Wendel} {et~al.} (2013)}]{wendel2013} Wendel, D. E., Olson, D. K., Hesse, M., 
et al., \href{http://dx.doi.org/10.1063/1.4833675}{2013, PhPl, 20, 2105}

\bibitem[{{Zenitani} \& {Hoshino} (2005)}]{zenitani2005} Zenitani, S. \& Hoshino, M., \href{https://ui.adsabs.harvard.edu/link_gateway/2005ApJ...618L.111Z/doi:10.1086/427873}{2005, 
ApJ, 618, L111}

\bibitem[Zhang (2018)]{Zhang2018b}
Zhang, B.,  \href{https://doi:10.1017/9781139226530}{2018,
The Physics of Gamma-Ray Bursts, Cambridge: Cambridge University Press.}


\bibitem[{{Zhang} \& {Yan} (2011)}]{zhang2011} Zhang, B. \& Yan, H., \href{https://ui.adsabs.harvard.edu/link_gateway/2011ApJ...726...90Z/doi:10.1088/0004-637X/726/2/90}{2011, ApJ, 726, 90}
  
\bibitem[{{Zhdankin et al.} (2018)}]{Zhdankin18}
Zhdankin, V., Uzdensky, D. A., Werner, G. R., Begelman, M. C., \href{https://doi.org/10.3847/2041-8213/aae88c}{2018, ApJL, 867, L18}
  
\end{thebibliography}





\bsp	
\label{lastpage}
\end{document}